\documentclass{aa}  

\usepackage{graphicx}
\usepackage{txfonts}
\usepackage{subcaption}
\usepackage{amssymb, amsmath}
\usepackage{placeins}
\usepackage{parskip}
\usepackage[colorlinks=true,allcolors=blue]{hyperref}
\begin{document}

    \title{Tomographic cross correlations between galaxy surveys and the CMB gravitational lensing potential. Effect of the redshift bin mismatch}
    \titlerunning{Tomography - redshift mismatch}

   \author{Chandra Shekhar Saraf
          \inst{1,2}
          \and
          Pawe\l{} Bielewicz\inst{3}
          }
    \authorrunning{C. S. Saraf et al.}
   \institute{Nicolaus Copernicus Astronomical Centre, Polish Academy of Sciences, ul.~Bartycka 18, Warsaw 00-716, Poland\\
              \email{cssaraf@camk.edu.pl}
         \and
            {Korea Astronomy and Space Science Institute, 776 Daedeok-daero, Yuseong-gu, Daejeon 34055, South Korea}\\
         \and
             National Centre for Nuclear Research, ul.~L.~Pasteura 7, Warsaw 02-093, Poland\\
             \email{pawel.bielewicz@ncbj.gov.pl}
             }

   \date{}

 
  \abstract
   {Upcoming surveys of the large-scale structure of our Universe will employ a large coverage area of about half of the sky and will significantly increase the observational depth. With these surveys, we will be able to cross-correlate cosmic microwave background (CMB) gravitational lensing and galaxy surveys divided into narrow redshift bins to map the evolution of the cosmological parameters with redshift.}
   {We study the effect of the redshift bin mismatch of objects that is due to photometric redshift errors in tomographic cross-correlation measurements.}
   {We used the code \texttt{FLASK} to create Monte Carlo simulations of the Vera C. Rubin Observatory Legacy Survey of Space and Time (LSST) and \textit{Planck} CMB lensing convergence. We simulated log-normal fields and divided galaxies into nine redshift bins with the Gaussian and modified Lorentzian photometric redshift errors. To estimate the parameters, we used angular power spectra of CMB lensing and galaxy density contrast fields and the maximum likelihood estimation method.}
   {We show that even with simple Gaussian errors with a standard deviation of $\sigma(z)=0.02(1+z)$, the galaxy auto-power spectra in tomographic bins are offset by $2-15\%$. The estimated cross-power spectra between galaxy clustering and CMB lensing are also biased, with smaller deviations $<5\%$. As a result, the $\sigma_{8}$ parameter deviates between $0.2-1.2\,\sigma$ due to the redshift bin mismatch of the objects. We propose a computationally fast and robust method based on the scattering matrix approach of Zhang et al. (2010), to correct for the redshift bin mismatch of the objects.}
   {The estimates of the parameters in tomographic studies such as the linear galaxy bias, the cross-correlation amplitude, and $\sigma_{8}$ are biased due to the redshift bin mismatch of the objects. The biases in these parameters are alleviated with our scattering matrix approach.}

   \maketitle
%
\section{Introduction}

The large-scale structure in the Universe has been a key contributor to testing the theories of gravity, the nature of dark energy, and dark matter. It gives us vital information about the growth of structure in the Universe and the history of cosmic expansion. Observations from galaxy survey experiments such as the Sloan Digital Sky Survey (SDSS; \citealt{Gunn2006}; \citealt{Strauss2002}), the Wide-field Infrared Survey Explorer (WISE; \citealt{Schlafly2019}; \citealt{Wright2010}), the Kilo-Degree Survey (KiDS; \citealt{Heymans2021}; \citealt{Jong2015}), Hyper Suprime-Cam (HSC; \citealt{Hikage2019}), the Two Micron All Sky Survey (2MASS; \citealt{2MASS2014}), and the Dark Energy Survey (DES; \citealt{DES2018}) have been the torchbearers in unveiling the shortcomings of the standard model of cosmology, the $\Lambda$CDM model. The upcoming galaxy surveys, including the Vera C. Rubin Observatory Legacy Survey of Space and Time (LSST; \citealt{Ivezi2019}; \citealt{LSST2009}), Euclid \citep{Euclid2011}, the Nancy Grace Roman Space Telescope \citep{WFIRST2013}, the Dark Energy Spectroscopic Instrument (DESI; \citealt{DESI2019}), and the Spectro-Photometer for the History of the Universe, Epoch of Reionization, and Ices Explorer (SPHEREx; \citealt{SPHEREX2014}) will play a crucial role in providing a deeper understanding of how our Universe works.

The cosmic microwave background (CMB) gives a view of the early Universe and carries information about the energy components of the Universe. The gravitational lensing of the CMB carries information about the growth of structure at redshift $1-3$. The CMB lensing signal has been valuable in studying the properties of dark energy and is a complementary probe to galaxy clustering and galaxy weak lensing. Cross correlations between the lensing map of the CMB and tracers of large-scale structure can be used to extract information about the evolution of the large-scale gravitational potential. The importance of cross correlations between CMB lensing convergence and galaxy positions in testing the validity of the standard cosmological model or the $\Lambda$CDM model was firmly established with many studies in the past few years (\citealt{Saraf2022}; \citealt{Mitayake2022}; \citealt{Robertson2021}; \citealt{Krolewski2021}; \citealt{Darwish2021} \citealt{Abbott2019}; \citealt{Bianchini2018}; \citealt{Singh2017}; \citealt{Bianchini2016}; \citealt{Bianchini2015}). 

Cross-correlation tomography performed with galaxy samples divided into narrow redshift bins also allows us to map the evolution of the cosmological parameters with redshift. Several tomographic cross-correlation studies (e.g. \citealt{Zhengyi2023}; \citealt{Yu2022}; \citealt{White2022}; \citealt{Pandey2022}; \citealt{Chang2022}; \citealt{Sun2022}; \citealt{Krolewski2021}; \citealt{Hang2021}; \citealt{Marques2020}; \citealt{Peacock&Bilicki2018}; \citealt{Giannantonio2016}) have identified differences in the value of cosmological parameters, such as the $\sigma_{8}$, $\Omega_{m}$, or the combined $S_{8}$ parameter (defined as $S_{8}\equiv\sigma_{8}\sqrt{\Omega_{m}/0.3}$), within the $\Lambda$CDM model. These low-redshift cross-correlation probes consistently measure a lower value for these parameters than the high-redshift CMB-only measurements from the \textit{Planck} satellite \citep{Planck2020VI}, resulting in the so-called $S_{8}$ and $\sigma_{8}-\Omega_{m}$ tensions. Other studies such as \cite{Bianchini2018}, \cite{Amon2018}, \cite{Blake2016}, \cite{Giannantonio2016}, and \cite{Pullen2016} reported consistent deviations in the $D_{g}$ and $E_{g}$ statistics when they tested the $\Lambda$CDM model with different galaxy surveys. The galaxy survey experiments in the future will play a pivotal role in increasing the significance of either tension or agreement in cosmological parameters and in quantifying possible deviations from the standard $\Lambda$CDM model.

In this paper, we present a tomographic cross-correlation study through $300$ Monte Carlo (MC) simulations of the Vera C. Rubin Observatory Legacy Survey of Space and Time (LSST) galaxy survey and \textit{Planck} CMB lensing convergence. We include root mean square scatter in redshifts $\frac{\sigma(z)}{1+z} = 0.02 \text{ and } 0.05$, but exclude catastrophic redshift errors and photometric calibration errors from our simulations. We superficially create the best possible observations in this way. We show that even for these idealistic observations, the photometric redshift uncertainties in the redshift distributions produce biased estimates of the cross correlation when the redshift bin mismatch of objects due to photometric redshift errors is not properly taken into account. Although a few attempts have been made to account for this effect (\citealt{Balaguera2018}; \citealt{Hang2021}), we propose a new fast and efficient method of estimating the tomographic cross correlation in an unbiased way using the scattering matrix formalism introduced by \cite{Zhang2010}.

The paper is organised as follows: section \ref{sec:simulations} presents the simulation setup and the theoretical background, and section \ref{sec:methodology} describes the method for computing the power spectra, reconstructing true redshift distributions, and estimating the parameters. In section \ref{sec:results}, we present the validation results of our simulations and estimation of galaxy auto-power spectra and cross-power spectra between galaxy over-density and CMB lensing convergence. We discuss the correction to the power spectra through the scattering matrix in section \ref{sec:scattering_matrix} and quantify the impact on the linear galaxy bias and on the amplitude of the cross-correlation parameters in section \ref{sec:parameter_estimation}. We study the apparent tension on the $\sigma_{8}$ parameter due to the redshift bin mismatch in section \ref{sec:sigma8}. Finally, we summarise our results in section \ref{sec:summary}.


\section{Simulations and theory}\label{sec:simulations}

\begin{figure*}
    \begin{subfigure}[b]{0.33\linewidth}
        \centering
        \includegraphics[width=\linewidth]{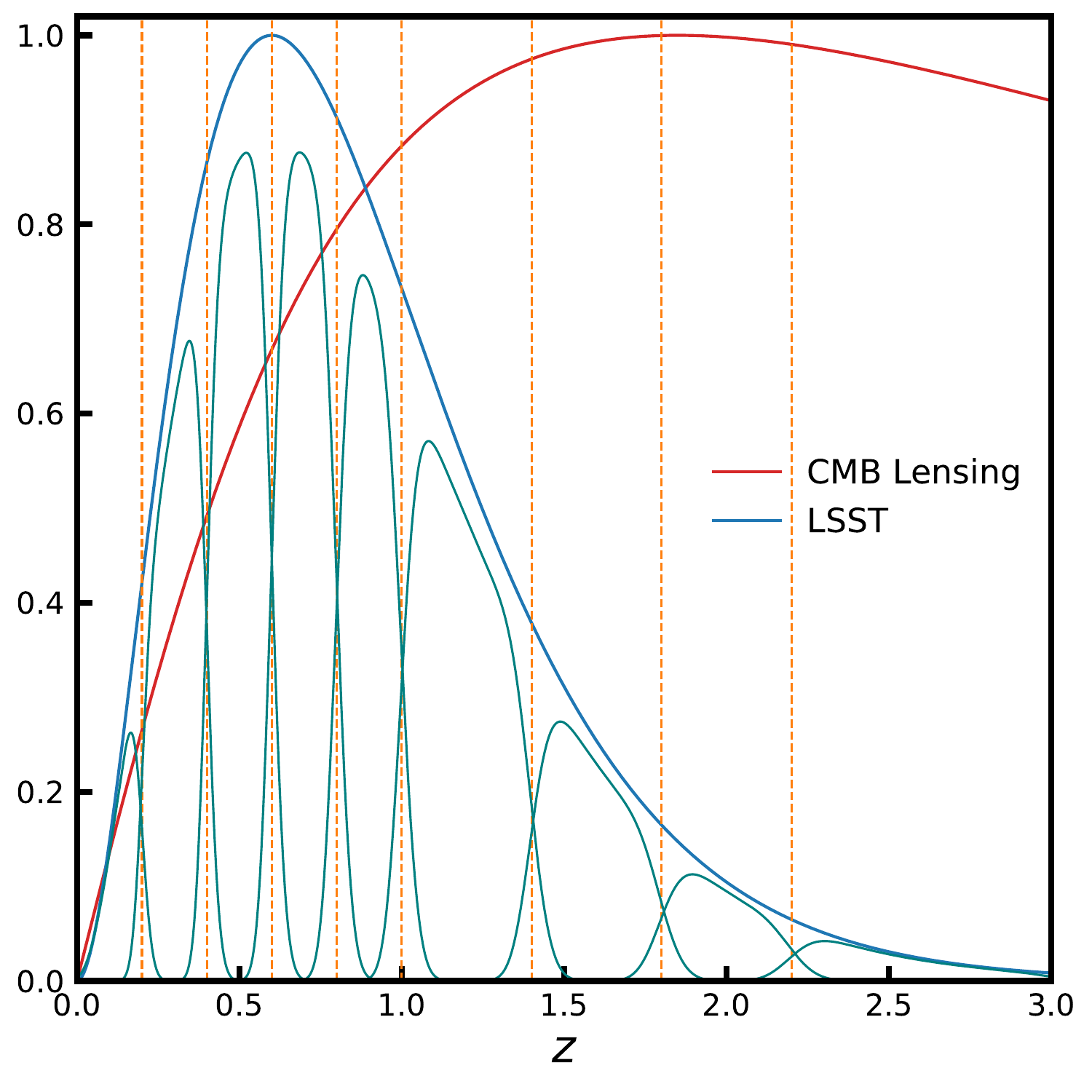}
        \caption{}
    \end{subfigure}%
    \begin{subfigure}[b]{0.33\linewidth}
        \centering
        \includegraphics[width=\linewidth]{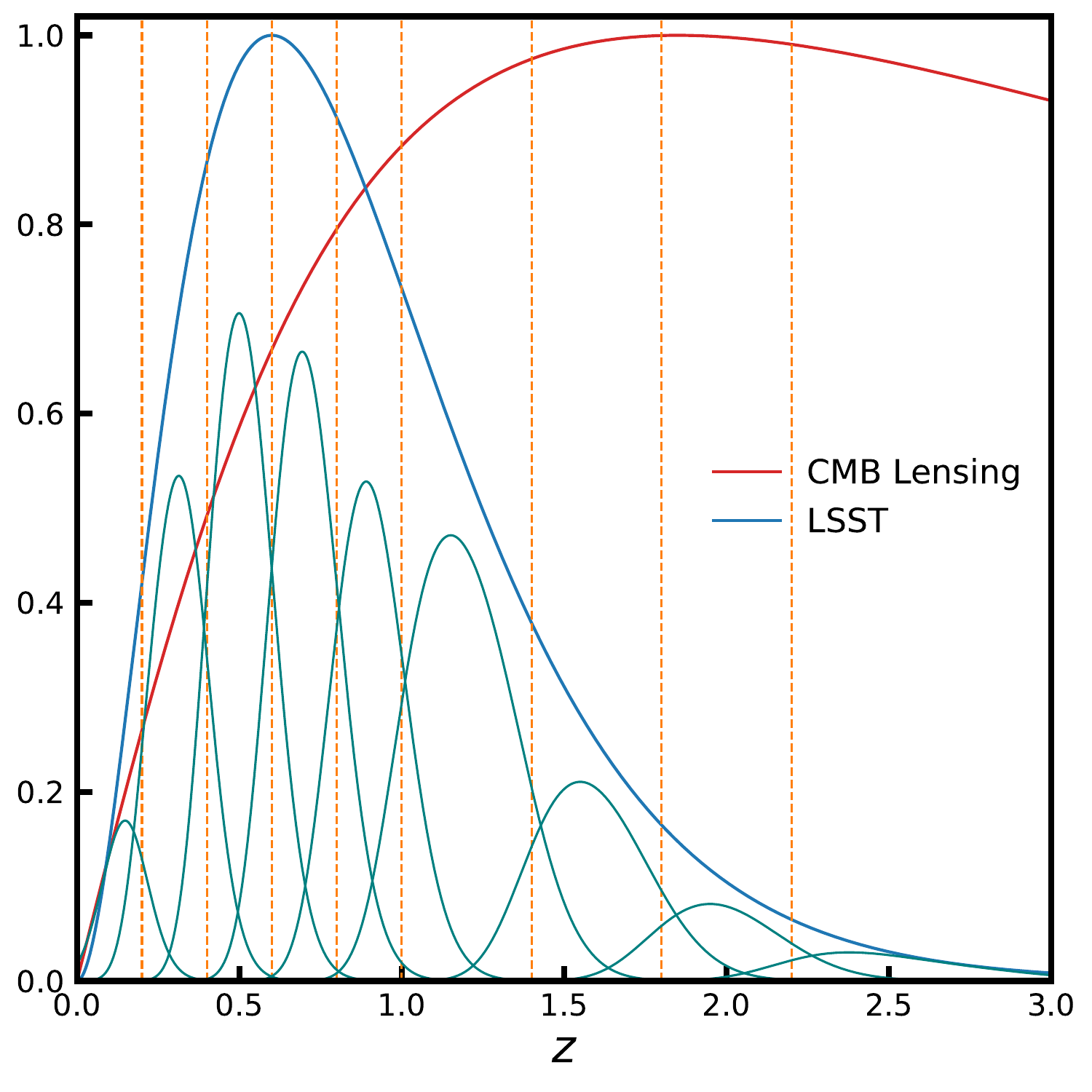}
        \caption{}
    \end{subfigure}%
    \begin{subfigure}[b]{0.33\linewidth}
        \centering
        \includegraphics[width=\linewidth]{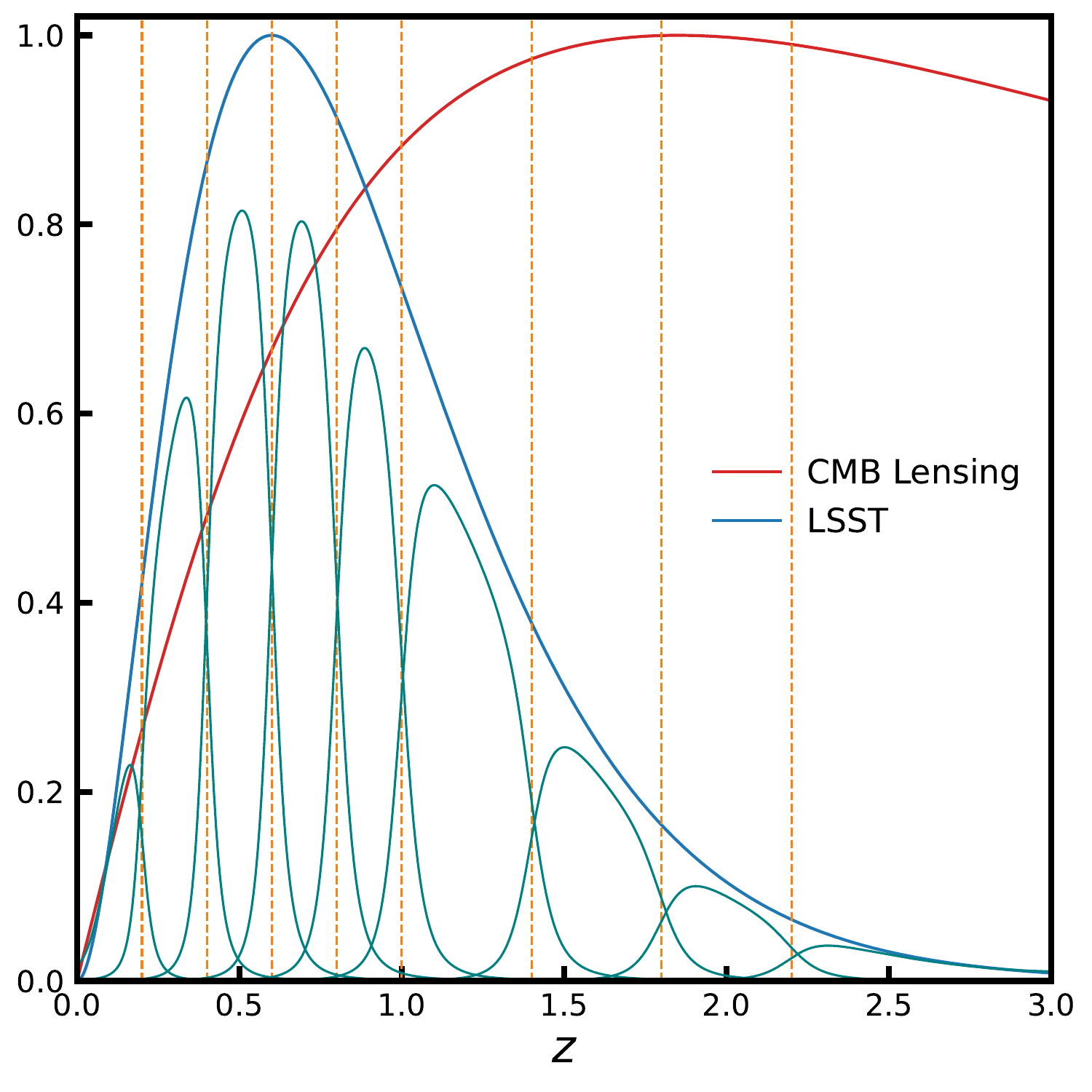}
        \caption{}
    \end{subfigure}%
    \caption{Effect of photometric redshift scatter on redshift distribution. True LSST redshift distribution (solid blue curve) divided into nine photometric redshift bins (shown by the solid green curves) for Gaussian redshift errors with (a) $\sigma_{0}=0.02$, (b) $\sigma_{0}=0.05$, and (c) for modified Lorentzian redshift errors with $\gamma_{0} = 0.02$. The red line denotes the CMB lensing kernel, and the dashed vertical orange lines mark the true redshift bins. The full LSST redshift distribution and CMB lensing kernel are normalised to unit maximum.}
    \label{fig:gal_dist_cmb_kernel}
\end{figure*}

We used the publicly available code \texttt{FLASK} \citep{Flask2016} to generate 300 tomographic MC realisations of correlated log-normal galaxy over-density and CMB lensing convergence fields. The galaxy density followed the LSST photometric redshift distribution profile (\citealt{Ivezi2019}; \citealt{LSST2009}) with a mean redshift at $0.9$ and a mean surface number density of $40 \text{ arcmin}^{-2}$. The simulated CMB lensing convergence field was consistent with \textit{Planck} observations \citep{Planck2020VIII}. The galaxy density field was induced with Poisson noise, and for the CMB convergence, we used the noise power spectrum provided in the \textit{Planck} 2018 data package\footnote{\url{https://pla.esac.esa.int/\#cosmology}}. Due to computational limitations, the sky area covered in our simulations was $2000\text{ deg}^{2}$. However, the results obtained in this study should remain valid for the planned area of the LSST survey if the errors are appropriately scaled by the fraction of sky coverage. The mock galaxy samples were divided into nine disjoint tomographic bins with redshift intervals of $(0.0,0.2,0.4,0.6,0.8,1.0,1.4,1.8,2.2,3.0]$, as marked by the dashed vertical lines in Fig. \ref{fig:gal_dist_cmb_kernel}.

The fiducial angular power spectra for each redshift bin we used to generate correlated maps were computed under the Limber approximation \citep{Limber1953}, 
\begin{equation}
	C_{\ell}^{xy} = \int\limits_{0}^{\infty}\frac{\mathrm{d}z}{c}\frac{H(z)}{\chi^{2}(z)}W^{x}(z)W^{y}(z)P\bigg(k=\frac{\ell+1/2}{\chi(z)},z\bigg),
	\label{eq:power_spectra}
\end{equation} 
where $\{x,y\}\in \{\kappa,g\}$, $\kappa\equiv$ convergence and $g\equiv$ galaxy over-density, $c$ is the speed of light, and $P(k,z)$ is the matter power spectrum generated using the public software CAMB\footnote{\url{https://camb.info/}} \citep{Lewis2000} using the \texttt{HALOFIT} prescription. The kernels $W^{x}(z)$ connects the observables to the underlying total matter distribution. Assuming a flat Universe \citep{Planck2020VI}, the lensing kernel $W^{\kappa}(z)$ and the galaxy kernel $W^{g}(z)$ are expressed as
\begin{align}
    W^{\kappa}(z) &= \frac{3\Omega_{m}}{2c}\frac{H_{0}^{2}}{H(z)}(1+z)\chi(z)\frac{\chi_{*}-\chi(z)}{\chi_{*}}\label{eq:lensing_kernel}\\
    W^{g}(z) &= b(z)\frac{\mathrm{d}N}{\mathrm{d}z}\label{eq:galaxy_kernel}.
\end{align}
Here, $H(z)$ is the Hubble parameter at redshift $z$, and $\chi(z)$ and $\chi_{*}$ are the comoving distances to redshift $z$ and the surface of the last scattering. $\Omega_{m}$ and $H_{0}$ are the present-day values of the matter density parameter and Hubble constant, respectively. $\frac{\mathrm{d}N}{\mathrm{d}z}$ stands for the normalised redshift distribution of galaxies, and $b(z)$ is the linear galaxy bias that relates the galaxy over-density to the total underlying matter density \citep{Fry1993}. We assumed that the magnification bias does not contribute to the galaxy kernel and leave the study of its effect for future work. The expressions for kernels $W^{\kappa}(z)$ and $W^{g}(z)$ given by Eqs.\,(\ref{eq:lensing_kernel}-\ref{eq:galaxy_kernel}) are valid in a flat Universe model (i.e. for zero curvature).

We used a redshift-dependent model of the galaxy bias (\citealt{Solarz2015}; \citealt{Moscardini1998}; \citealt{Fry1996}),
\begin{equation}
    b(z) = 1+\frac{b_{0}-1}{D(z)},
\end{equation}
where we took $b_{0}\equiv b(z=0) =1.3$, and $D(z)$ is the linear growth function normalised to unity at $z=0$
\begin{equation}
    D(z) = \exp\bigg\{ -\int\limits_{0}^{z}\frac{[\Omega_{m}(z')]^{\gamma}}{1+z'}\mathrm{d}z' \bigg\},
    \label{eq:growth function}
\end{equation}
where $\gamma=0.55$ is the growth index for General Relativity \citep{Linder2005}.

For every simulation, the code \texttt{FLASK} produces CMB convergence map and galaxy number count maps for each tomographic bin along with the catalogue of galaxy redshifts. We call these maps true datasets. {We generated photometric redshifts, $z_{p}$, for galaxies in every simulation by drawing positive random values from a Gaussian distribution $\mathcal{N}(z_{t},\sigma_{0}(1+z_{t}))$ (hereafter, Case-I) and from modified Lorentzian function $L(z_{p}; z_{t},\gamma_{0}(1+z_{t}),a)$ (hereafter, Case-II) given by}
\begin{equation}
    L(z_{p};z_{t},\gamma_{0}(1+z_{t}),a) \propto \bigg[ 1+\frac{1}{2a}\bigg(\frac{z_{p}-z_{t}}{\gamma_{0}(1+z_{t})}\bigg)^{2} \bigg]^{-a}.
    \label{eq:mod_lor_distribution}
\end{equation}
{We adopted two different values of $\sigma_{0}$, $0.02 \text{ and } 0.05$, for the Gaussian distribution, and we took $\gamma_{0}=0.02,\,a=2$ for the modified Lorentzian distribution to study the dependence of our results on the strength of the redshift scatter.} The galaxies were again divided into nine tomographic bins based on their photometric redshifts. In Fig. \ref{fig:gal_dist_cmb_kernel}, we show the true LSST redshift distribution (solid blue curve) divided into nine photometric redshift bins (solid green lines) and the CMB lensing kernel (solid red curve) in the redshift range $0\leq z\leq 3$. The green lines represent how the disjoint true redshift bins transform after the photometric redshift errors are introduced. The dashed vertical orange lines mark the boundaries of true redshift bins.

We built galaxy over-density maps from photometric number count maps (hereafter, photometric datasets) with a HEALPix\footnote{\url{https://healpix.jpl.nasa.gov/}} \citep{Gorski2005} resolution parameter $N_{\text{side}}=1024$ using
\begin{equation}
    g(\hat{{n}}) = \frac{n(\hat{{n}})-\overline{n}}{\overline{n}},
    \label{eq:gal_overdensity}
\end{equation}
where $n(\hat{{n}})$ is the number of galaxies at an angular position $\hat{{n}}$, and $\overline{n}$ is the mean number of galaxies per pixel. In Table \ref{tab:phy_prop_all_bins}, we present based on one realisation the comparison of the mean number of objects per pixel (for $N_{\text{side}}=1024$) and median redshift between the true and photometric datasets.

For the simulations and analyses presented in this paper, we adopted the flat $\Lambda$CDM cosmology with the best-fit \textit{Planck} + \textit{WP} + highL + lensing cosmological parameters as described in \cite{Planck2020VI}. Here, \textit{WP} refers to \textit{WMAP} polarisation data at low multipoles, highL are the high-resolution CMB data from the Atacama Cosmology Telescope (ACT), and South Pole Telescope (SPT) and lensing refer to the inclusion of \textit{Planck} CMB lensing data in the parameter likelihood.

\begin{table*}
    \renewcommand{\arraystretch}{1.2}
	\centering
	\caption{Physical properties of true and photometric datasets from a single realisation.}
	\label{tab:phy_prop_all_bins}
	\begin{tabular}{ccccccccc} 
		\hline\hline
            $z$ & $\overline{n}$ (true) & \multicolumn{3}{c}{$\overline{n}$ (photo)} & $\overline{z}$ (true) & \multicolumn{3}{c}{$\overline{z}$ (photo)}\\
            \cline{3-5}\cline{7-9}
             &  & \multicolumn{2}{c}{Gaussian} & Modified Lorentzian &  & \multicolumn{2}{c}{Gaussian} & Modified Lorentzian\\
            \cline{3-4}\cline{7-8}
		      & & $\sigma_{0} = 0.02$ & $\sigma_{0} = 0.05$ & & & $\sigma_{0} = 0.02$ & $\sigma_{0} = 0.05$ & \\
		      \hline
		      $[0.0,0.2)$ & 14.35 & 14.77 & 17.02 & 15.64 & 0.144 & 0.144 & 0.145 & 0.148 \\
            $[0.2,0.4)$ & 57.02 & 57.04 & 57.11 & 57.11 & 0.311 & 0.312 & 0.315 & 0.315 \\
            $[0.4,0.6)$ & 81.78 & 81.60 & 80.68 & 81.15 & 0.502 & 0.503 & 0.506 & 0.507 \\
            $[0.6,0.8)$ & 82.79 & 82.58 & 81.54 & 82.08 & 0.698 & 0.699 & 0.702 & 0.704 \\
            $[0.8,1.0)$ & 70.58 & 70.42 & 69.63 & 70.03 & 0.896 & 0.898 & 0.900 & 0.902 \\
            $[1.0,1.4)$ & 93.31 & 93.19 & 92.60 & 92.93 & 1.178 & 1.180 & 1.183 & 1.185 \\
            $[1.4,1.8)$ & 44.37 & 44.40 & 44.57 & 44.50 & 1.572 & 1.575 & 1.577 & 1.580 \\
            $[1.8,2.2)$ & 18.44 & 18.50 & 18.83 & 18.68 & 1.968 & 1.972 & 1.974 & 1.977 \\
            $[2.2,3.0)$ & 9.58 & 9.68 & 10.22 & 9.98 & 2.472 & 2.471 & 2.471 & 2.475 \\
		\hline
	\end{tabular}
    \tablefoot{$z$ marks the redshift intervals for the tomographic bins, $\overline{n}$ is the mean number of objects per pixel and $\overline{z}$ represents the median redshift of the tomographic bin.}
\end{table*}


\section{Method}\label{sec:methodology}
In this section, we outline the method for computing power spectra from maps and for estimating the true redshift distribution from the photometric redshift distribution.

\subsection{Estimating the power spectra}\label{sec:power_spectra}
We extracted the full sky power spectra from partial sky power spectra for every tomographic bin using the pseudo-$C_{\ell}$ method based on the \texttt{MASTER} algorithm \citep{Hivon2002}, taking into account mode coupling induced by incomplete sky coverage and pixelisation effects. We estimated the full sky power spectra in linearly spaced multipole bins with $\Delta\ell = 30$ between $50\leq\ell\leq 1500$. The noise-subtracted mean full sky power spectrum over $\text{N}_{\text{sim}}$ realisations was computed as \citep{Saraf2022}
\begin{equation}
	 \overline{C}_{L}^{xy}\equiv\langle\hat{C}_{L}^{xy}\rangle = \frac{1}{\text{N}_{\text{sim}}}\sum_{i=1}^{\text{N}_{\text{sim}}}\hat{C}_{L}^{xy,i} - \langle{N}_{L}^{xy}\rangle_{\text{MC}},
	\label{eq:average_spectra}
\end{equation}
where $\hat{C}_{L}^{xy,i}$ represents the full sky power spectrum estimate for the $i$th simulation, and $\langle{N}_{L}^{xy}\rangle_{\text{MC}}$ is the average noise power spectrum from the Monte Carlo simulations. The errors associated with the mean power spectrum were computed from the diagonal of the power spectrum covariance matrix $\text{Cov}_{LL'}^{xy}$ as
\begin{equation}
	\Delta \overline{C}_{L}^{xy} = \bigg(\frac{\text{Cov}_{LL}^{xy}}{\text{N}_{\text{sim}}}\bigg)^{1/2},
	\label{eq:err_simul}
\end{equation}
where
\begin{equation}
	\text{Cov}_{LL'}^{xy} = \frac{1}{\text{N}_{\text{sim}}-1}\sum_{i=1}^{M}(\hat{C}_{L}^{xy,i}-\overline{C}_{L}^{xy})(\hat{C}_{L'}^{xy,i}-\overline{C}_{L'}^{xy}).
	\label{eq:cov_simul}
\end{equation}

\subsection{Estimating the true redshift distribution}
The true redshift distribution $\frac{\mathrm{d}N(z_{t})}{\mathrm{d}z_{t}}$ was estimated from the observed photometric redshift distribution $\frac{\mathrm{d}N(z_{p})}{\mathrm{d}z_{p}}$ and some quantification of errors on the photometric redshifts (from a cross validation with some spectroscopic survey or posteriors from machine-learning methods). These errors are often expressed by conditional probabilities \mbox{$p(z_{p}-z_{t}|z_{t})$} and $p(z_{t}-z_{p}|z_{p})$, which we call photometric redshift error distributions. The true and photometric redshift distributions are then related through \citep{Seth&Rossi2010}
\begin{equation}
    \frac{\mathrm{d}N(z_{t},z_{p})}{\mathrm{d}z_{t}\mathrm{d}z_{p}} = \frac{\mathrm{d}N(z_{t})}{\mathrm{d}z_{t}}p(z_{p}-z_{t}|z_{t}) = \frac{\mathrm{d}N(z_{p})}{\mathrm{d}z_{p}}p(z_{t}-z_{p}|z_{p}).
\end{equation}
Thus, depending on whether we have estimates of $p(z_{p}-z_{t}|z_{t})$ or $p(z_{t}-z_{p}|z_{p})$, the method for estimating the true redshift distribution is called deconvolution or convolution, respectively.

\subsubsection{Convolution method}\label{sec:convolution}
{When $p(z_{t}-z_{p}|z_{p})$ is known, we can estimate $\frac{\mathrm{d}N(z_{t})}{\mathrm{d}z_{t}}$ for each tomographic bin $i$ using the convolution
\begin{equation}
    \frac{\mathrm{d}N^{i}(z_{t})}{\mathrm{d}z_{t}} = \int\mathrm{d}z_{p}\frac{\mathrm{d}N(z_{p})}{\mathrm{d}z_{p}}W^{i}(z_{p})p^{i}(z_{t}-z_{p}|z_{p}),
    \label{eq:true_dist_conv}
\end{equation}
where $\frac{\mathrm{d}N(z_{p})}{\mathrm{d}z_{p}}$ is the observed photometric redshift distribution of the objects, and $W^{i}(z_{p})$ is the window function defining the $i$th redshift bin given by a step function,

\begin{equation}
    W^{i}(z) = \begin{cases}
        1,& \text{if}\quad z^{i}_{\text{min}}\leq z<z^{i+1}_{\text{min}}\\
        0,& \text{otherwise}.
    \end{cases}
    \label{eq:window_function}
\end{equation}

Generally, $p^{i}(z_{t}-z_{p}|z_{p})$ is fitted with parametric functions such as a Gaussian with an assumed zero mean (\citealt{Sun2022}; \citealt{Marques2020}) or a modified Lorentzian (\citealt{Hang2021}; \citealt{Peacock&Bilicki2018}). \cite{Seth&Rossi2010} showed that the quantity $p(z_{t}-z_{p}|z_{p})$ is biased and is not centred on zero. {In our study, we fit the error distributions $p(z_{t}-z_{p}|z_{p})$ with a modified Lorentzian function (for Case-II) and a sum of three Gaussians (for Case-I)},
\begin{equation}
    \mathcal{N}(x) = \sum\limits_{i=1}^{3} A_{i}\exp\bigg[\frac{(x-\mu_{i})^{2}}{2\sigma_{i}^{2}}\bigg],
\end{equation}
where $A,\mu,\sigma$ control the amplitude, mean, and width of the individual Gaussians. The sum of the Gaussians can account for the bias in $p(z_{t}-z_{p}|z_{p})$ and for other characteristic features of the error distributions, such as non-Gaussian wings and a higher peak in the centre. We also checked the fit with a higher number of Gaussians, but did not find any improvement in the quality of fit beyond three Gaussians. For each of the tomographic bins, we fit for $A_{i},\mu_{i},\sigma_{i}$, and estimated the true redshift distribution using Eq.\,(\ref{eq:true_dist_conv}).

\subsubsection{Deconvolution method}\label{sec:deconvolution}
The true redshift distribution $\frac{\mathrm{d}N(z_{t})}{\mathrm{d}z_{t}}$ can be estimated by a deconvolution method when $p(z_{p}-z_{t}|z_{t})$ is known,
\begin{equation}
    \frac{\mathrm{d}N(z_{p})}{\mathrm{d}z_{p}} = \int\mathrm{d}z_{t}\frac{\mathrm{d}N(z_{t})}{\mathrm{d}z_{t}}p(z_{p}-z_{t}|z_{t}).
    \label{eq:true_dist_deconv}
\end{equation}
{We fit $p(z_{p}-z_{t}|z_{t})$ with a single Gaussian for Case-I and with a modified Lorentzian function for Case-II. We find the mean to be consistent with zero, in agreement with the unbiased nature of \mbox{$p(z_{p}-z_{t}|z_{t})$} \citep{Seth&Rossi2010}. For Case-I, we also fit the error distribution with a sum of Gaussians to find no significant improvement in the fit quality.} \cite{Padma2005} proposed a deconvolution method based on Tikhonov regularisation, which lacks a general method for quantifying the impact of the penalty function on the reconstruction of $\frac{\mathrm{d}N(z_{t})}{\mathrm{d}z_{t}}$. We used a different approach for the deconvolution. It is based on the convolution theorem and kernel-based regularisation \citep{Meister2009}. The true redshift distribution in our approach was estimated as
\begin{equation}
    \frac{\mathrm{d}N(z_{t})}{\mathrm{d}z_{t}} = \mathcal{F}^{-1}\bigg[\frac{\mathcal{F}[\frac{\mathrm{d}N(z_{p})}{\mathrm{d}z_{p}}]}{\mathcal{F}[p(z_{p}-z_{t}|z_{t})]}\bigg],
    \label{eq:true_dist_deconv_fourier_method}
\end{equation}
where $\mathcal{F}$ and $\mathcal{F}^{-1}$ represent the Fourier and inverse Fourier transforms, respectively. We show the performance of our deconvolution method through a toy example in Appendix \ref{appndx:validate_deconvolution}.

We estimated the true redshift distribution for the entire redshift range, $0<z\leq 3$. Then, the true redshift distribution for each tomographic bin, $i$, can be expressed as
\begin{equation}
    \frac{\mathrm{d}N^{i}}{\mathrm{d}z_{t}} = \frac{\mathrm{d}N}{\mathrm{d}z_{t}}W^{i}(z_{t}),
    \label{eq:true_dist_binned}
\end{equation}
{where $W^{i}(z_{t})$ is the window function given by Eq.\,(\ref{eq:window_function}).}

The corresponding photometric redshift distribution for every tomographic bin $i$ is given by 
\begin{equation}
    \frac{\mathrm{d}N^{i}(z_{p})}{\mathrm{d}z_{p}} = \int\mathrm{d}z_{t}\frac{\mathrm{d}N^{i}(z_{t})}{\mathrm{d}z_{t}}p^{i}(z_{p}-z_{t}|z_{t}),
    \label{eq:true_dist_deconv_bin}
\end{equation}
where $p^{i}(z_{p}-z_{t}|z_{t})$ is the error distribution for bin $i$. Eq.\,(\ref{eq:true_dist_deconv}) does not follow convolution strictly near $z=0$ because negative redshifts are unphysical. The reconstructed true redshift distribution is therefore inaccurate close to redshift $z=0$, and we expect these inaccuracies to affect the first two tomographic bins to some extent.

\subsection{Galaxy bias and cross-correlation amplitude}\label{sec:likeli_params}
For every tomographic bin, we estimated two parameters: the linear galaxy bias $b$, and the amplitude of the cross-power spectrum $A$. We assumed the linear galaxy bias to be constant in each tomographic bin. This is a fairly good assumption for redshift bins, which are narrow relative to the redshift dependence of the bias. The amplitude of the cross-power spectrum acts as a rescaling of the observed cross-power spectrum to the fiducial theoretical power spectrum. From an unbiased estimation of the cross-power spectrum, we expect $A=1$ in the $\Lambda$CDM cosmology.

The galaxy auto-power spectrum scales as $b^{2}$, while the cross-power spectrum depends on the product $b\times A$. To break this degeneracy, we performed a maximum likelihood estimation on the joint data vector $\hat{C}_{L} = (\hat{C}_{L}^{\kappa g},\hat{C}_{L}^{gg})$, using the likelihood function
\begin{equation}
\begin{split}
	\mathcal{L}&(\hat{C}_{L}|b,A) = \frac{1}{\sqrt{(2\pi)^{N_{L}}det(\text{Cov}_{LL'})}} \times\\
	& \times \text{exp}\bigg\lbrace -\frac{1}{2}[\hat{C}_{L}-C_{L}(b,A)](\text{Cov}_{LL'})^{-1}[\hat{C}_{L'}-C_{L'}(b,A)]\bigg\rbrace,
\end{split}
\label{eq:joint_likeli}
\end{equation}
where $C_{L}(b,A)$ is the joint theoretical power spectrum template, which is defined as $C_{L}(b,A) = (AC_{L}^{\kappa g}(b),C_{L}^{gg}(b))$, and the covariance matrix is given as
\begin{equation}
	\text{Cov}_{LL'} = 
	\begin{bmatrix}
		\text{Cov}_{LL'}^{\kappa g,\kappa g} & \text{Cov}_{LL'}^{\kappa g,gg} \\
            \vspace{0.1mm}\\
		\text{Cov}_{LL'}^{\kappa g,gg} & \text{Cov}_{LL'}^{gg,gg}	.
	\end{bmatrix}
	\label{eq:joint_cov_full}
\end{equation}
The individual elements of $\text{Cov}_{LL'}$ are approximated by the expression \citep{Saraf2022}
\begin{equation}
	\begin{split}
	\text{Cov}_{LL'}^{AB,CD} = \frac{1}{(2\ell_{L'}+1)\Delta\ell f_{\text{sky}}}&\bigg[\sqrt{C_{L}^{AC}C_{L'}^{AC}C_{L}^{BD}C_{L'}^{BD}}\\
+&\sqrt{C_{L}^{AD}C_{L'}^{AD}C_{L}^{BC}C_{L'}^{BC}}\bigg]\delta_{LL'},
	\end{split}
	\label{eq:error_covariance}
\end{equation}
where $\{A, B, C, D\}\in\{\kappa,g\}$, $\Delta\ell$ is the multipole binwidth, and we assumed a common sky coverage fraction between the CMB convergence and the galaxy density fields. Eq.\,(\ref{eq:joint_cov_full}) was used to estimate the parameters from a single realisation. In section \ref{sec:parameter_estimation}, we use the average power spectra from $300$ simulations to estimate the parameters. We then divide the covariance matrix $\text{Cov}_{LL'}$ by the total number of simulations $\text{N}_{\text{sim}}$ to obtain the covariance matrix for the average.

We used flat priors ${b \in [0,10]}$ and ${A \in [-5,5]}$ to estimate the parameters, and the remaining cosmological parameters were kept constant with values from our fiducial background cosmology described in section \ref{sec:simulations}. To effectively sample the parameter space, we used the publicly available software package \texttt{EMCEE} \citep{EMCEE2013}. The best-fit value of the parameters are medians of their posterior distributions, with $\pm 1\,\sigma$ errors being the $16$th and $84$th percentile, respectively.


\section{Results}\label{sec:results}

In this section, we present the results of estimating the true redshift distribution and power spectra for datasets simulated by the code FLASK. We searched for any systematics in our estimates from the datasets with photometric redshift errors. The results of the tests for datasets without errors are presented in Appendix \ref{appndx:validate_true_flask}.

\subsection{Estimating the true redshift distribution}

\begin{figure*}[!ht]
    \begin{subfigure}[b]{0.33\linewidth}
        \centering
        \includegraphics[width=\linewidth]{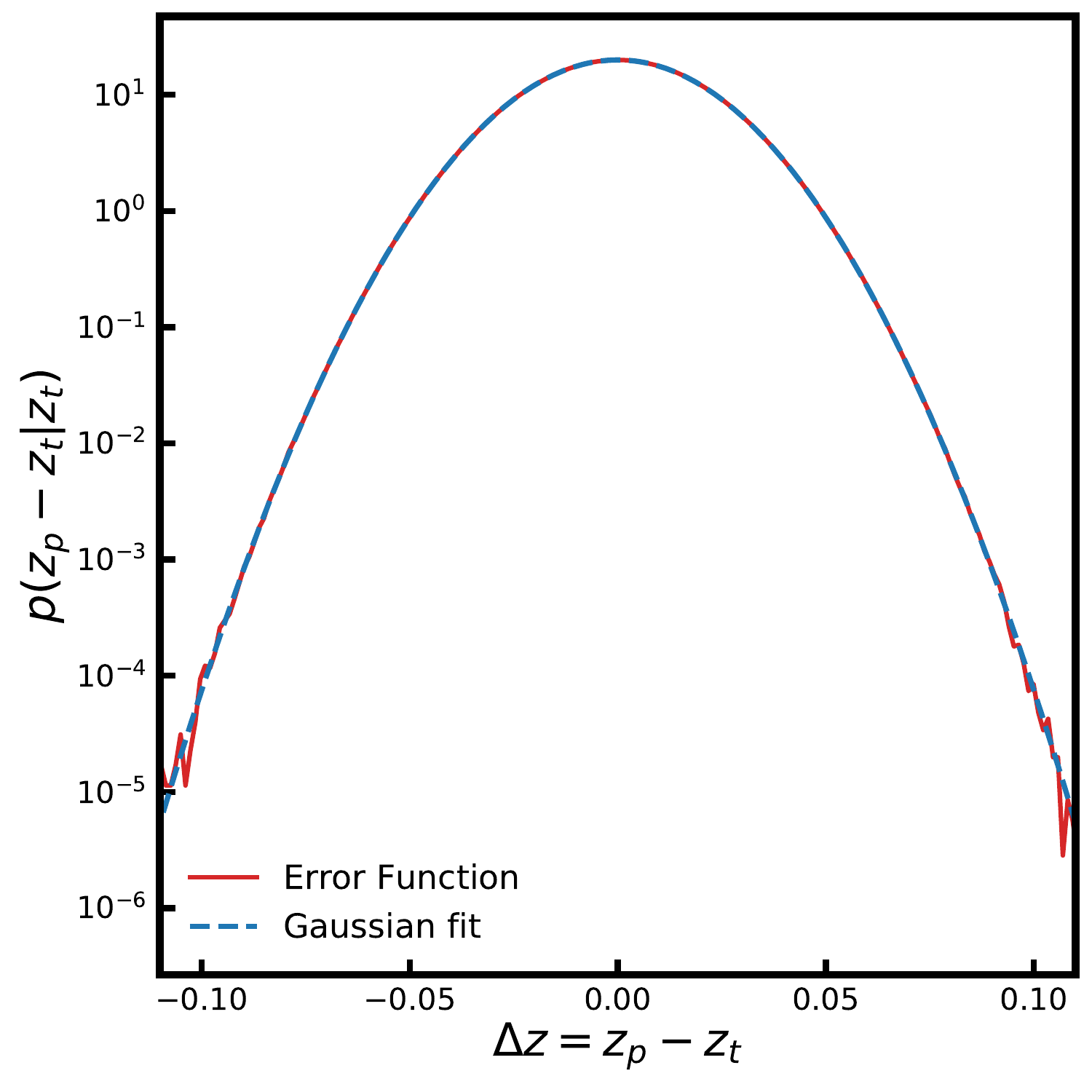}
        \caption{}
    \end{subfigure}%
    \begin{subfigure}[b]{0.33\linewidth}
        \centering
        \includegraphics[width=\linewidth]{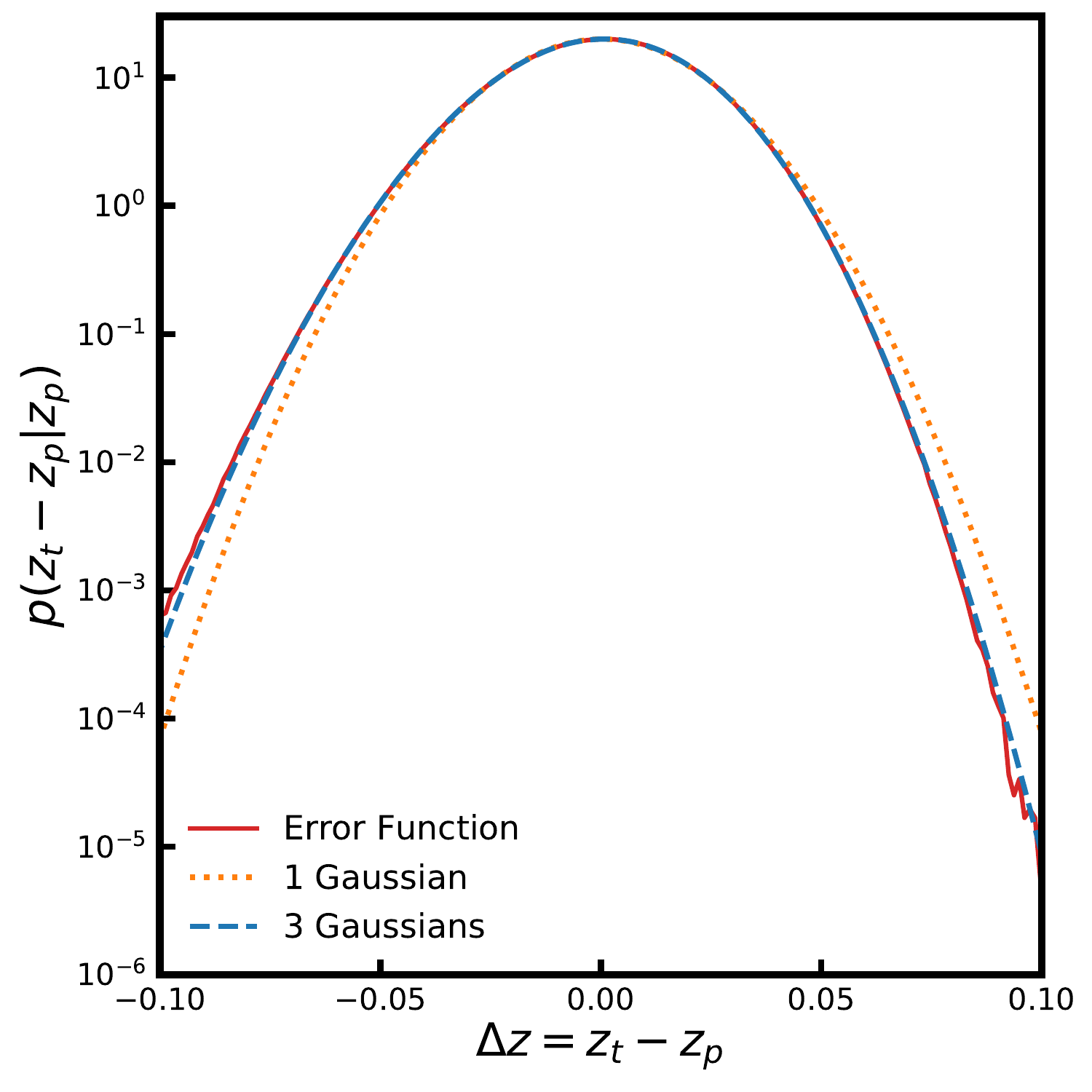}
        \caption{}
    \end{subfigure}%
    \begin{subfigure}[b]{0.33\linewidth}
        \centering
        \includegraphics[width=\linewidth]{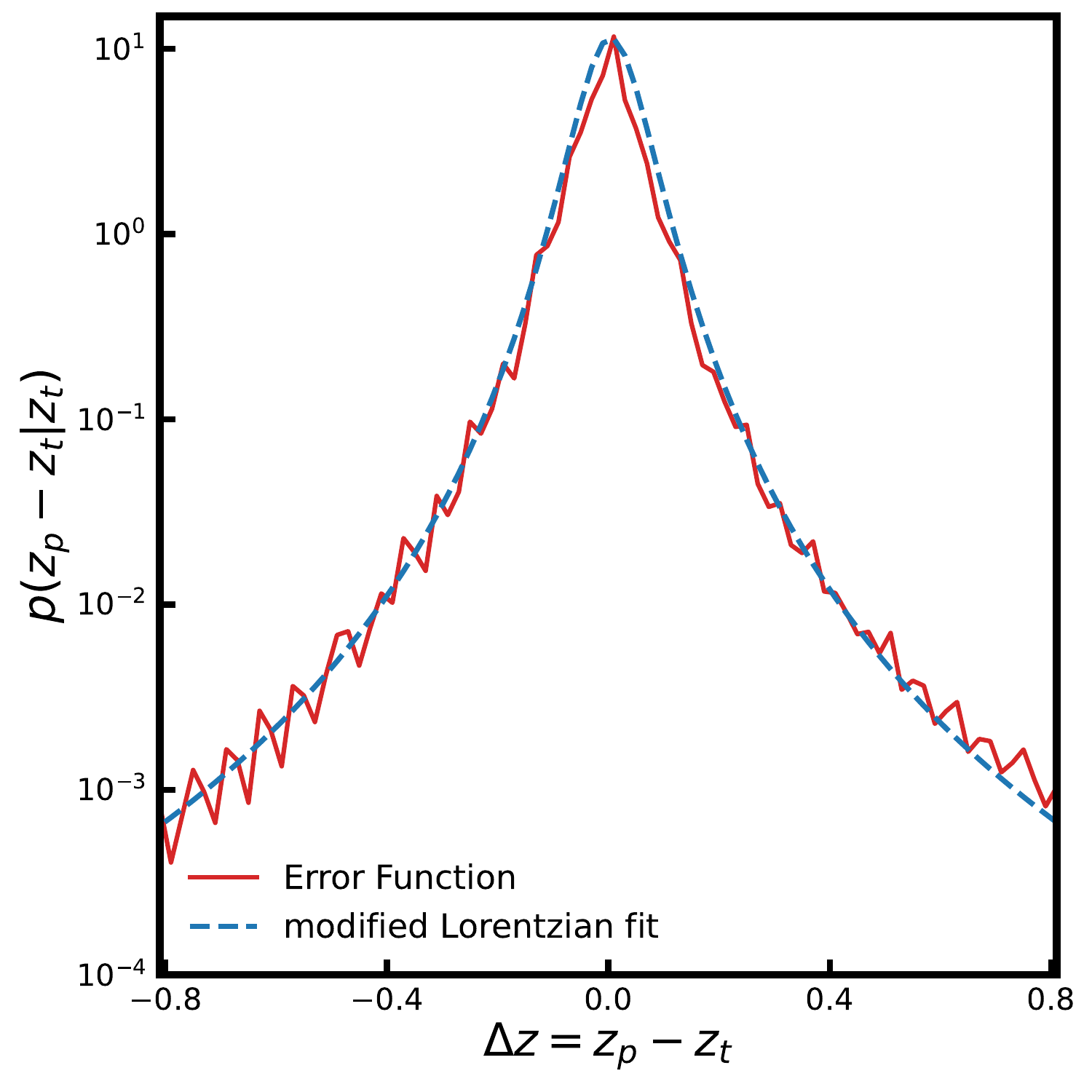}
        \caption{}
    \end{subfigure}
    \caption{Modelling the redshift error distribution with parametric functions. (a) Fit made to the error distribution $p(z_{p}-z_{t}|z_{t})$ with $\sigma_{0}=0.02$ (\textit{solid red line}) using a single Gaussian (\textit{blue dashed line}). (b) \textit{Right}: Fit made to the error distribution $p(z_{t}-z_{p}|z_{p})$ with $\sigma_{0}=0.02$ (\textit{solid red line}) using a single Gaussian (\textit{dotted orange line}), and sum of three Gaussians (\textit{dashed blue line}). (c) Fit made to the error distribution $p(z_{p}-z_{t}|z_{t})$ with $\gamma_{0}=0.02$ (\textit{solid red line}) using a single modified Lorentzian function (\textit{dashed blue line}).}
    \label{fig:err_func_fit_true_and_photo}
\end{figure*}
The estimation of the true redshift distribution requires fitting the error function (i.e. either $p(z_{t}-z_{p}|z_{p})$ or $p(z_{p}-z_{t}|z_{t})$) with parametric functions. {We used a single Gaussian for $p(z_{p}-z_{t}|z_{t})$ and the sum of three Gaussian to fit $p(z_{t}-z_{p}|z_{p})$ for Case-I. For Case-II, we used a single modified Lorentzian function to fit both $p(z_{t}-z_{p}|z_{p})$ and $p(z_{p}-z_{t}|z_{t})$}. {Fig. \ref{fig:err_func_fit_true_and_photo}a and Fig. \ref{fig:err_func_fit_true_and_photo}b show the quality of fit to the error functions with $\sigma_{0}=0.02$ and compares the single Gaussian with the three Gaussian fit of $p(z_{t}-z_{p}|z_{p})$ for Case-I.} The sum of three Gaussians provides a significantly better fit by accurately capturing the non-Gaussian wings of the error function. {Fig. \ref{fig:err_func_fit_true_and_photo}c shows the fit made to the error distribution $p(z_{p}-z_{t}|z_{t})$ with a single modified Lorentzian function (Case-II).} The uncertainties in these fits remain within $0.25\%$, and we do not expect these sub-percent uncertainties to bias the power spectra or the parameter estimates.

{In Fig. \ref{fig:rel_err_true_dist_nsim_300}, we show the relative error, averaged over 300 simulations, of the redshift distribution reconstructed using the deconvolution method for the Gaussian error distribution with $\sigma_{0} = 0.02, 0.05$ and for the modified Lorentzian error distribution with $\gamma_{0}=0.02$.} The reconstructed redshift distribution is within $\sim 2\%$ for the entire redshift range. The strongest deviations occur near boundaries and are possibly due to sharp cuts in the redshift distribution at $z=0$ and $z=3$ (section \ref{sec:deconvolution}).

\begin{figure*}
    \begin{subfigure}[b]{0.33\linewidth}
        \centering
        \includegraphics[width=\linewidth]{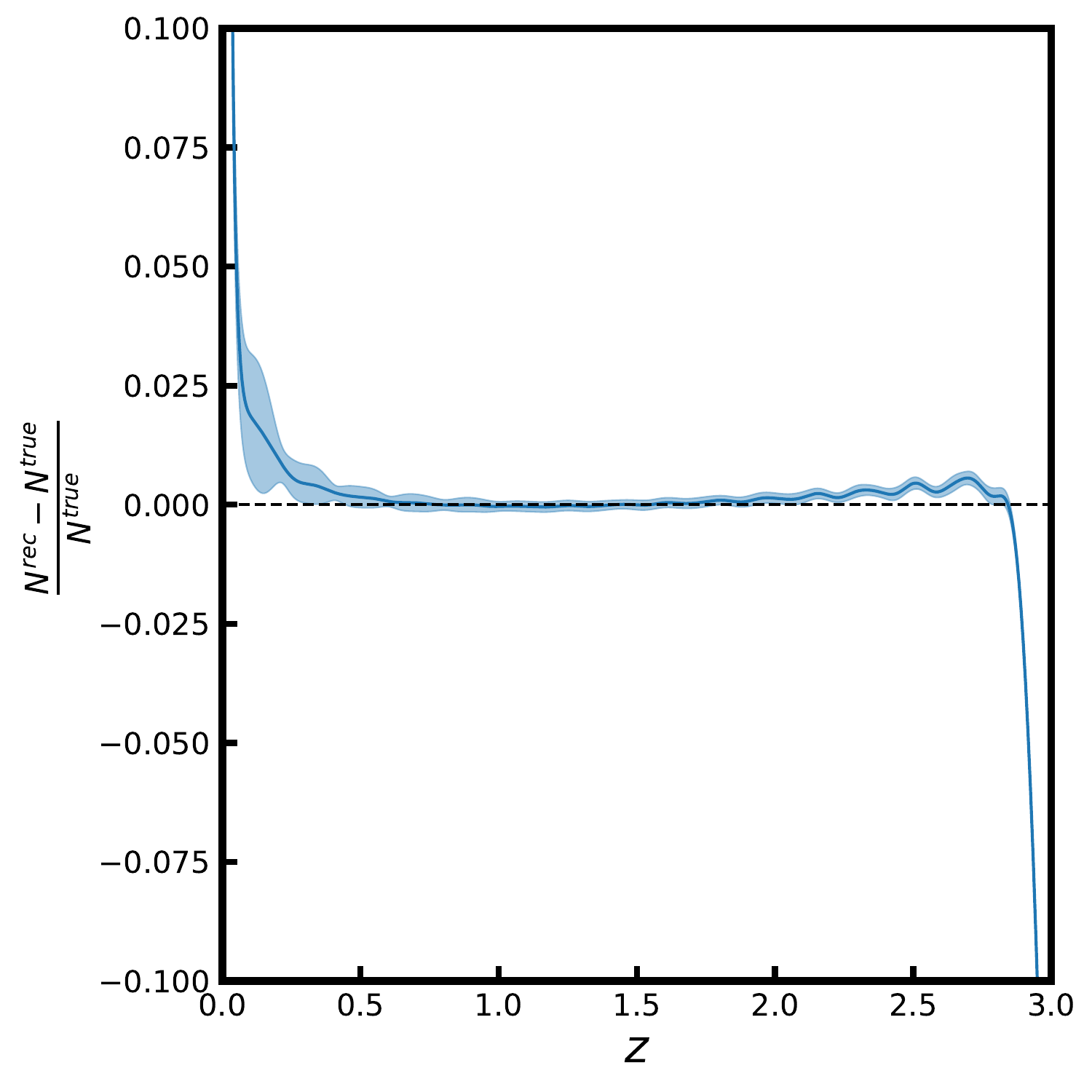}
        \caption{}
    \end{subfigure}%
    \begin{subfigure}[b]{0.33\linewidth}
        \centering
        \includegraphics[width=\linewidth]{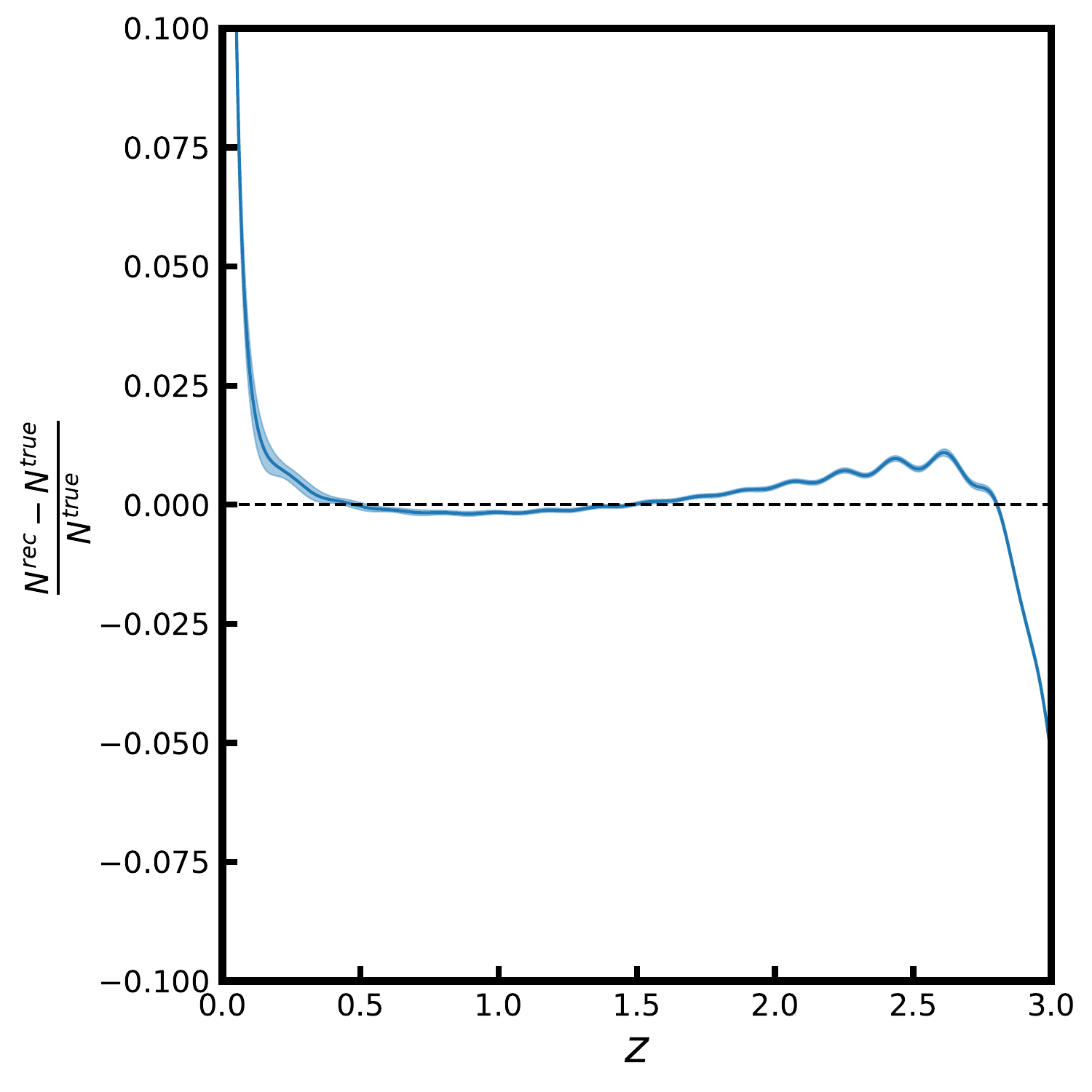}
        \caption{}
    \end{subfigure}%
    \begin{subfigure}[b]{0.33\linewidth}
        \centering
        \includegraphics[width=\linewidth]{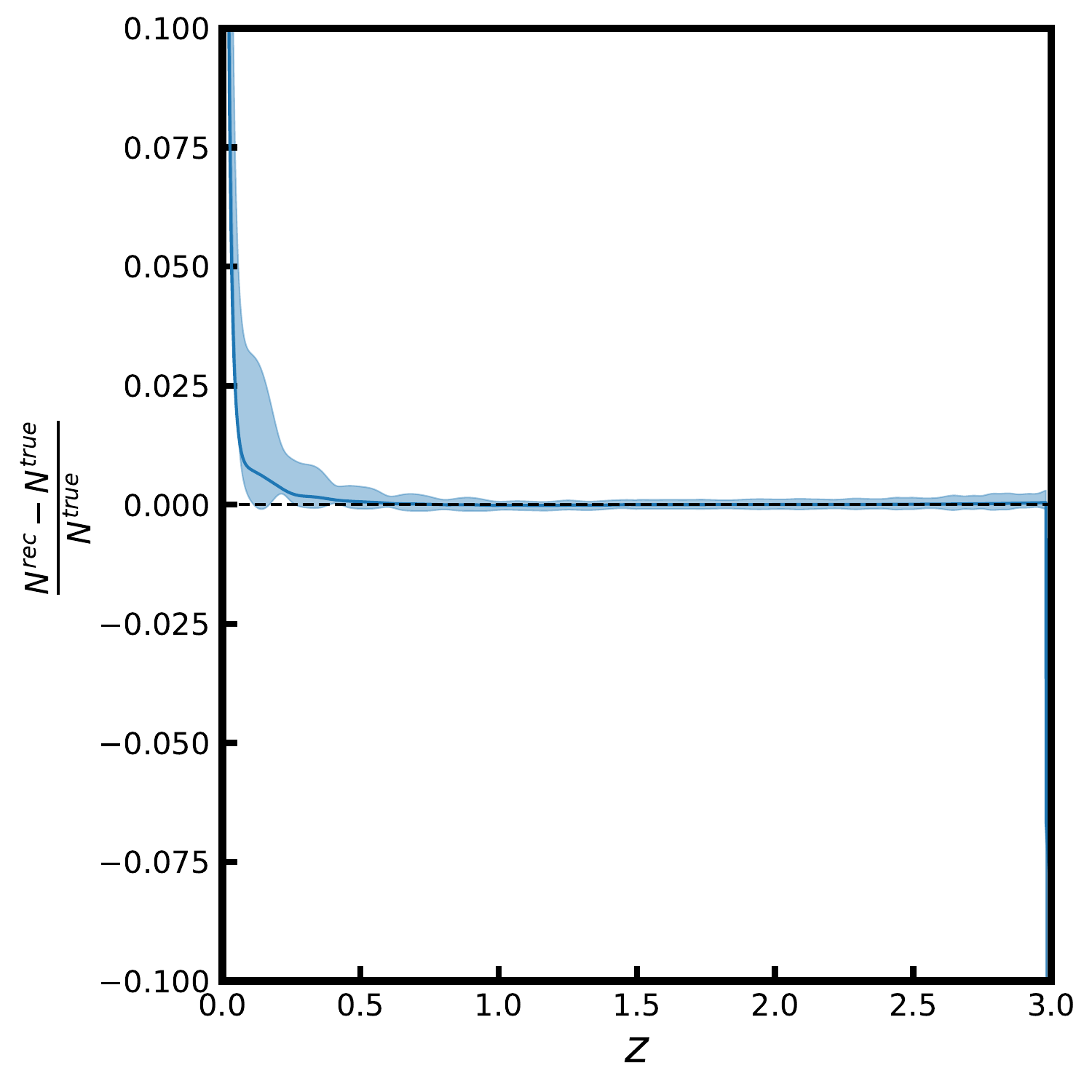}
        \caption{}
    \end{subfigure}
    \caption{{Average relative difference between the redshift distributions reconstructed using the deconvolution method and the true distribution from 300 simulations for (a) Gaussian error distribution with $\sigma_{0}=0.02$, (b) a Gaussian error distribution with $\sigma_{0}=0.05$, and (c) a modified Lorentzian error distribution with $\gamma_{0}=0.02$. The shaded region represents the $1\sigma$ deviations in reconstruction between 300 simulations.}}
    \label{fig:rel_err_true_dist_nsim_300}
\end{figure*}

\subsection{Power spectra from photometric datasets}\label{sec:results_power_spectra}

\begin{figure*}
    \begin{subfigure}[b]{0.33\linewidth}
        \centering
        \includegraphics[width=\linewidth]{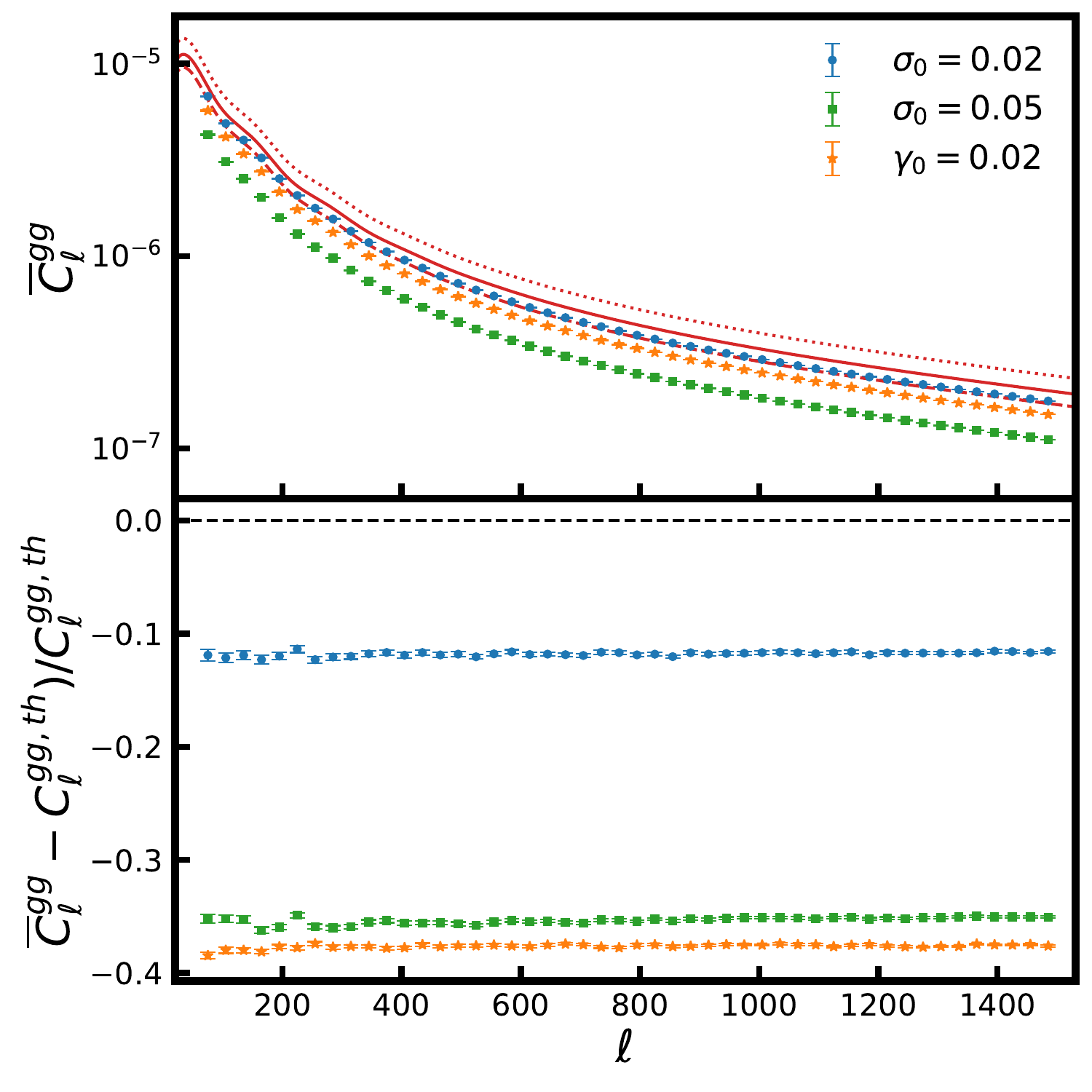}
    \end{subfigure}%
    \begin{subfigure}[b]{0.33\linewidth}
        \centering
        \includegraphics[width=\linewidth]{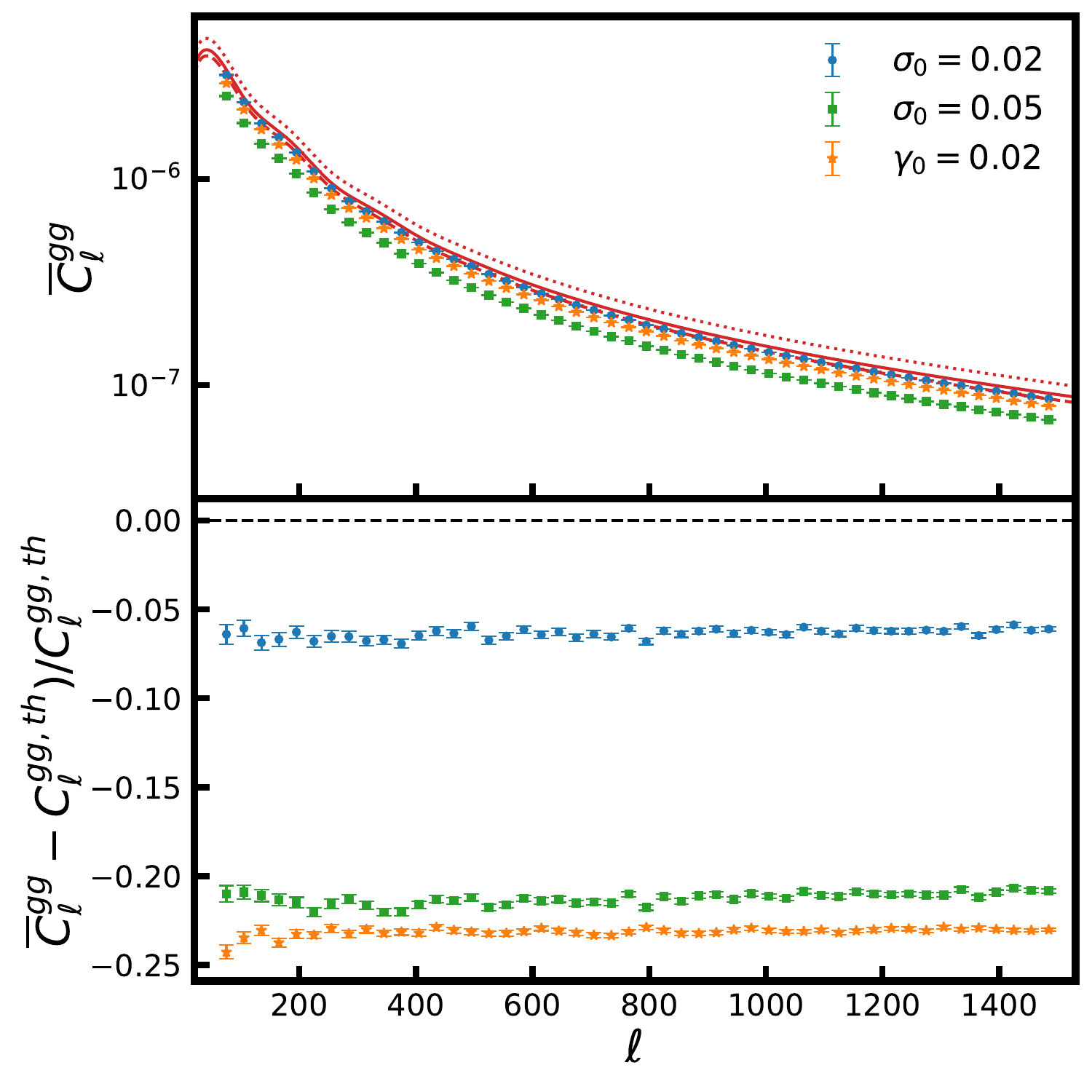}
    \end{subfigure}%
    \begin{subfigure}[b]{0.33\linewidth}
        \centering
        \includegraphics[width=\linewidth]{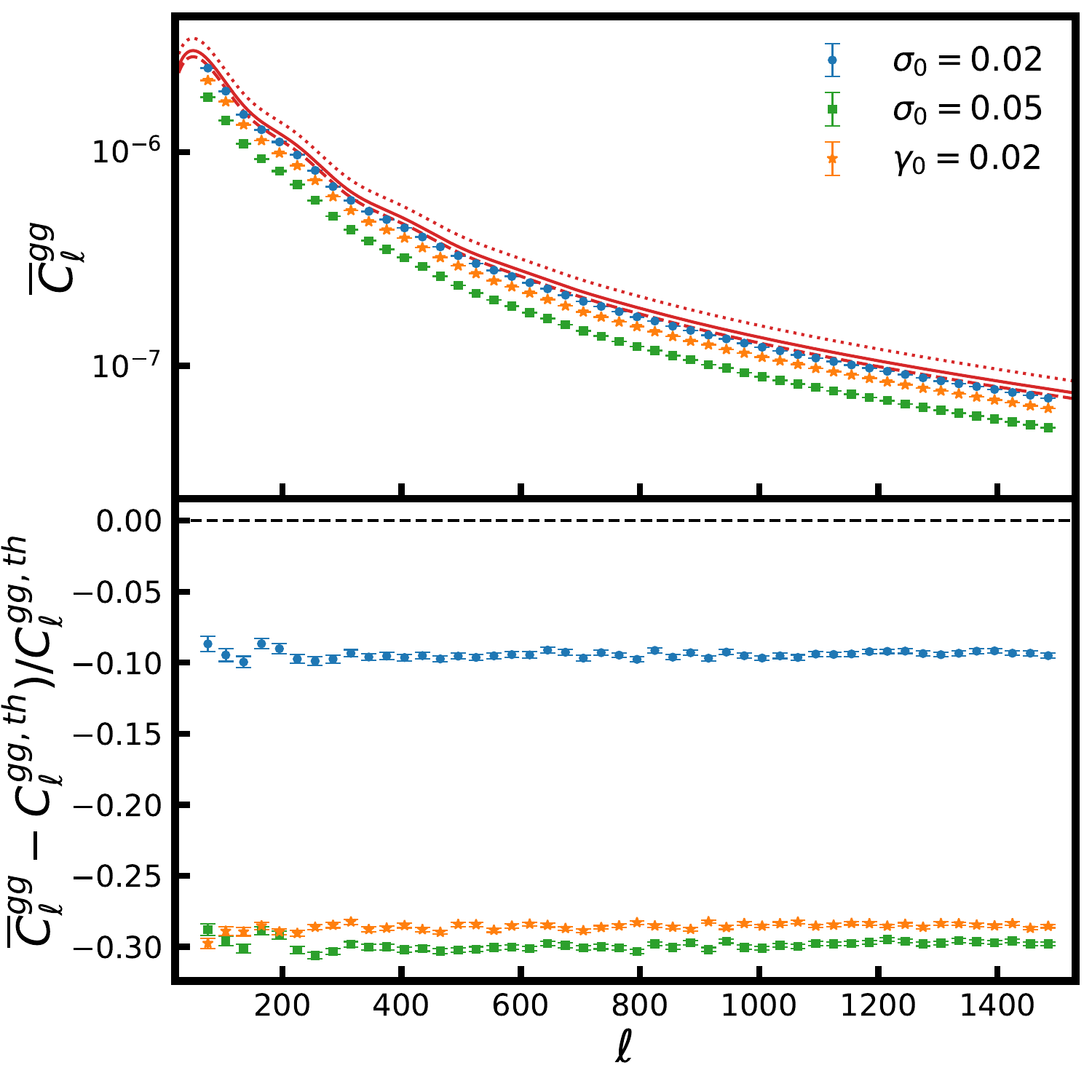}
    \end{subfigure}\\[1ex]
    \begin{subfigure}[b]{0.33\linewidth}
        \centering
        \includegraphics[width=\linewidth]{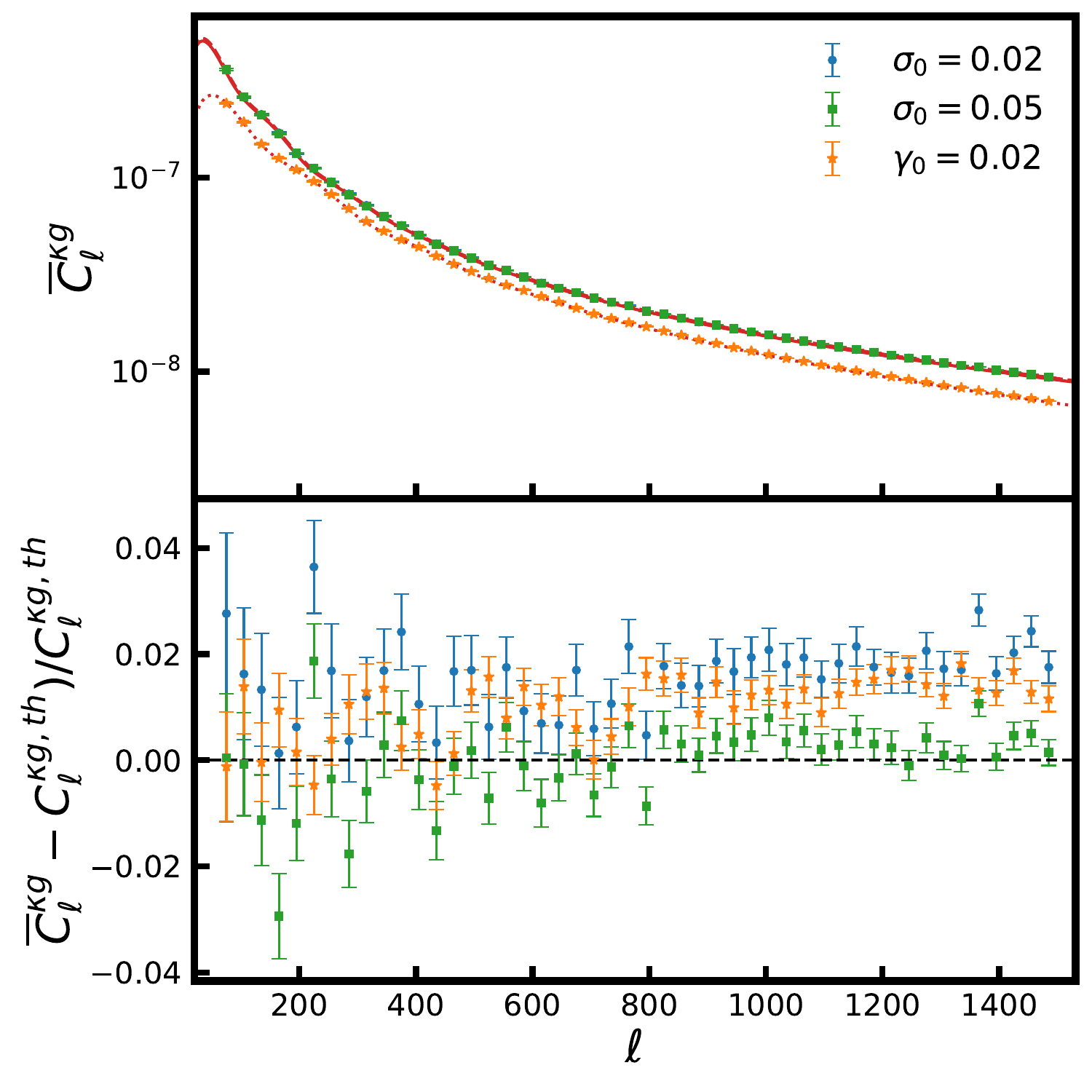}
        \captionsetup{labelformat=empty}
        \caption{\large{Bin 5 ($0.8\leq z<1.0$)}}
    \end{subfigure}%
    \begin{subfigure}[b]{0.33\linewidth}
        \centering
        \includegraphics[width=\linewidth]{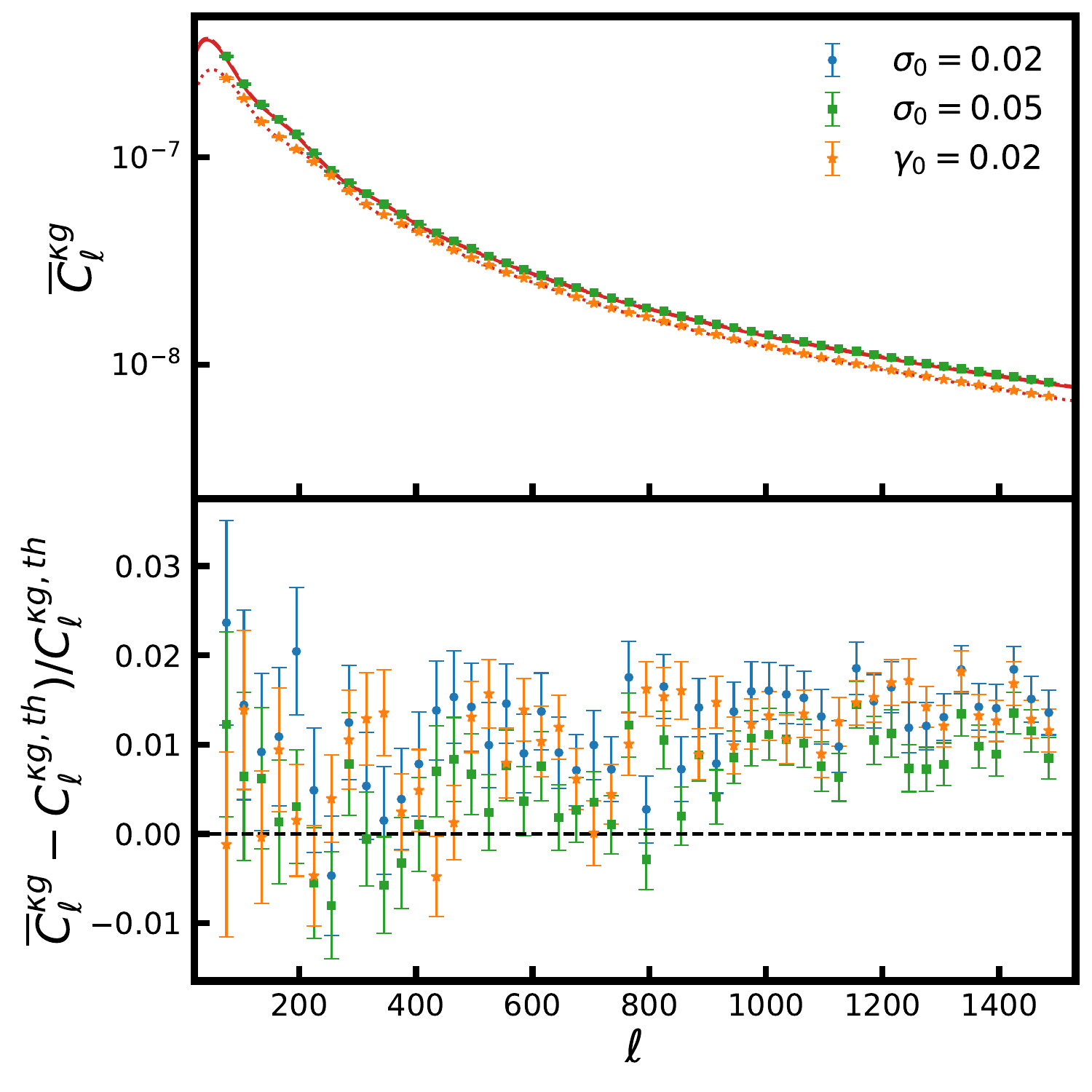}
        \captionsetup{labelformat=empty}
        \caption{\large{Bin 6 ($1.0\leq z<1.4$)}}
    \end{subfigure}%
    \begin{subfigure}[b]{0.33\linewidth}
        \centering
        \includegraphics[width=\linewidth]{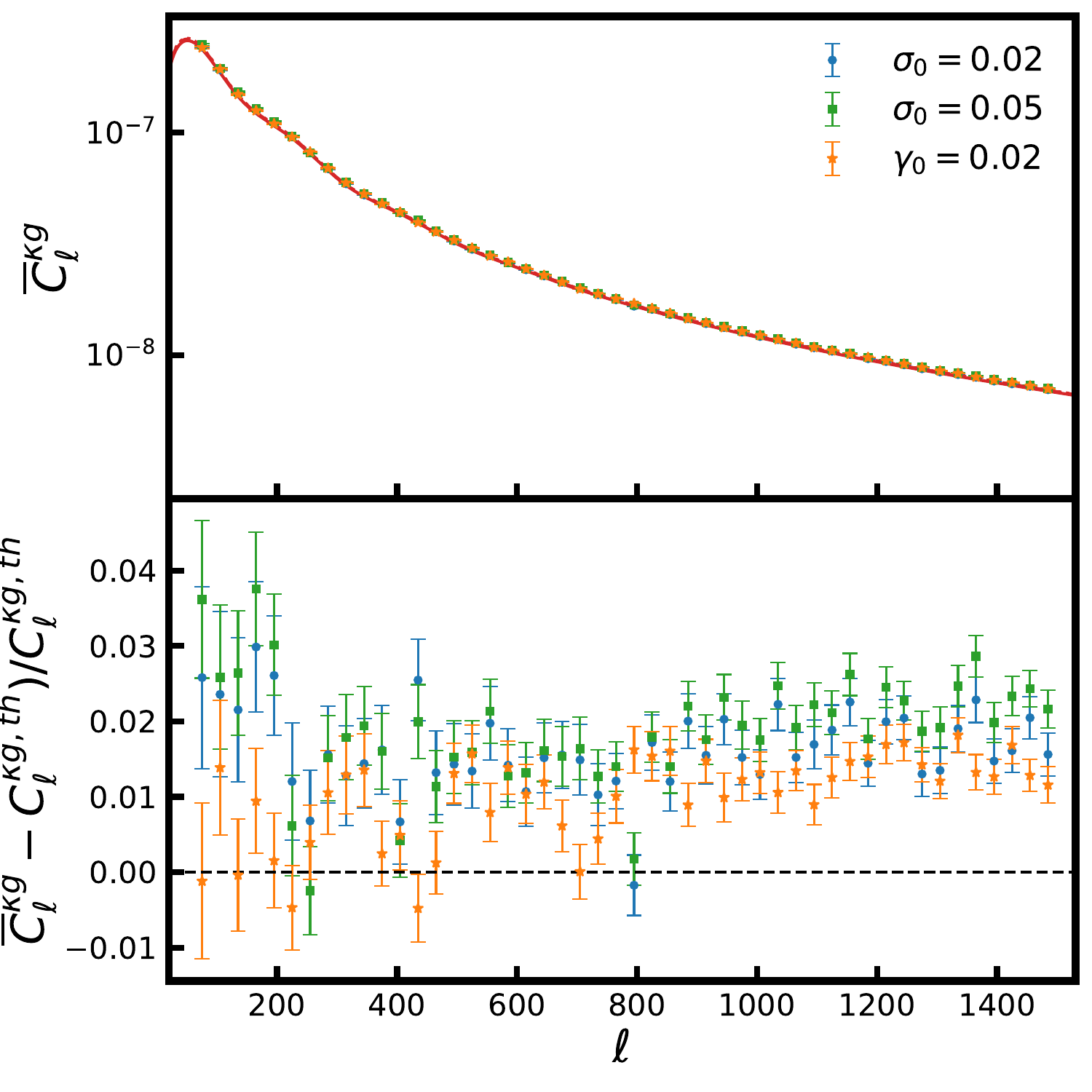}
        \captionsetup{labelformat=empty}
        \caption{\large{Bin 7 ($1.4\leq z<1.8$)}}
    \end{subfigure}
    \caption{Comparison of estimated power spectra for different photometric redshift errors. \textit{Top:} Average galaxy auto-power spectrum. \textit{Bottom:} Cross-power spectrum reconstructed from 300 realisations of photometric datasets for three tomographic bins. {The blue, green, and orange symbols show the reconstructed power spectra for Case-I ($\sigma_{0} = 0.02,\, \sigma_{0} = 0.05$) and Case-II ($\gamma_{0}=0.02$), respectively. The solid, dashed, and dotted red lines are the corresponding theoretical power spectra computed with redshift distributions estimated by the convolution method using Eq.\,(\ref{eq:true_dist_conv}).} The error bars are computed from the covariance matrix of simulations using Eq.\,(\ref{eq:err_simul}).}
    \label{fig:plot_photo_from_true_conv_err_nsim_300}
\end{figure*}

We present the power spectra for simulations with photometric redshift errors estimated using the method described in section \ref{sec:power_spectra}. {The noise-subtracted average galaxy auto-power spectra $(\overline{C}_{\ell}^{gg})$ and cross-power spectra between CMB lensing and galaxy over-density $(\overline{C}_{\ell}^{\kappa g})$ extracted from photometric datasets for three tomographic bins are shown in Fig. \ref{fig:plot_photo_from_true_conv_err_nsim_300}. The blue circles and green squares show the reconstructed power spectra for Case-I with $\sigma_{0}=0.02$ and $\sigma_{0}=0.05$, respectively, and the orange stars show the average power spectra for Case-II with $\gamma_{0}=0.02$. The red lines (solid, dashed, and dotted) are the theoretical power spectra computed using the redshift distributions estimated by the more common convolution method (section \ref{sec:convolution}) for $\sigma_{0}=0.02,\,\sigma_{0}=0.05  \text{ and } \gamma_{0}=0.02$, respectively.} We also present the relative difference between the extracted average power spectra and their theoretical expectations. The relative difference between estimated and theoretical power spectra for the other tomographic bins are shown in Appendix \ref{sec_apndx:power_spectra_true_err_conv}.

{The estimated galaxy auto-power spectra are smaller than expectations in every tomographic bin, with offsets varying between $2-15\%$ for $\sigma_{0}=0.02$, between $15-40\%$ for $\sigma_{0}=0.05$, and between $20-40\%$ for $\gamma_{0}=0.02$. We note that even with a smaller redshift scatter of $\gamma_{0} = 0.02$, we obtain large offsets in the galaxy auto-power spectra, which is due to the broad tails of the modified Lorentzian error distribution. The cross-power spectra show comparatively smaller biases of $<5\%$ in every tomographic bin for both Case-I and Case-II.} We find similar offsets when the true redshift distributions (and hence, the theoretical power spectra) were computed using the deconvolution method (following Eq.\,(\ref{eq:true_dist_deconv_bin})). This shows that the offsets in the power spectra are not related to the method of choice. {Furthermore, since the level of offsets in the power spectra depends on the strength of photometric redshift scatter (Case-I) and type of redshift error distribution (Case-II), this confirms that the origin of these offsets is rooted in the leakage of objects from one redshift bin to the next due to photometric redshift errors.} These deviations will also affect the parameter estimation from the power spectra.


\section{Correction for leakage through scattering matrix}\label{sec:scattering_matrix}
As a result of the errors in photometric redshifts, a fraction of galaxies observed in a given photometric redshift bin come from other redshift bins. The leakage of objects across redshift bins changes the strength of the correlation in a tomographic analysis and the results in the non-zero correlation between different redshift bins. In this section, we attempt to counter the redshift bin mismatch of objects through a scattering matrix. When we divide galaxies into $n$ tomographic bins, then \cite{Zhang2010} showed that noise-subtracted galaxy auto-power spectra between the $i$th and $j$th photometric bins, $C_{ij}^{gg,\text{ph}}$, are related to the noise-subtracted galaxy auto-power spectra from the $k$th true redshift bin $C^{gg,\text{tr}}_{kk}$ by
\begin{equation}
    C_{ij}^{gg,\text{ph}}(\ell) = \sum\limits_{k}P_{ki}P_{kj}C_{kk}^{gg,\text{tr}}(\ell),
    \label{eq:scattering_relation_gg}
\end{equation}
when there are no cross correlations between true redshift bins. Eq.\,(\ref{eq:scattering_relation_gg}) has a generalised form for the case when true redshift bins have non-zero correlations, but using disjoint true redshift bins significantly reduces the complexity. The elements of the scattering matrix $P_{ij}$ are defined as the ratio $N_{i\to j}/N_{j}^{\text{ph}}$, where $N_{i\to j}$ is the number of galaxies coming from the $i$th true redshift bin to the $j$th photometric bin, and $N_{j}^{\text{ph}}$ is the total number of galaxies in the $j$th photometric bin. This definition also produces a natural normalisation $\sum\limits_{i}P_{ij} = 1$. A similar relation for the cross-power spectra between the galaxy over-density in redshift bin $i$ and the CMB lensing convergence can be obtained as
\begin{equation}
    C_{i}^{\kappa g,\text{ph}}(\ell) = \sum\limits_{k}P_{ki}C_{kk}^{\kappa g,\text{tr}}(\ell).
    \label{eq:scattering_relation_kg}
\end{equation}
When we collect $P_{ij}$ as elements of the scattering matrix $P$, then we can compactly write
\begin{align}
    C^{gg,\text{ph}} &= P^{\text{T}}C^{gg,\text{tr}}P\label{eq:scattering_relation_gg_matrix}\\
    C^{\kappa g,\text{ph}} &= P^{\text{T}}C^{\kappa g,\text{tr}},\label{eq:scattering_relation_kg_matrix}
\end{align}
where $P^{\text{T}}$ denotes the transpose of matrix $P$. Eqs.\,(\ref{eq:scattering_relation_gg_matrix}-\ref{eq:scattering_relation_kg_matrix}) show that the redistribution of galaxies across redshift bins due to photometric redshift errors results in a non-trivial relation between the photometric and true power spectra, weighted by the elements of the scattering matrix. Thus, to properly mitigate the effects of leakage on power spectra, a precise estimation of the scattering matrix is necessary.

\cite{Zhang2017} proposed an algorithm for solving problems similar to Eq.\,(\ref{eq:scattering_relation_gg_matrix}) based on the non-negative matrix factorization (NMF) method. This algorithm simultaneously approximates the matrices $P$ and $C^{gg,\text{tr}}$. However, the NMF method proves to be computationally challenging for cases with many tomographic and multipole bins. Here, we propose an alternative method for the fast and efficient computation of the scattering matrix based on the true and photometric redshift distributions. We first estimate the true redshift distribution, $\frac{\mathrm{d}N(z_{t})}{\mathrm{d}z_{t}}$, for the entire redshift range following Eq.\,(\ref{eq:true_dist_deconv_fourier_method}), and then, we use Eq.\,(\ref{eq:true_dist_deconv_bin}) to compute the redshift distribution $\frac{\mathrm{d}N^{i}(z_{p})}{\mathrm{d}z_{p}}$ for every tomographic bin $i$. The elements of the scattering matrix $P_{ij}$ can then be computed directly with the relation
\begin{equation}
    P_{ij} = \frac{\int\limits_{z_{min}^{j}}^{z_{min}^{j+1}}\mathrm{d}z_{p} \frac{\mathrm{d}N^{i}}{\mathrm{d}z_{p}}} {\int\limits_{z_{min}^{j}}^{z_{min}^{j+1}}\mathrm{d}z_{p} \frac{\mathrm{d}N}{\mathrm{d}z_{p}}},
    \label{eq:scattering_matrix_element_equation_paper2}
\end{equation}
where $\frac{\mathrm{d}N}{\mathrm{d}z_{p}}$ is the observed photometric redshift distribution of the galaxies, and $z_{min}^{j}$ is the lower limit of the $j$th redshift bin. Our method of computing the elements of the scattering matrix is significantly faster than the NMF method and is only subject to an accurate estimation of the error distribution and the true redshift distribution.

{Fig. \ref{fig:performance_scattering_matrix} shows our estimation of the scattering matrix with our proposed method using redshift distributions for the three cases of $\sigma_{0} = 0.02,\, \sigma_{0} = 0.05 \text{, and } \gamma_{0} = 0.02$.} The average value of the scattering matrix $\langle P \rangle$ and its standard deviation $\sigma(P)$ averaged over $300$ simulations is shown in the top and middle panels of Fig. \ref{fig:performance_scattering_matrix}, respectively. We note that the scattering matrix elements corresponding to the first true redshift bin have the strongest standard deviation. This behaviour is expected as the objects near $z=0$ do not strictly follow convolution, as discussed in section \ref{sec:deconvolution}. The accuracy of the estimated scattering matrix can be verified using the true scattering matrix $P^{\text{True}}$, which is computed based on exactly counting the number of objects that move from one redshift bin to the next in the catalogue generated by the code \texttt{FLASK}. In the bottom panel of Fig. \ref{fig:performance_scattering_matrix}, we show the difference between the scattering matrix computed with our method and $P^{\text{True}}$, averaged over $300$ realisations. We find that $|P - P^{\text{True}}|<0.006$ for all elements of the scattering matrix. The strongest differences occur in the first and last tomographic bins, that is, near the boundaries of the redshift range we simulated in our analysis. Hence, the overall precision and accuracy in the estimation of the scattering matrix is found to be good and can be used to correct the redshift bin mismatch for the power spectra.

Given an estimate of the true redshift distribution deconvoluted from the observed photometric redshift distribution and scattering matrix, the effect of the redshift bin mismatch on the objects can be corrected for in two ways: either by transforming the true theoretical power spectra $C^{\text{th,tr}}$ into $C^{\text{th,ph}}$ using Eqs.\,(\ref{eq:scattering_relation_gg_matrix}) and (\ref{eq:scattering_relation_kg_matrix}) and comparing them to the estimated photometric power spectra $\hat{C}^{\text{ph}}$, or by inverting Eqs.\,(\ref{eq:scattering_relation_gg_matrix}) and (\ref{eq:scattering_relation_kg_matrix}) to transform the estimated photometric power spectra $\hat{C}^{\text{ph}}$ into true power spectra $\hat{C}^{\text{tr}}$ and comparing them to the theoretical true power spectra $C^{\text{th,tr}}$. We used the former approach for the figures that compare the power spectra and the latter approach to estimate the parameters.

\begin{figure*}
    \begin{subfigure}[b]{0.33\linewidth}
        \centering
        \includegraphics[width=0.9\linewidth]{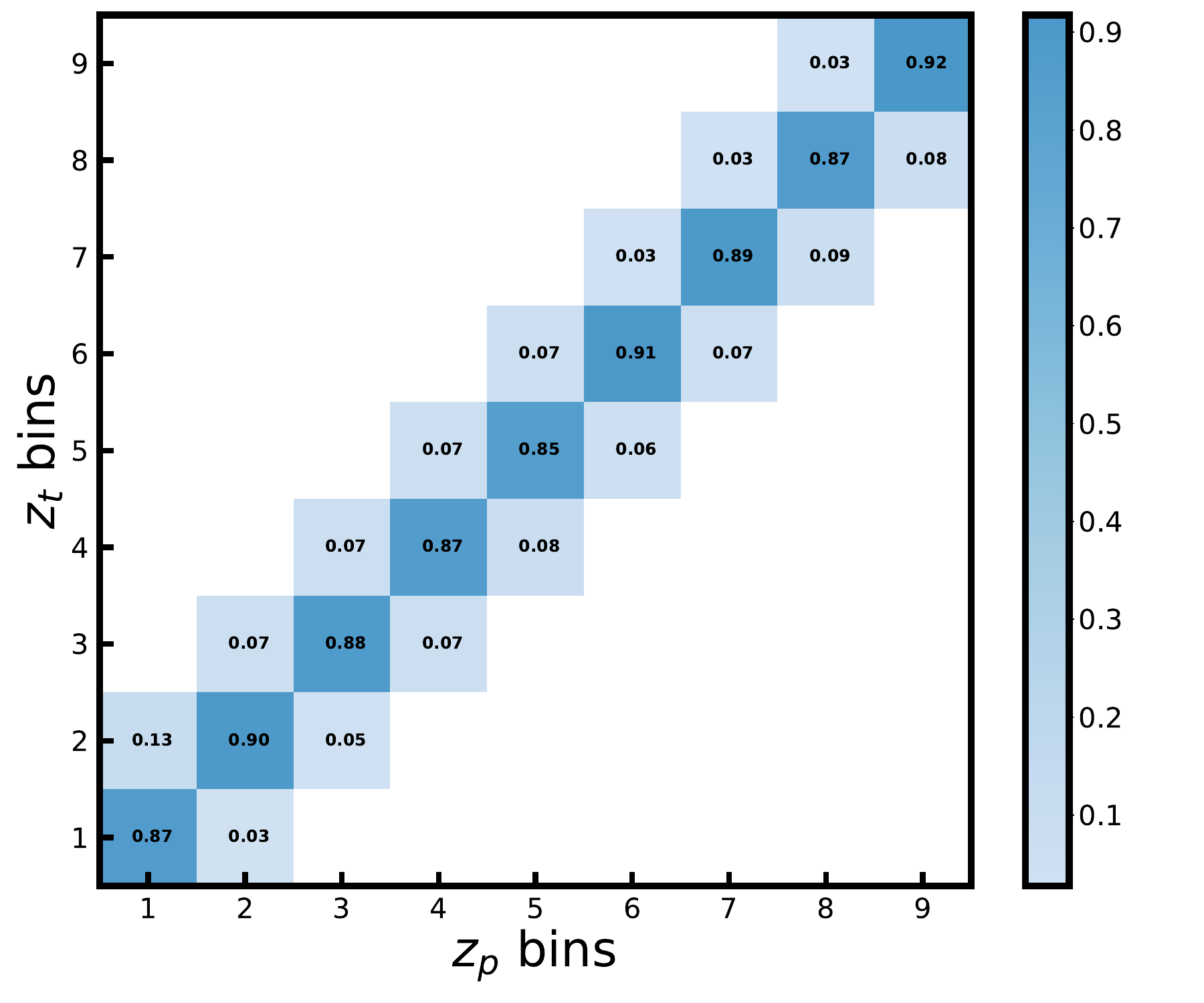}
        \caption{$\langle P \rangle$}
    \end{subfigure}%
    \begin{subfigure}[b]{0.33\linewidth}
        \centering
        \includegraphics[width=0.9\linewidth]{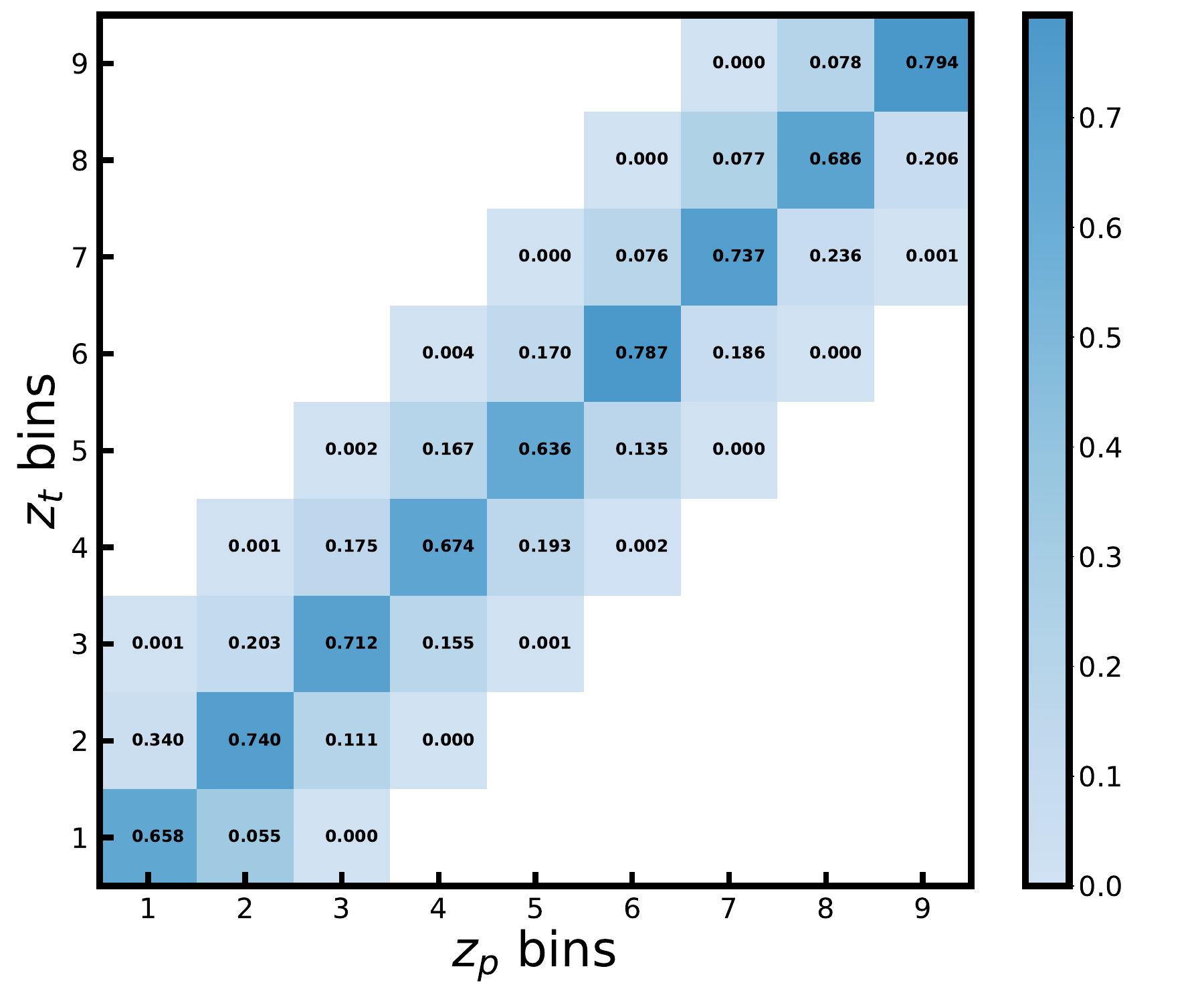}
        \caption{$\langle P \rangle$}
    \end{subfigure}
    \begin{subfigure}[b]{0.33\linewidth}
        \centering
        \includegraphics[width=0.9\linewidth]{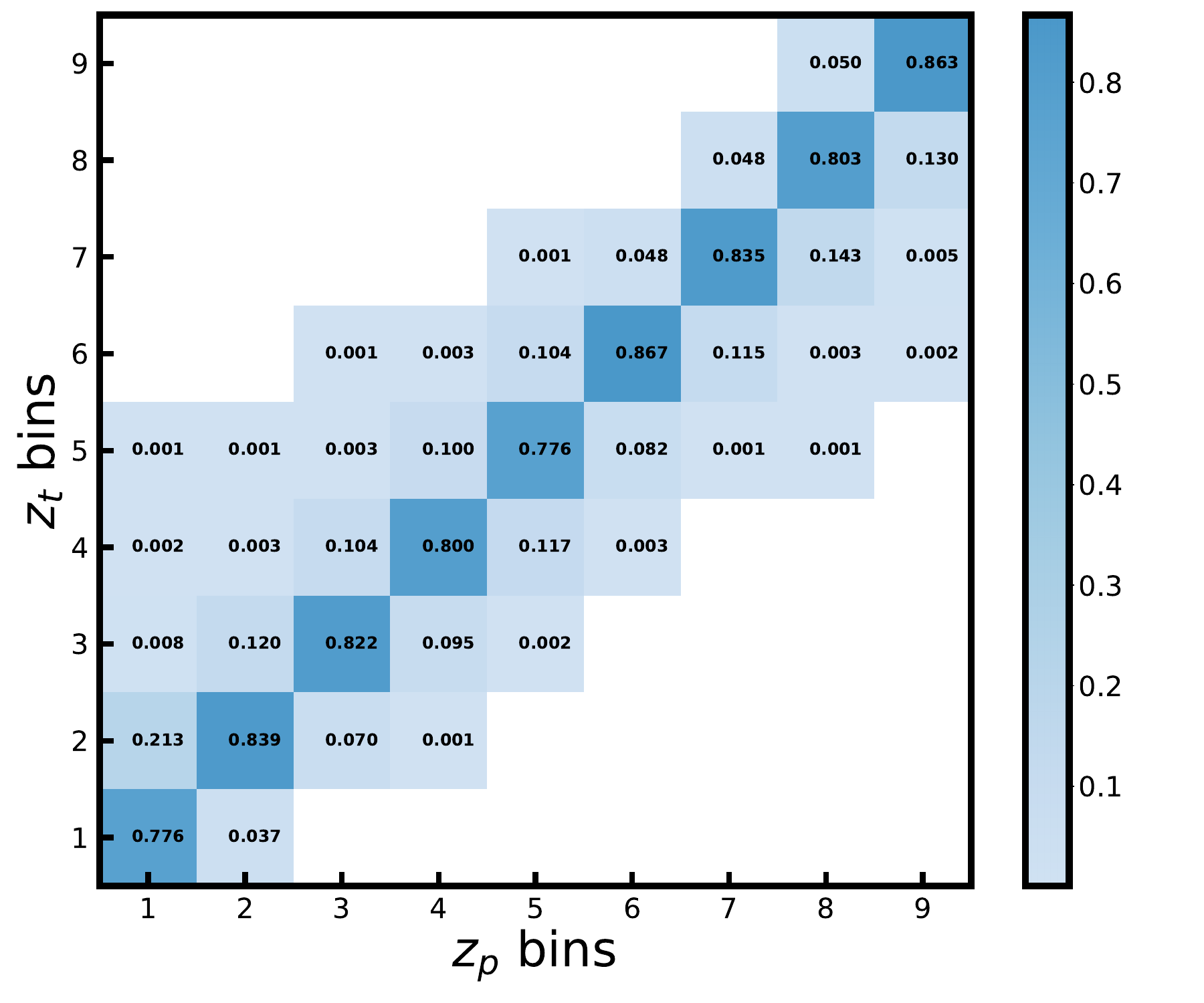}
        \caption{$\langle P \rangle$}
    \end{subfigure}\\
    \begin{subfigure}[b]{0.33\linewidth}
        \centering
        \includegraphics[width=0.9\linewidth]{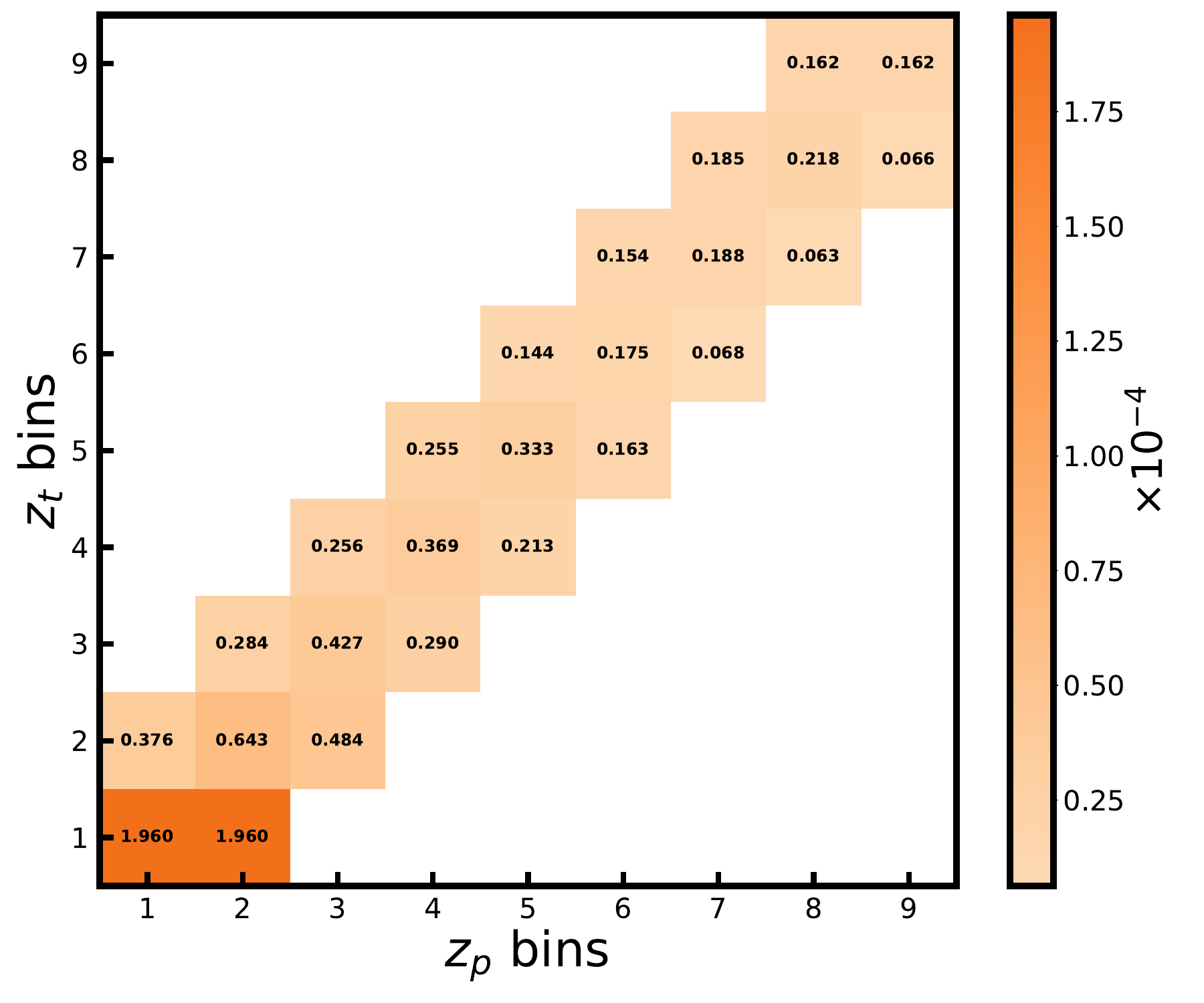}
        \caption{$\sigma(P)$}
    \end{subfigure}%
    \begin{subfigure}[b]{0.33\linewidth}
        \centering
        \includegraphics[width=0.9\linewidth]{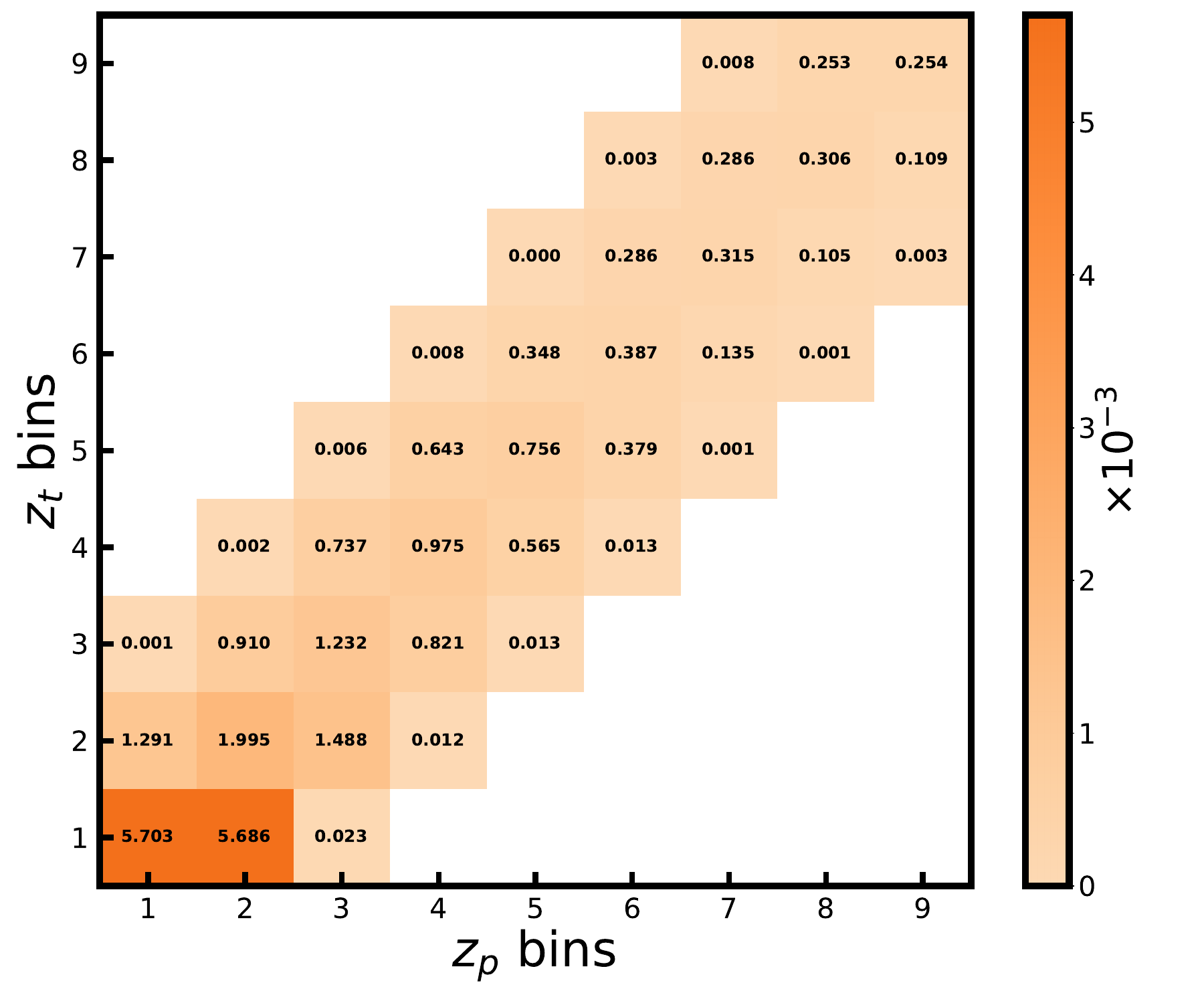}
        \caption{$\sigma(P)$}
    \end{subfigure}
    \begin{subfigure}[b]{0.33\linewidth}
        \centering
        \includegraphics[width=0.9\linewidth]{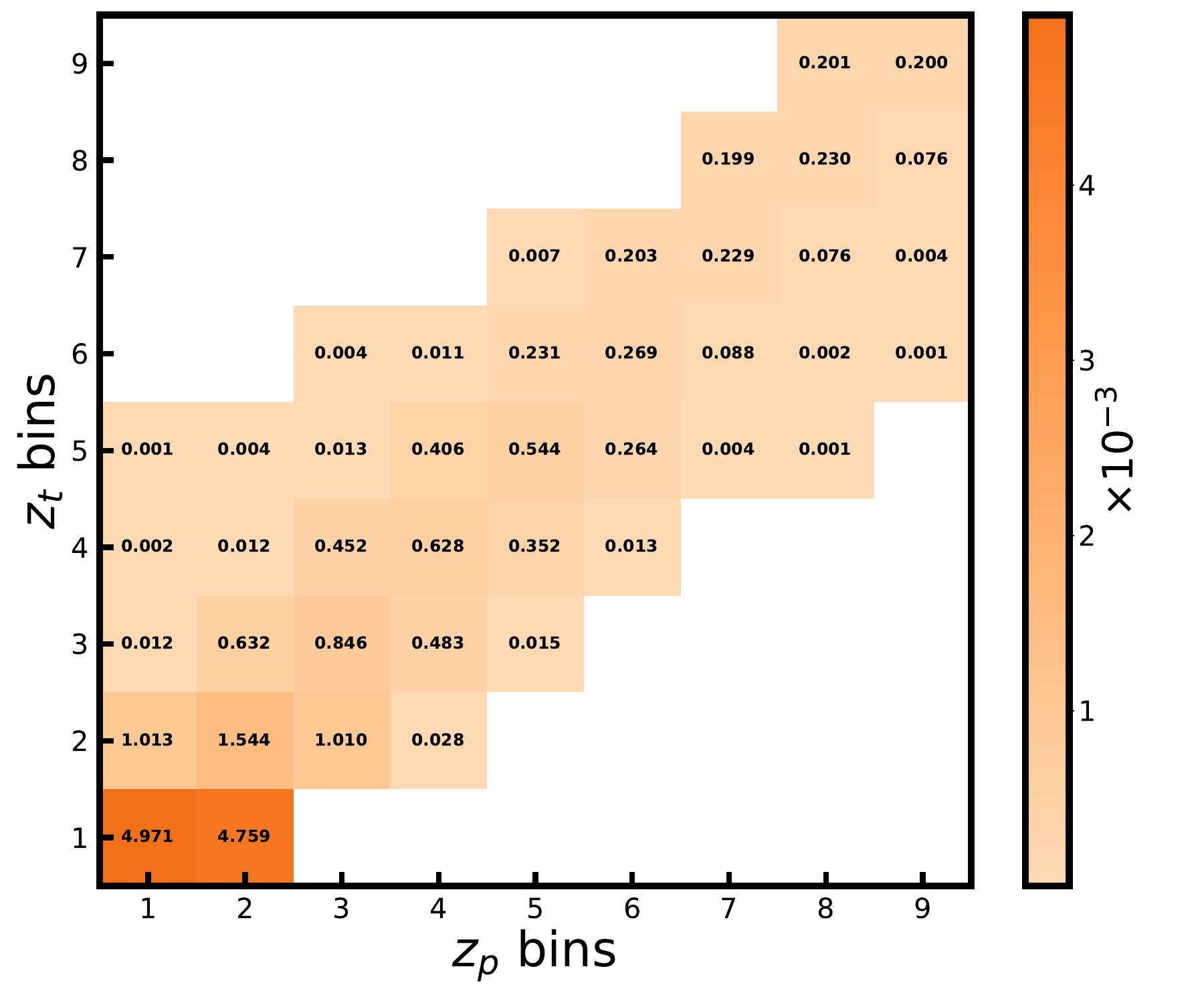}
        \caption{$\sigma(P)$}
    \end{subfigure}\\
    \begin{subfigure}[b]{0.33\linewidth}
        \centering
        \includegraphics[width=0.9\linewidth]{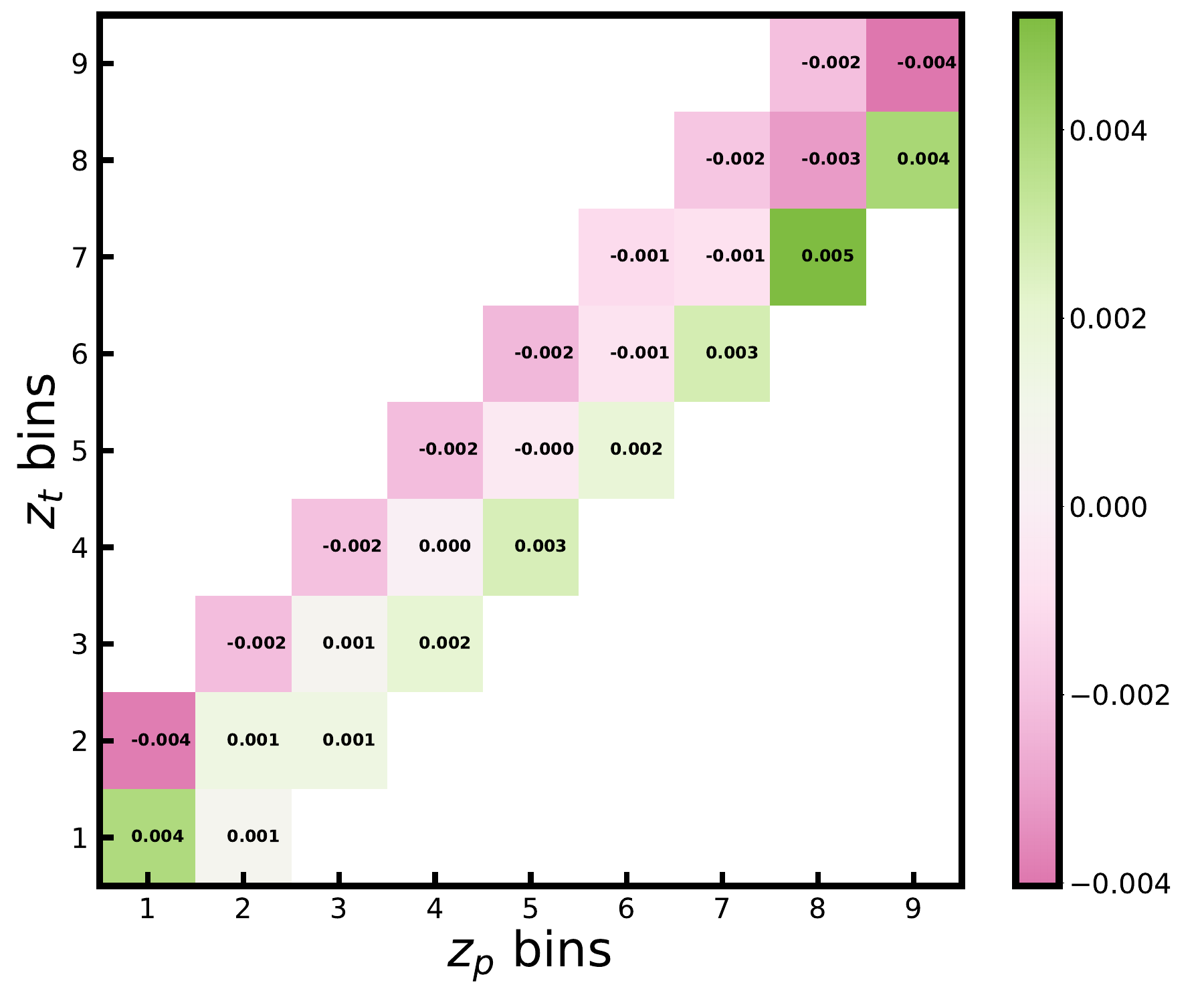}
        \caption{$\langle P - P^{\text{True}}\rangle$}
    \end{subfigure}%
    \begin{subfigure}[b]{0.33\linewidth}
        \centering
        \includegraphics[width=0.9\linewidth]{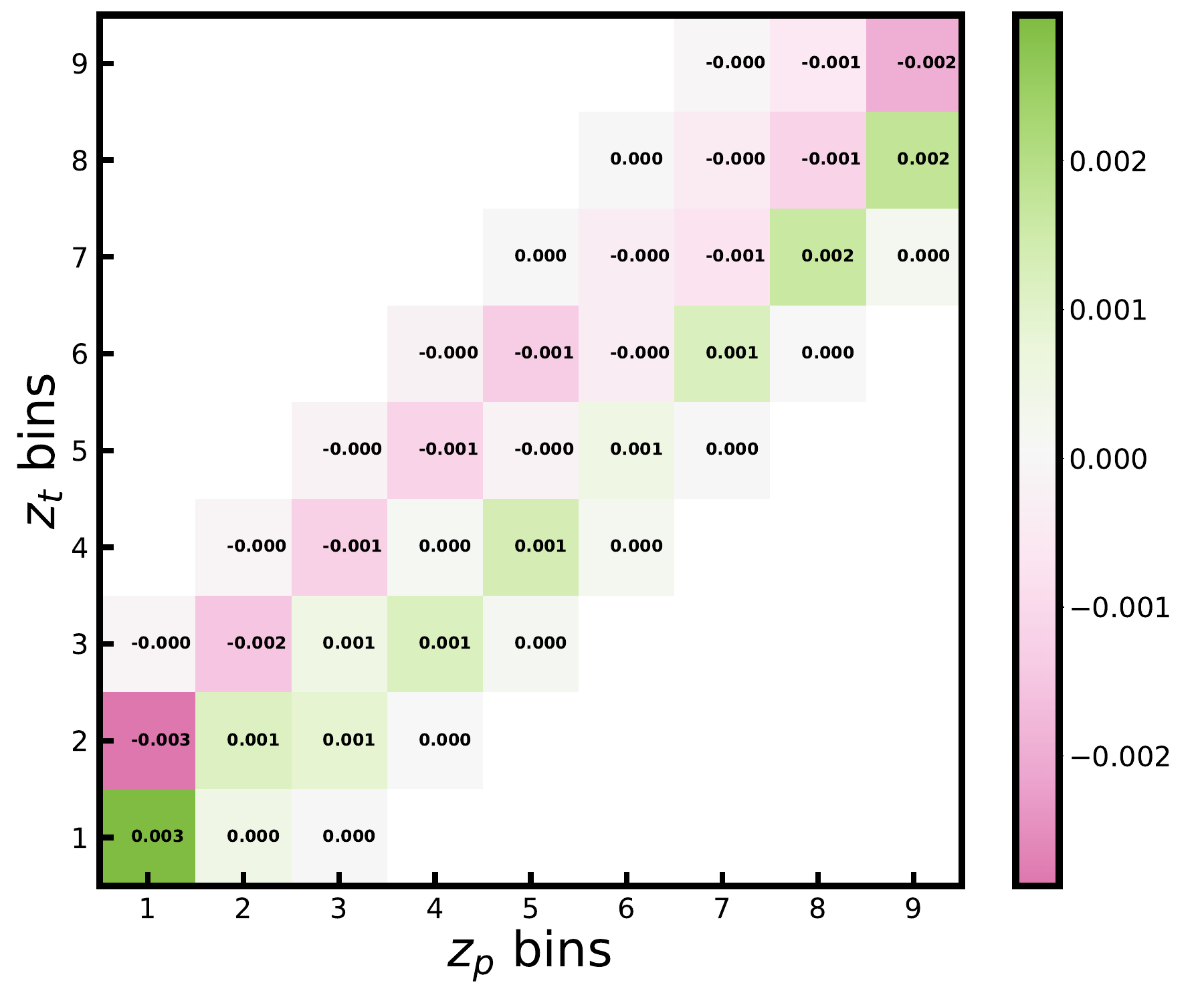}
        \caption{$\langle P - P^{\text{True}}\rangle$}
    \end{subfigure}
    \begin{subfigure}[b]{0.33\linewidth}
        \centering
        \includegraphics[width=0.9\linewidth]{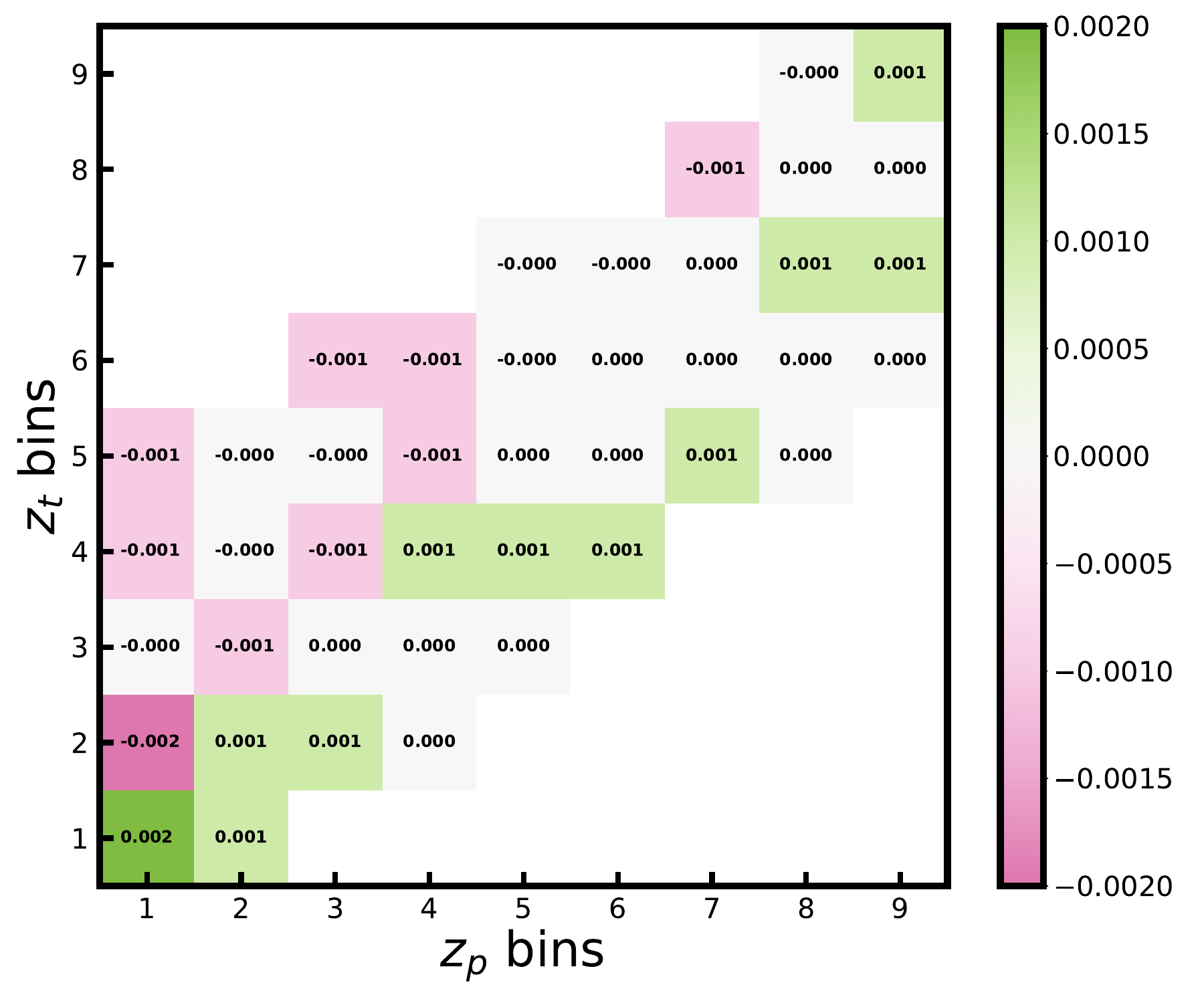}
        \caption{$\langle P - P^{\text{True}}\rangle$}
    \end{subfigure}
    \caption{Performance of the scattering matrix computed with the method described in section \ref{sec:scattering_matrix}. \textit{Top}: Average scattering matrix computed from 300 realisations. \textit{Middle}: Its standard deviation. \textit{Bottom}: Difference between the scattering matrix computed from our method and the true scattering matrix computed from simulated catalogues. {\textit{Left column}: Scattering matrix computed for $\sigma_{0} = 0.02$. \textit{Middle column}: Scattering matrix computed for $\sigma_{0} = 0.05$. \textit{Right column}: Scattering matrix computed for $\gamma_{0} = 0.02$.}}
    \label{fig:performance_scattering_matrix}
\end{figure*}

\begin{figure*}
    \begin{subfigure}[b]{0.33\linewidth}
        \centering
        \includegraphics[width=\linewidth]{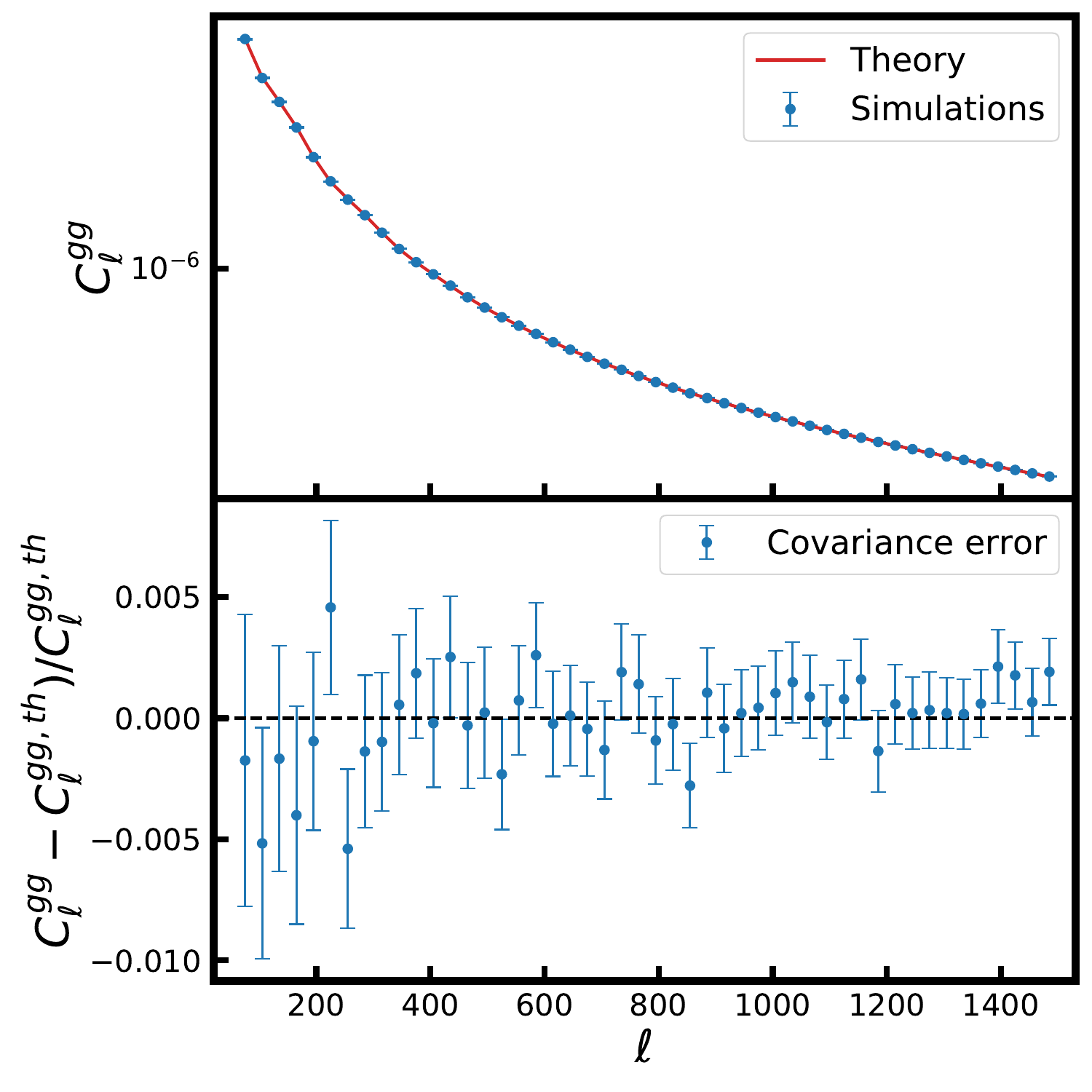}
    \end{subfigure}%
    \begin{subfigure}[b]{0.33\linewidth}
        \centering
        \includegraphics[width=\linewidth]{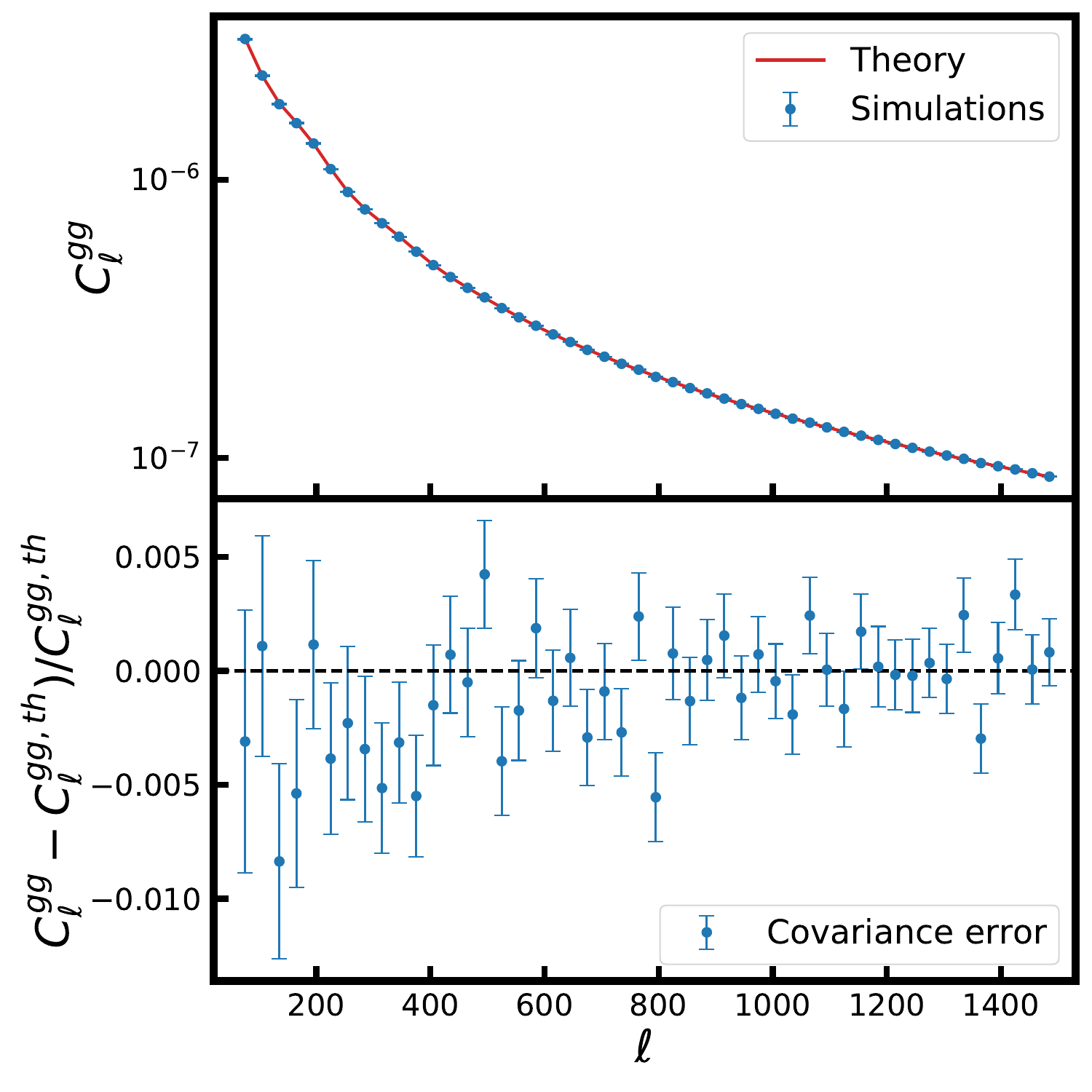}
    \end{subfigure}%
    \begin{subfigure}[b]{0.33\linewidth}
        \centering
        \includegraphics[width=\linewidth]{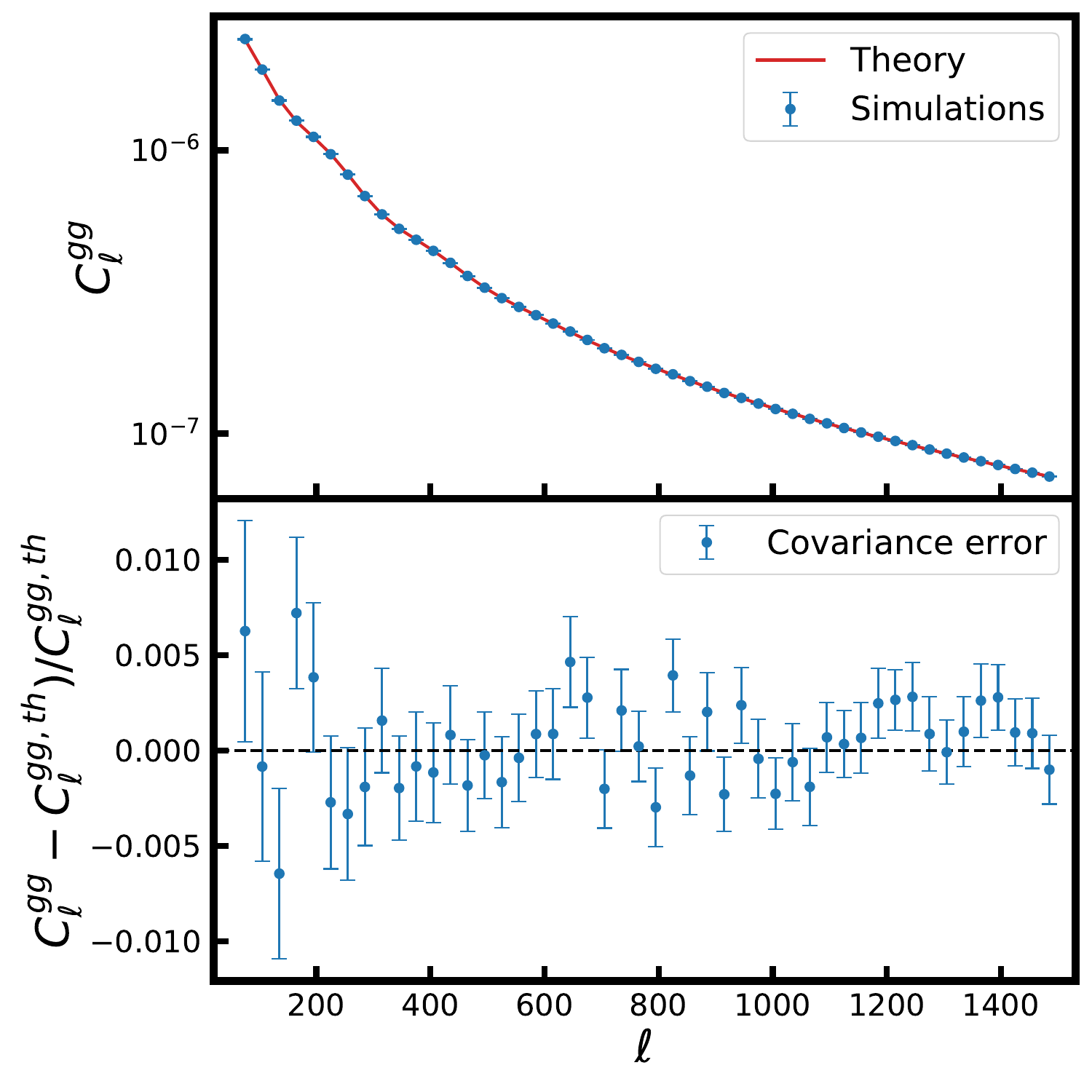}
    \end{subfigure}\\[1ex]
    \begin{subfigure}[b]{0.33\linewidth}
        \centering
        \includegraphics[width=\linewidth]{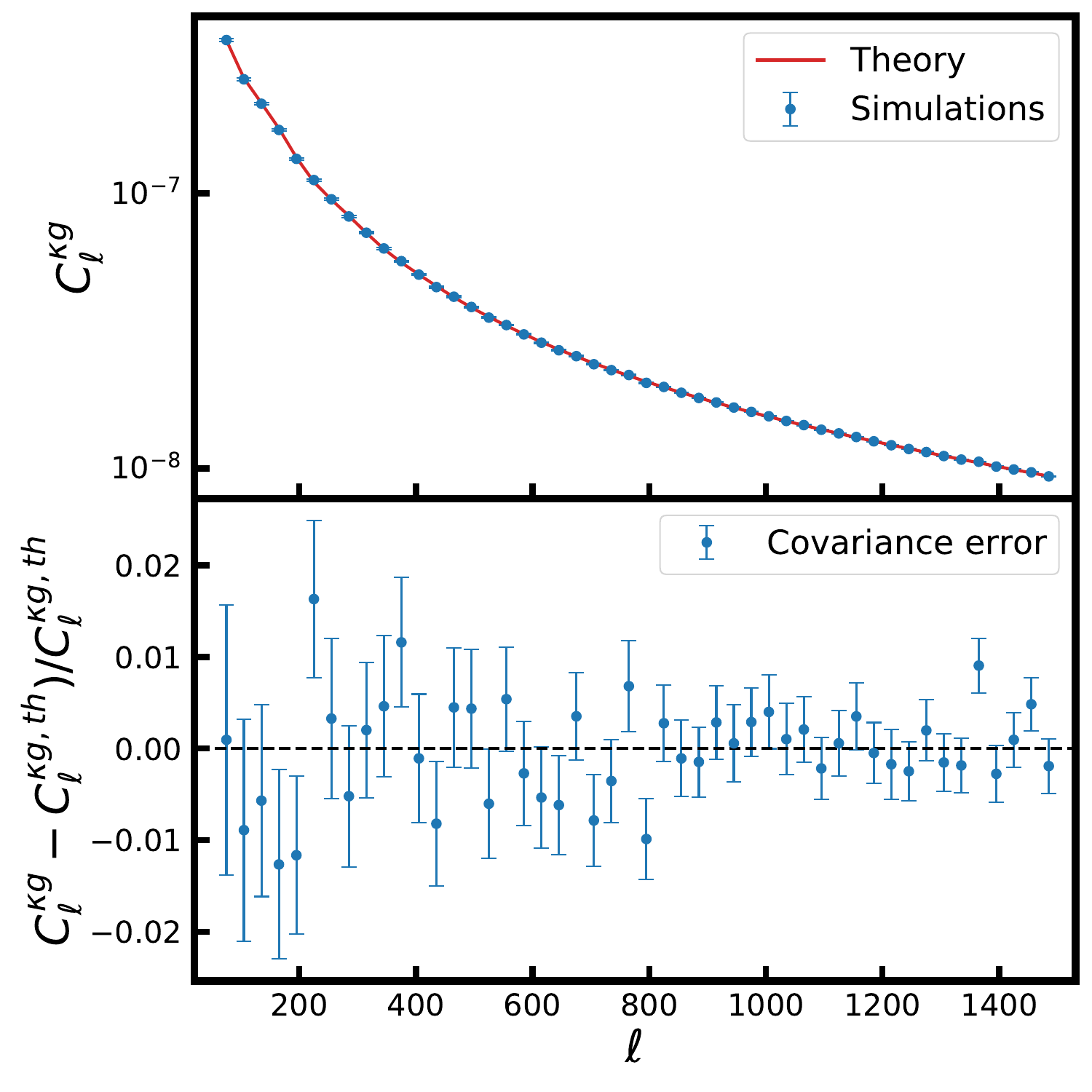}
        \captionsetup{labelformat=empty}
        \caption{\large{Bin 5 ($0.8\leq z<1.0$)}}
    \end{subfigure}%
    \begin{subfigure}[b]{0.33\linewidth}
        \centering
        \includegraphics[width=\linewidth]{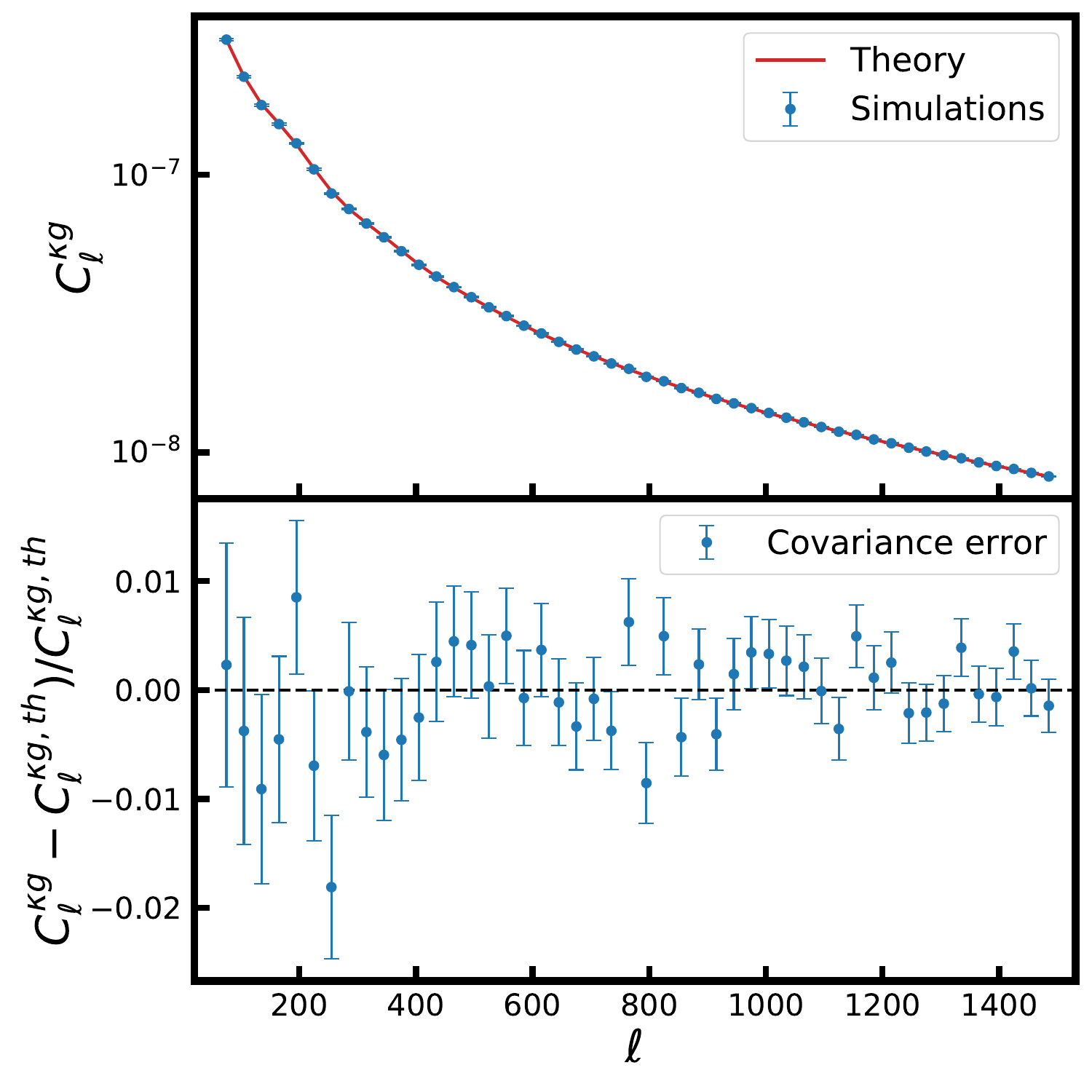}
        \captionsetup{labelformat=empty}
        \caption{\large{Bin 6 ($1.0\leq z<1.4$)}}
    \end{subfigure}%
    \begin{subfigure}[b]{0.33\linewidth}
        \centering
        \includegraphics[width=\linewidth]{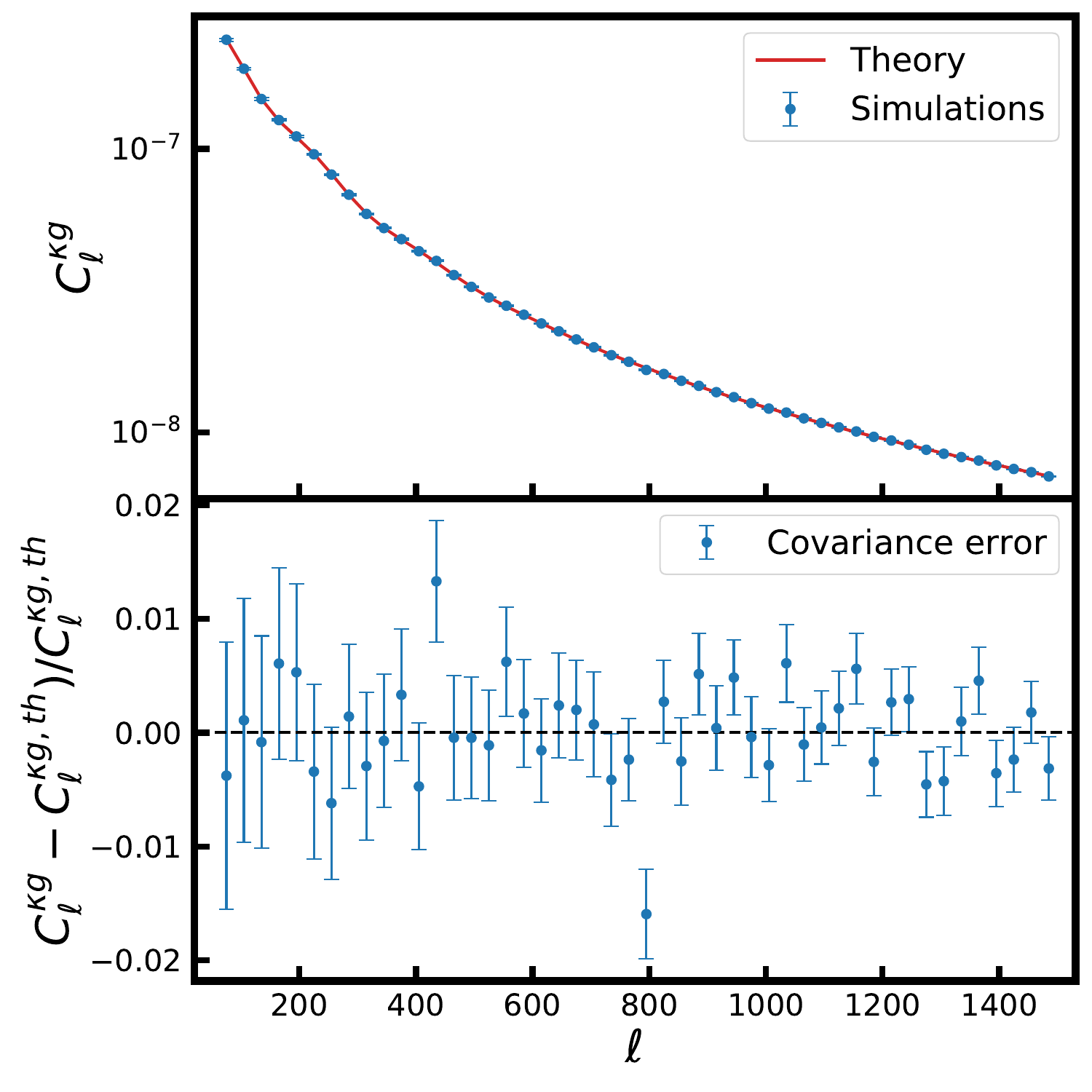}
        \captionsetup{labelformat=empty}
        \caption{\large{Bin 7 ($1.4\leq z<1.8$)}}
    \end{subfigure}
    \caption{Comparison of estimated angular power spectra with theoretical power spectra after correction for redshift bin mismatch. \textit{Top:} Average galaxy auto-power spectrum. \textit{Bottom:} Cross-power spectrum reconstructed from $300$ realisations of photometric datasets with $\sigma_{0}=0.02$, shown for three tomographic bins. The red line represents the theoretical power spectrum corrected for the redshift bin mismatch using Eqs.\,(\ref{eq:scattering_relation_gg_matrix}) and (\ref{eq:scattering_relation_kg_matrix}). The error bars are computed from the covariance matrix of simulations using Eq.\,(\ref{eq:err_simul}).}
    \label{fig:plot_photo_from_scat_mat_dist_nsim_300_0.02}
\end{figure*}

\begin{figure*}
    \begin{subfigure}[b]{0.33\linewidth}
        \centering
        \includegraphics[width=\linewidth]{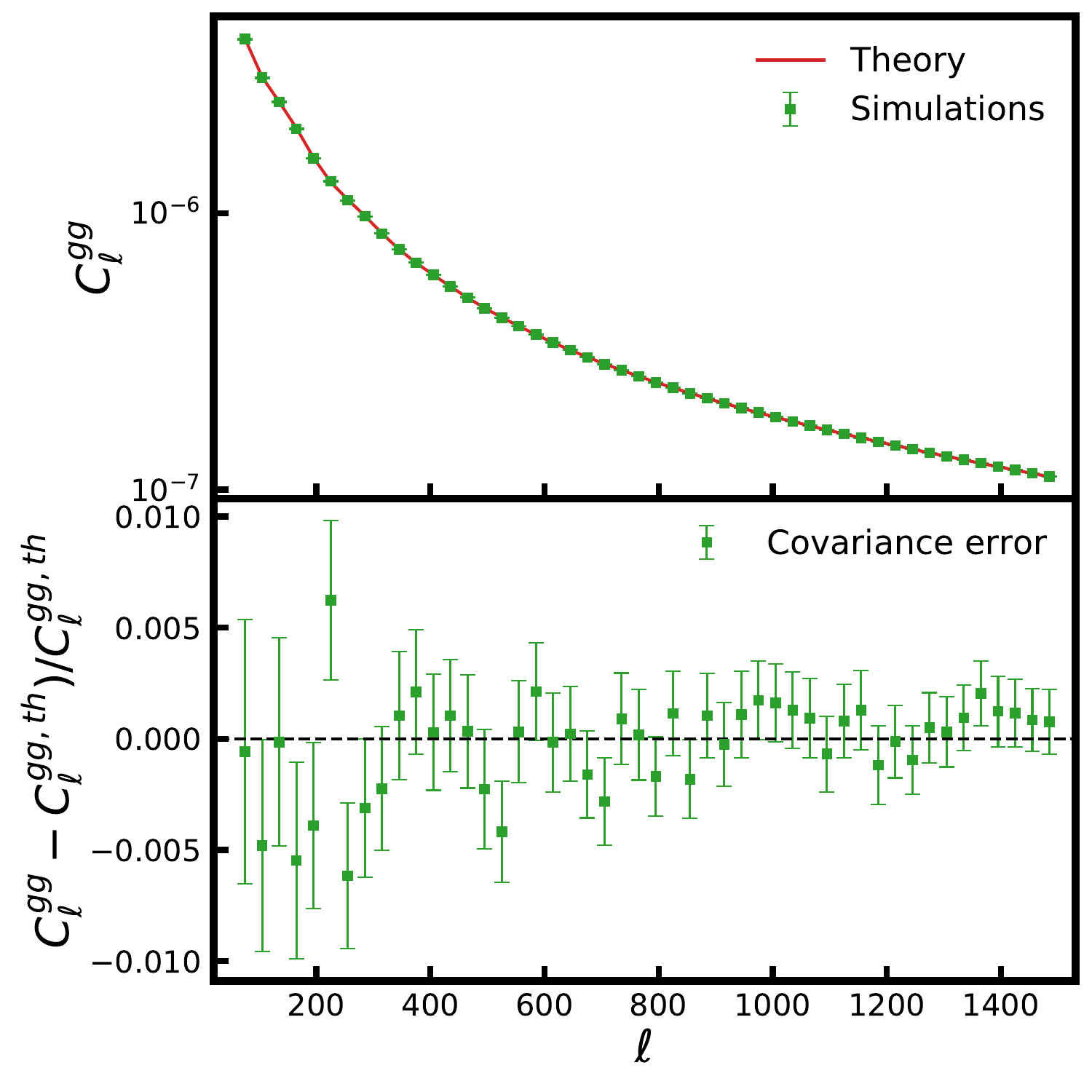}
    \end{subfigure}%
    \begin{subfigure}[b]{0.33\linewidth}
        \centering
        \includegraphics[width=\linewidth]{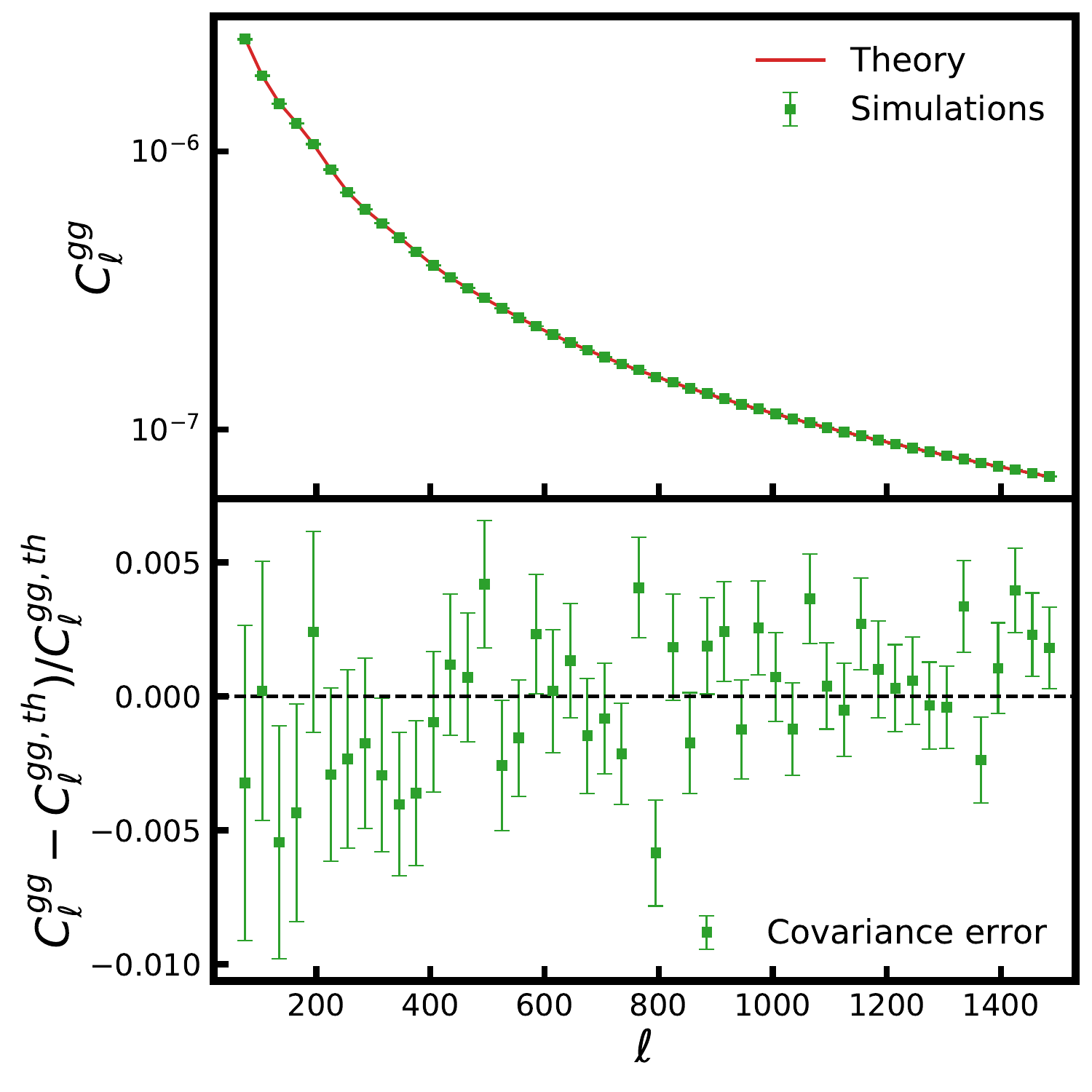}
    \end{subfigure}%
    \begin{subfigure}[b]{0.33\linewidth}
        \centering
        \includegraphics[width=\linewidth]{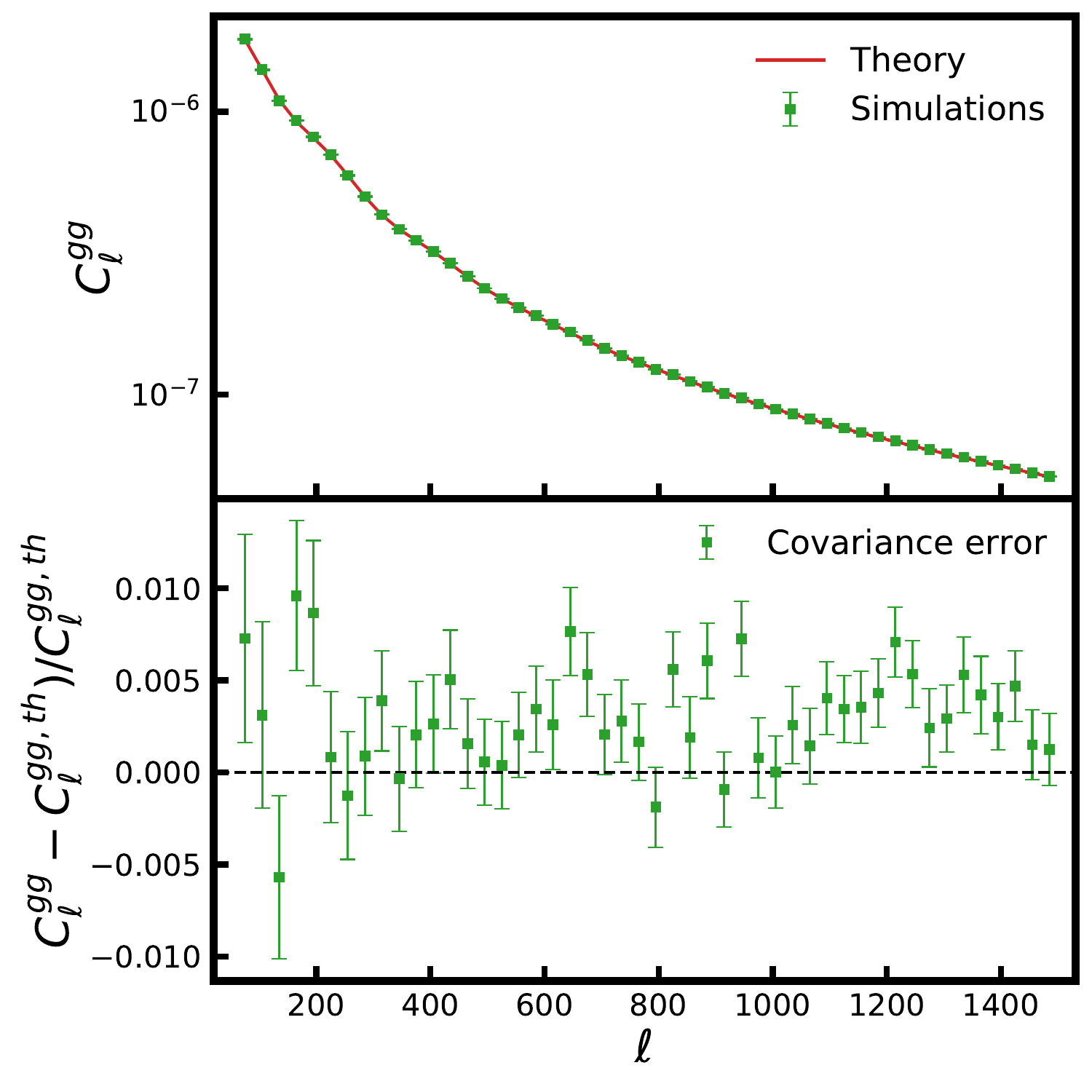}
    \end{subfigure}\\[1ex]
    \begin{subfigure}[b]{0.33\linewidth}
        \centering
        \includegraphics[width=\linewidth]{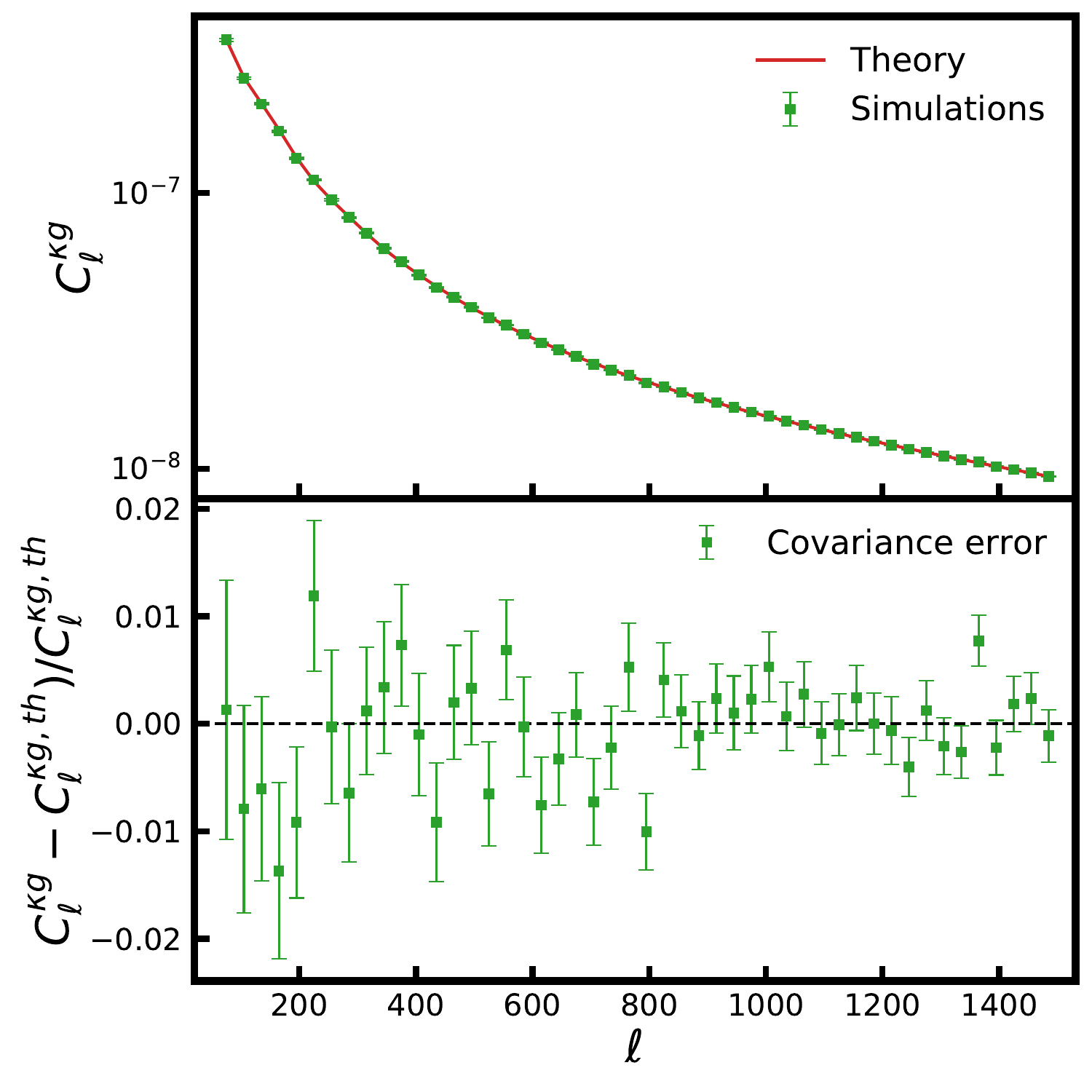}
        \captionsetup{labelformat=empty}
        \caption{\large{Bin 5 ($0.8\leq z<1.0$)}}
    \end{subfigure}%
    \begin{subfigure}[b]{0.33\linewidth}
        \centering
        \includegraphics[width=\linewidth]{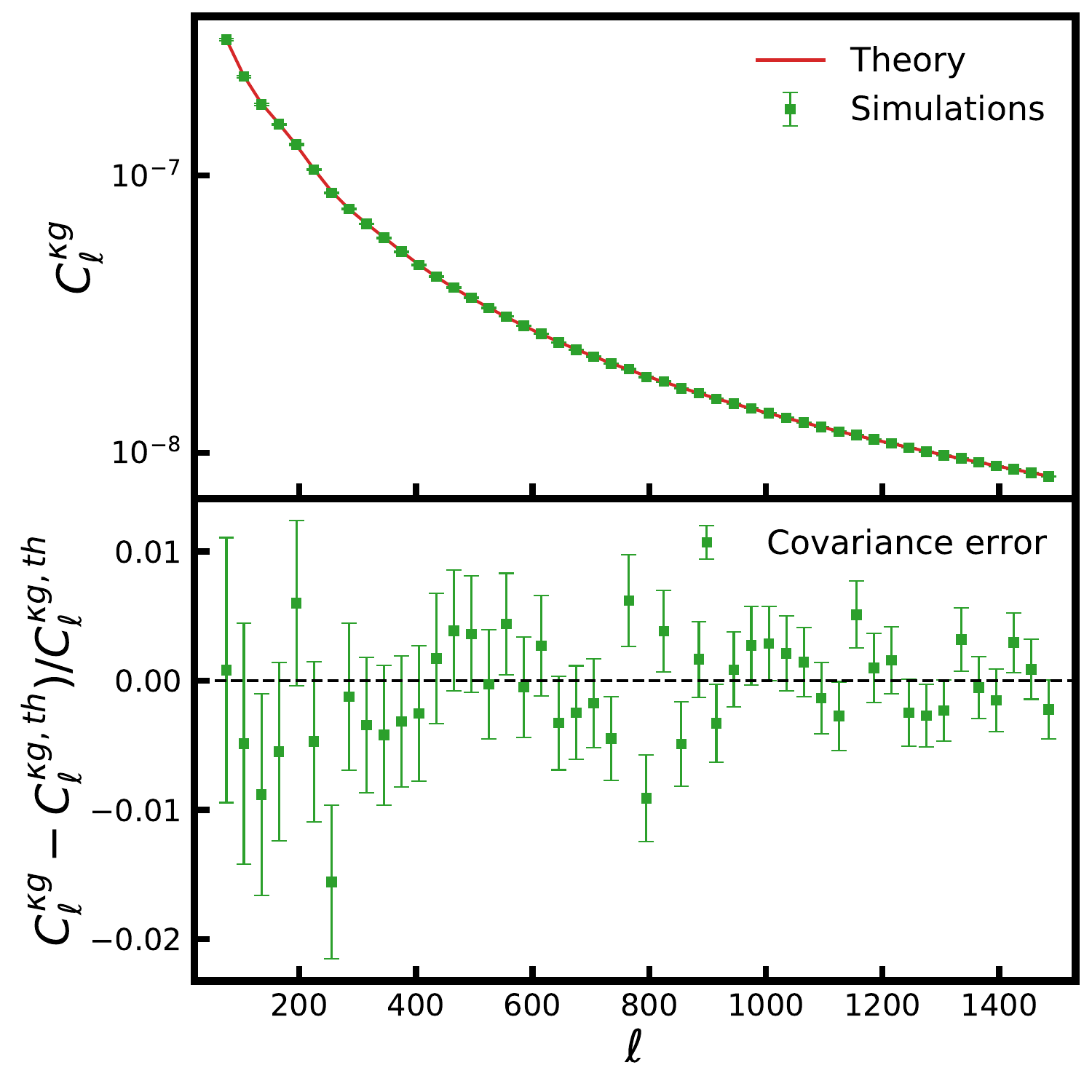}
        \captionsetup{labelformat=empty}
        \caption{\large{Bin 6 ($1.0\leq z<1.4$)}}
    \end{subfigure}%
    \begin{subfigure}[b]{0.33\linewidth}
        \centering
        \includegraphics[width=\linewidth]{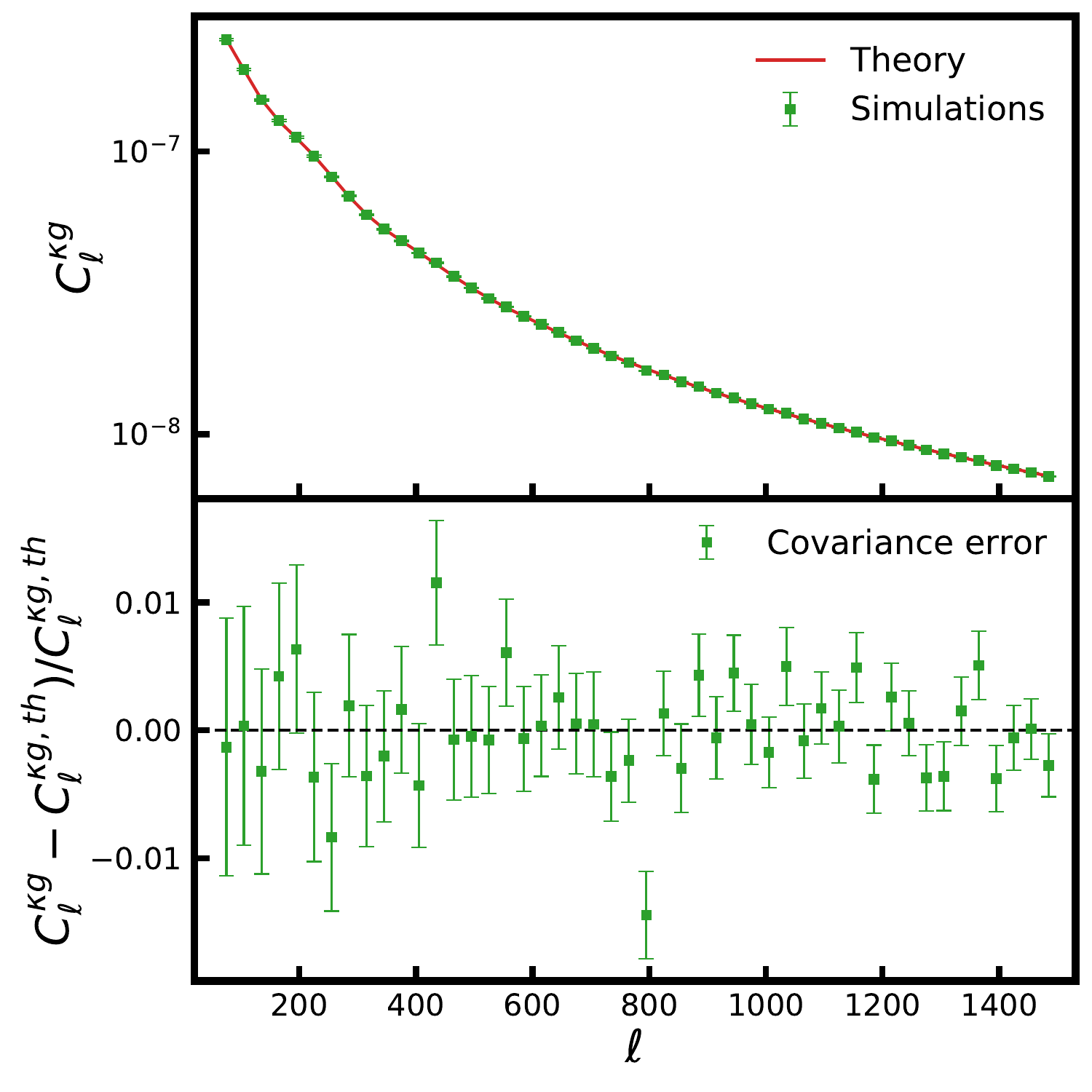}
        \captionsetup{labelformat=empty}
        \caption{\large{Bin 7 ($1.4\leq z<1.8$)}}
    \end{subfigure}
    \caption{Comparison of estimated angular power spectra with theoretical power spectra after correction for redshift bin mismatch. \textit{Top:} Average galaxy auto-power spectrum. \textit{Bottom:} Cross-power spectrum reconstructed from $300$ realisations of photometric datasets with $\sigma_{0}=0.05$, shown for three tomographic bins. The red line represents the theoretical power spectrum corrected for the redshift bin mismatch using Eqs.\,(\ref{eq:scattering_relation_gg_matrix}) and (\ref{eq:scattering_relation_kg_matrix}). The error bars are computed from the covariance matrix of simulations using Eq.\,(\ref{eq:err_simul}).}
    \label{fig:plot_photo_from_scat_mat_dist_nsim_300_0.05}
\end{figure*}

\begin{figure*}
    \begin{subfigure}[b]{0.33\linewidth}
        \centering
        \includegraphics[width=\linewidth]{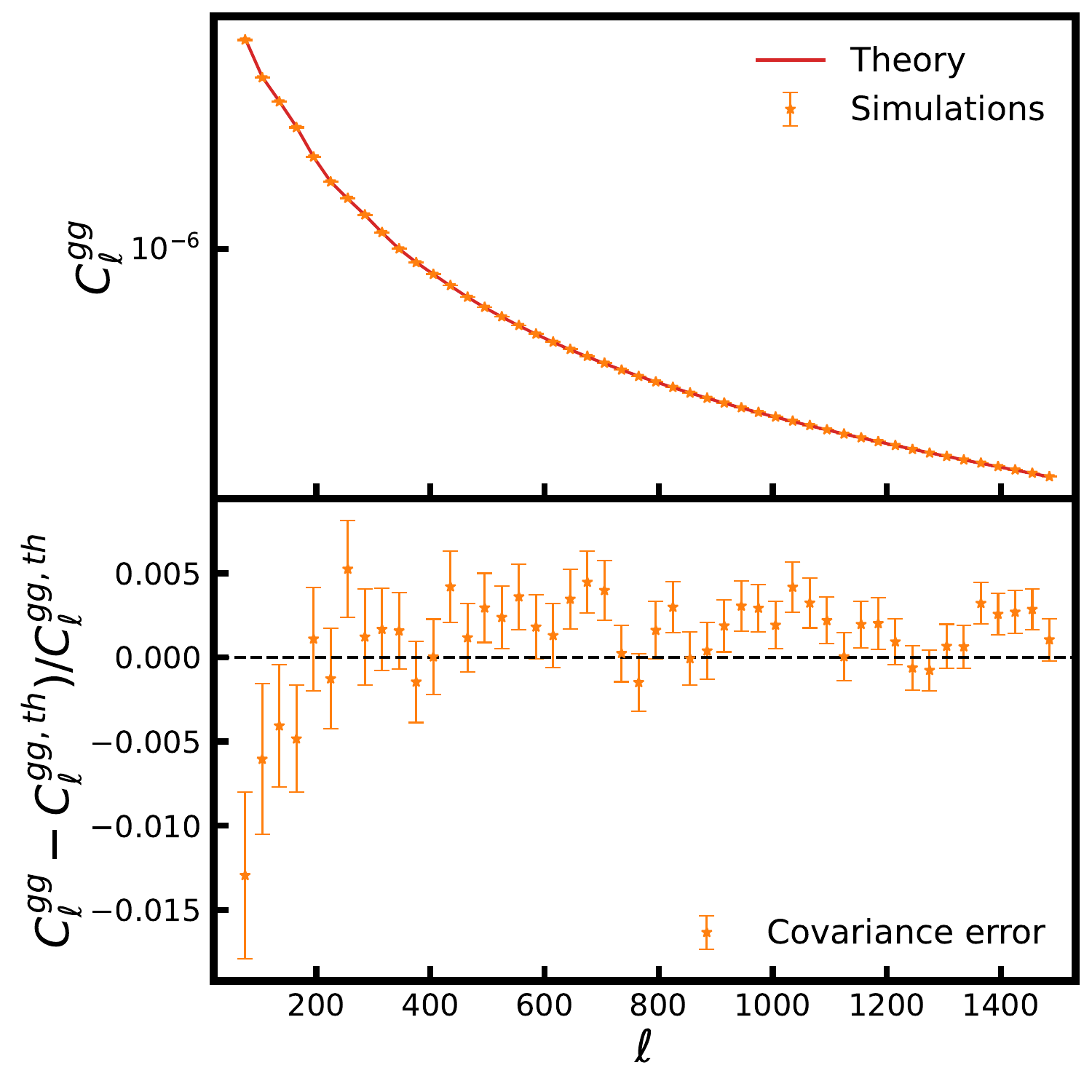}
    \end{subfigure}%
    \begin{subfigure}[b]{0.33\linewidth}
        \centering
        \includegraphics[width=\linewidth]{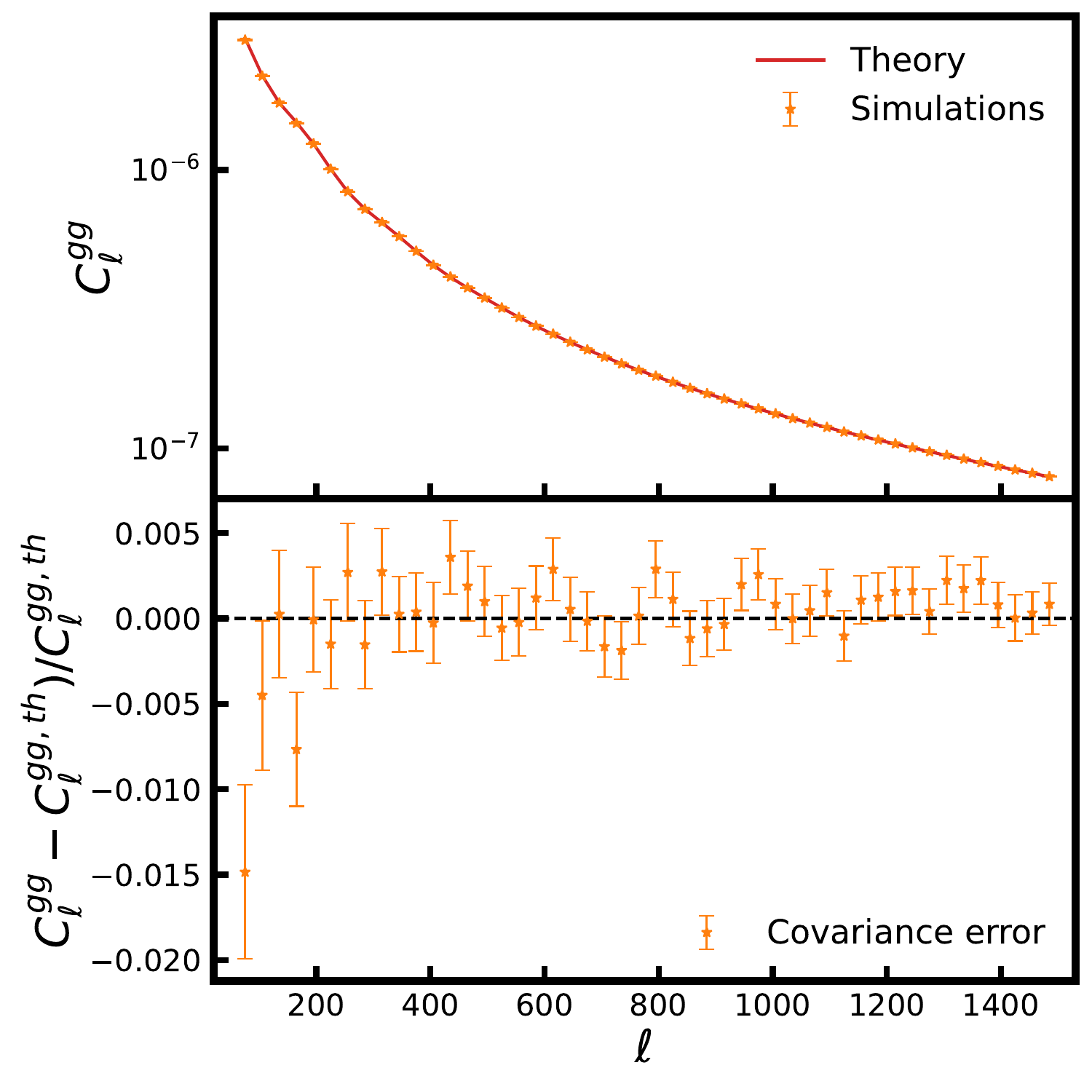}
    \end{subfigure}%
    \begin{subfigure}[b]{0.33\linewidth}
        \centering
        \includegraphics[width=\linewidth]{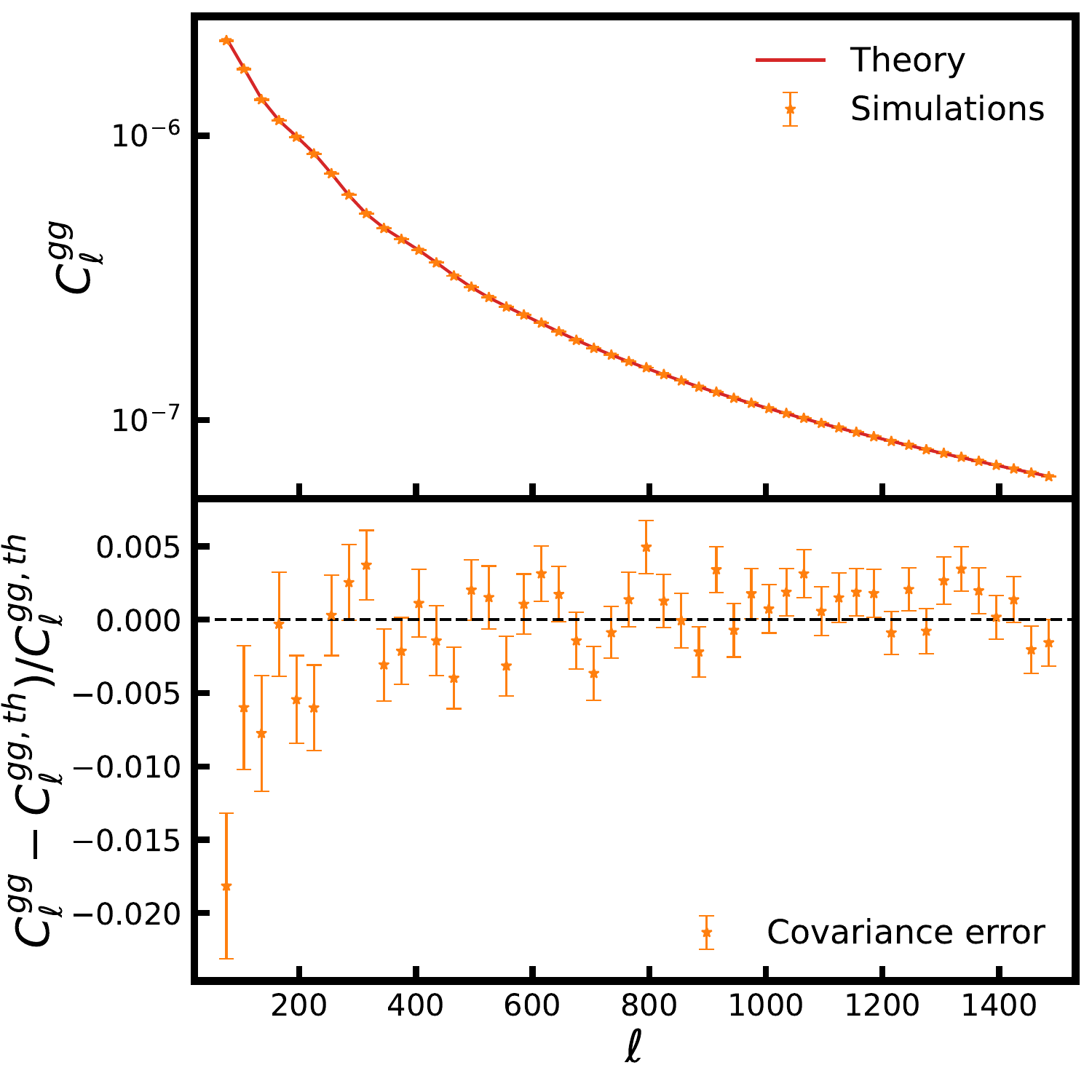}
    \end{subfigure}\\[1ex]
    \begin{subfigure}[b]{0.33\linewidth}
        \centering
        \includegraphics[width=\linewidth]{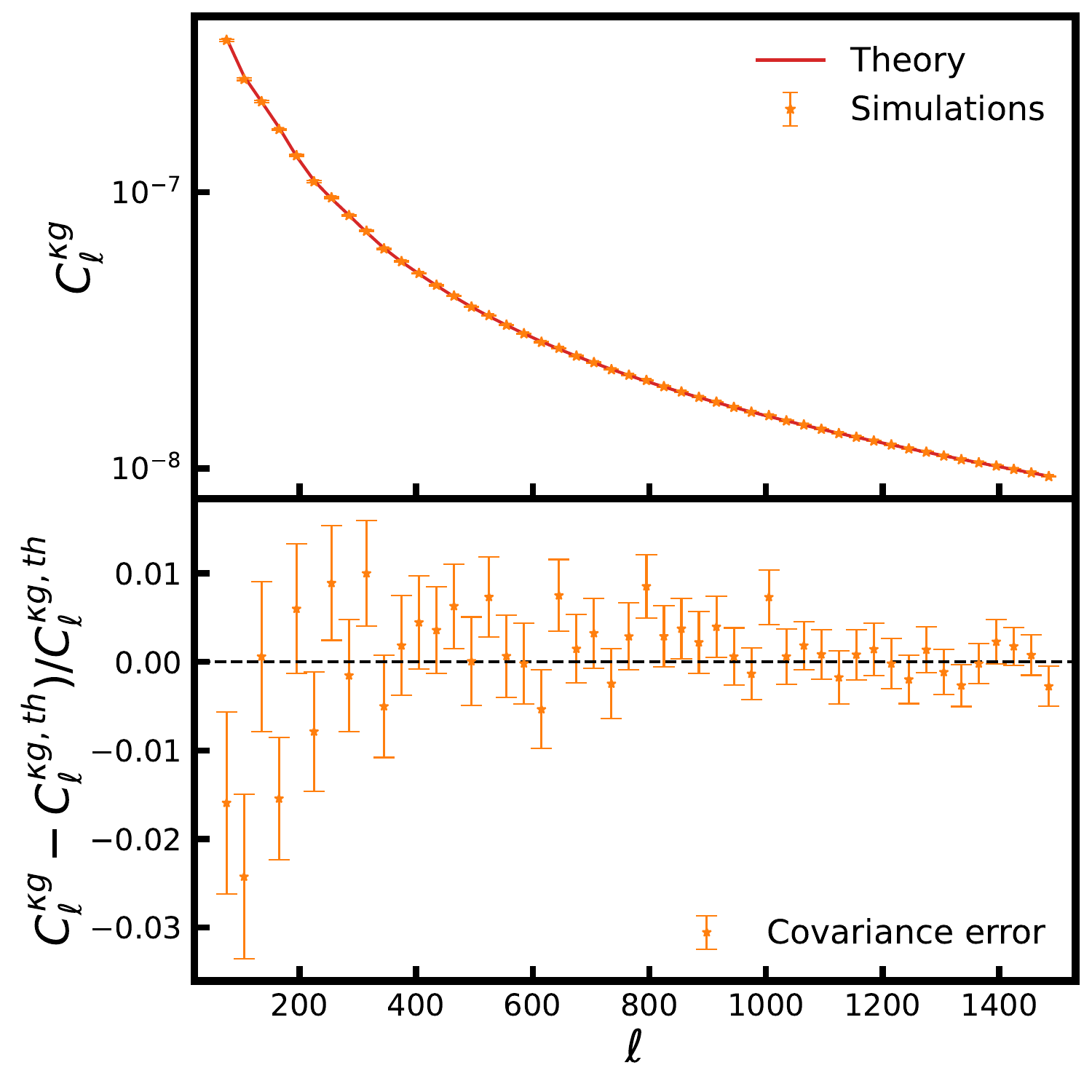}
        \captionsetup{labelformat=empty}
        \caption{\large{Bin 5 ($0.8\leq z<1.0$)}}
    \end{subfigure}%
    \begin{subfigure}[b]{0.33\linewidth}
        \centering
        \includegraphics[width=\linewidth]{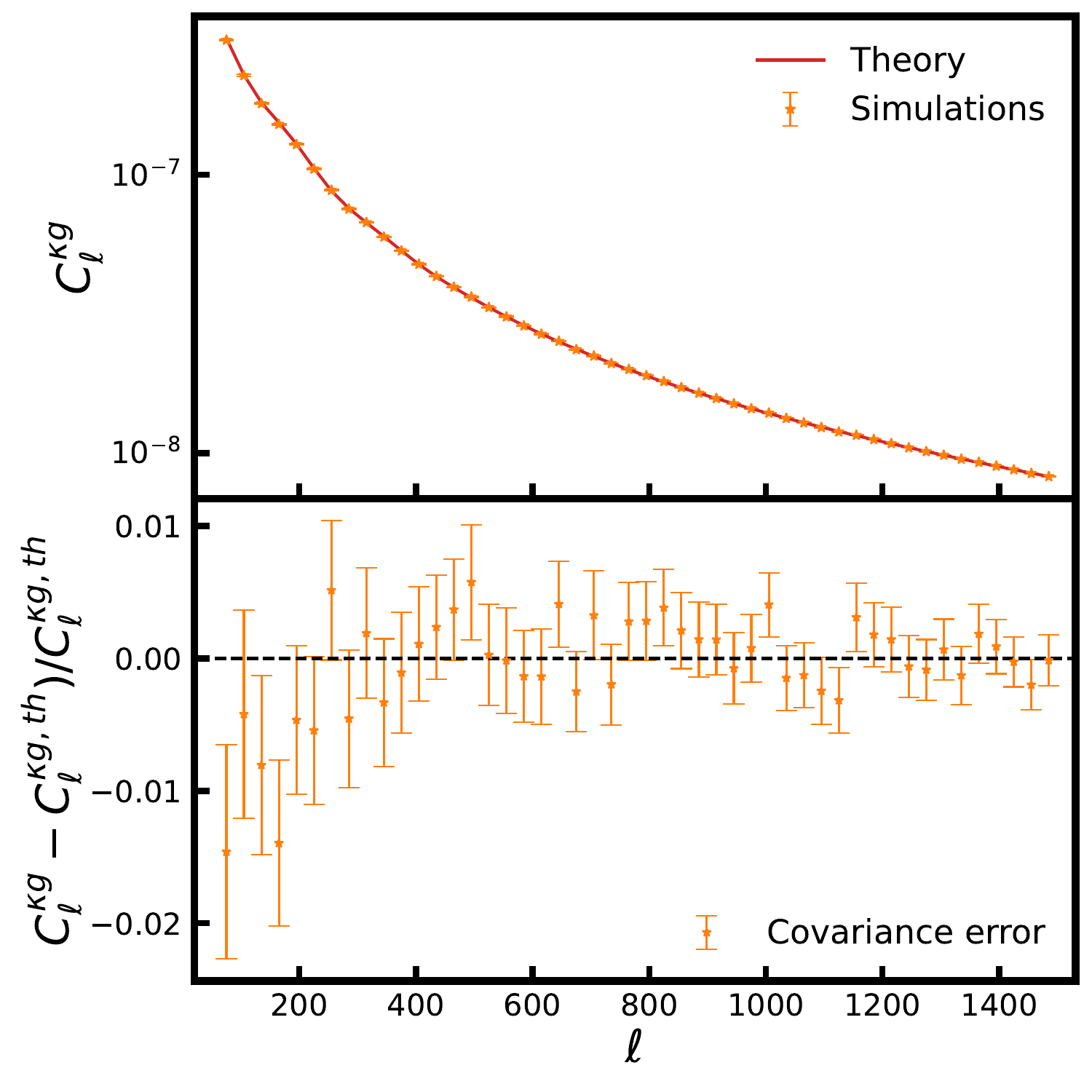}
        \captionsetup{labelformat=empty}
        \caption{\large{Bin 6 ($1.0\leq z<1.4$)}}
    \end{subfigure}%
    \begin{subfigure}[b]{0.33\linewidth}
        \centering
        \includegraphics[width=\linewidth]{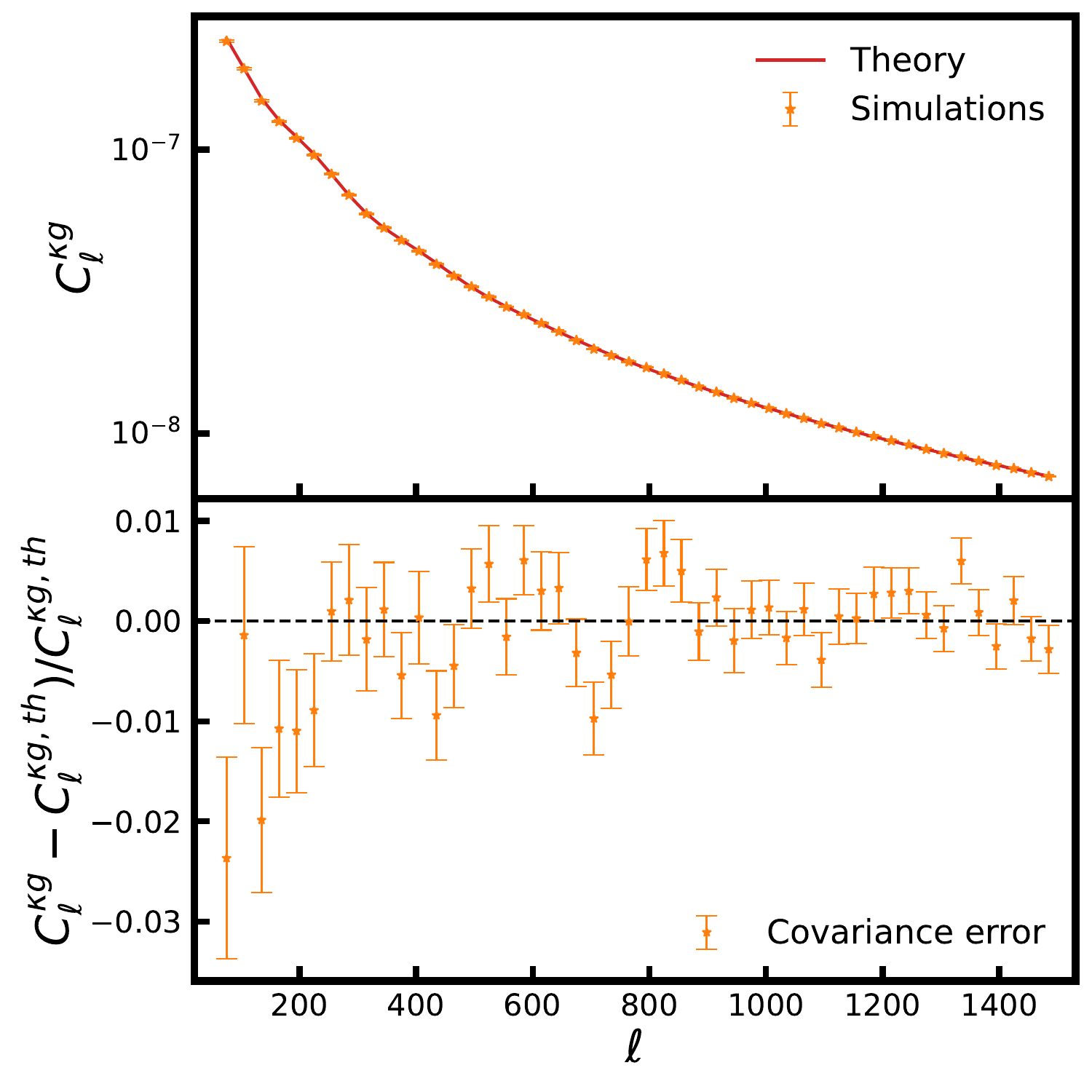}
        \captionsetup{labelformat=empty}
        \caption{\large{Bin 7 ($1.4\leq z<1.8$)}}
    \end{subfigure}
    \caption{Comparison of estimated angular power spectra with theoretical power spectra after correction for redshift bin mismatch. \textit{Top:} Average galaxy auto-power spectrum. \textit{Bottom:} Cross-power spectrum reconstructed from $300$ realisations of photometric datasets with a modified Lorentzian error distribution, shown for three tomographic bins. The red line represents the theoretical power spectrum corrected for the redshift bin mismatch using Eqs.\,(\ref{eq:scattering_relation_gg_matrix}) and (\ref{eq:scattering_relation_kg_matrix}). The error bars are computed from the covariance matrix of simulations using Eq.\,(\ref{eq:err_simul}).}
    \label{fig:plot_photo_from_scat_mat_dist_nsim_300_gamma_0.02}
\end{figure*}

{In Figs. \ref{fig:plot_photo_from_scat_mat_dist_nsim_300_0.02}-\ref{fig:plot_photo_from_scat_mat_dist_nsim_300_gamma_0.02}, we compare for the three tomographic bins shown in Fig. \ref{fig:plot_photo_from_true_conv_err_nsim_300} the noise-subtracted average estimated photometric power spectra with the theoretical power spectra corrected for the bin mismatch leakage using Eqs.\,(\ref{eq:scattering_relation_gg_matrix}) and (\ref{eq:scattering_relation_kg_matrix}).} The power spectra for other tomographic bins are presented in Appendix \ref{sec_apndx:power_spectra_scat_mat}. The theoretical power spectra after correction for leakage agree completely with the estimated power spectra in all bins, except for the first and last tomographic bin. The disparity in the first and last bins results directly from the inaccuracy of the convolution method near the lower and upper bounds of the redshift distribution considered in the analysis. Nevertheless, we note that even for these tomographic bins, the agreement with corresponding theoretical power spectra improves.


\section{Parameter estimation}\label{sec:parameter_estimation}

In previous sections, we observed that the power spectra in every tomographic bin become biased by the leakage of objects across redshift bins, which can be corrected for by an accurate estimation of the scattering matrix. In this section, we study the impact of the leakage on the estimation of the redshift-dependent linear galaxy bias $b$ and on the amplitude of the cross correlation $A$ from tomographic bins, estimated using the maximum likelihood estimation method discussed in section \ref{sec:likeli_params}.

Before accounting for the leakage, we estimated the linear galaxy bias and cross-correlation amplitude for every tomographic bin using the average galaxy power spectra and the average cross-power spectra estimated from the photometric datasets. The theoretical power spectrum templates for a tomographic bin $i$ were computed using Eq.\,(\ref{eq:power_spectra}) with the redshift distributions given by Eq.\,(\ref{eq:true_dist_conv}). To estimate the parameters $b$ and $A$ after correcting for the redshift bin mismatch, we transformed the extracted photometric power spectra ($\overline{C}^{gg,\text{ph}}$ and $\overline{C}^{\kappa g,\text{ph}}$) into true power spectra by inverting Eqs.\,(\ref{eq:scattering_relation_gg_matrix}) and (\ref{eq:scattering_relation_kg_matrix}). The theoretical power spectrum templates for the likelihood estimation after correction for leakage were computed by substituting Eq.\,(\ref{eq:true_dist_binned}) for the redshift distribution in Eq.\,(\ref{eq:power_spectra}). It is important to note that the photometric power spectra in the tomographic analysis are a combination of true power spectra as represented in Eqs.\,(\ref{eq:scattering_relation_gg_matrix}) and (\ref{eq:scattering_relation_kg_matrix}). Thus, the linear galaxy bias in a photometric redshift bin is also a combination of the linear galaxy bias from the true redshift bins. The parameters can also be directly estimated over the estimated photometric power spectra by properly defining the covariance matrix in the likelihood function. However, transforming the estimated photometric power spectra into true power spectra to estimate the parameter avoids complexities based on defining the covariance matrix and also reduces the computation time.

{In Figs. \ref{fig:plot_likeli_nsim_300_0.02}-\ref{fig:plot_likeli_nsim_300_gamma_0.02}, we compare the posterior distribution of the parameters estimated from the average power spectra with $\sigma_{0} = 0.02,\,\sigma_{0} = 0.05 \text{, and } \gamma_{0} = 0.02$ for three tomographic bins before and after leakage correction.} The lighter and darker shaded contour represents the $68\%$ and $95\%$ confidence intervals. The red lines mark the true values of the galaxy bias and cross-correlation amplitude. {The best-fit values of the linear galaxy bias and cross-correlation amplitude from all tomographic bins with $\pm 1\,\sigma$ errors, estimated before and after correction for leakage, are quoted in Tables \ref{tab:likeli_params_comp_0.02}-\ref{tab:likeli_params_comp_gamma_0.02} for $\sigma_{0} = 0.02,\,\sigma_{0} = 0.05 \text{, and } \gamma_{0} = 0.02$.} Column $b_{\text{true}}$ contains the true values of bias for every tomographic bin. {The linear galaxy bias for every tomographic bin estimated from photometric datasets is smaller than their expected value for both Case-I and Case-II, whereas the amplitude of cross correlation is consistently higher than the expected value of unity.} However, both parameters become consistent with their expected values after we corrected for the redshift bin mismatch of the objects. {In Fig. \ref{fig:lsst_tomography_gal_bias_parameter} we show the variations in the redshift evolution of the linear galaxy bias parameter due to the effects from the redshift bin mismatch of the objects for Case-I and Case-II.} The dashed black line marks the fiducial evolution of the galaxy bias used in our simulations. The linear galaxy bias shows marginal deviations from its true values after correction with the scattering matrix, and the amplitude of the cross correlation is perfectly consistent with its expected value of unity within $1\,\sigma$.

\begin{figure*}
    \begin{subfigure}[b]{0.33\linewidth}
        \centering
        \includegraphics[width=\linewidth]{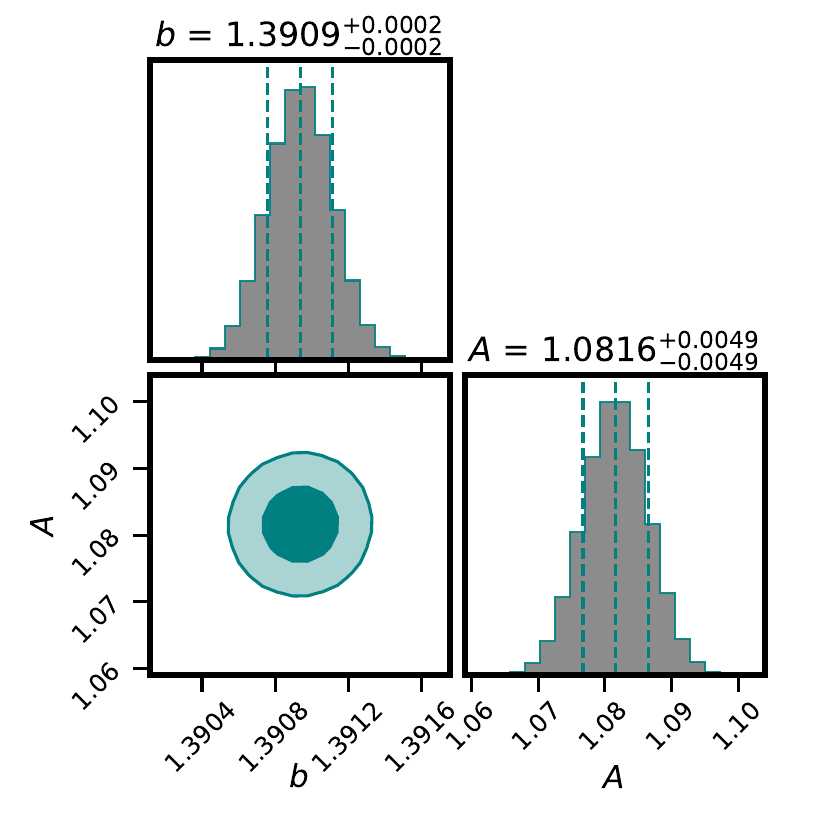}
        \captionsetup{labelformat=empty}
        \caption{\large{Bin 5 ($0.8\leq z<1.0$)}}
    \end{subfigure}%
    \begin{subfigure}[b]{0.33\linewidth}
        \centering
        \includegraphics[width=\linewidth]{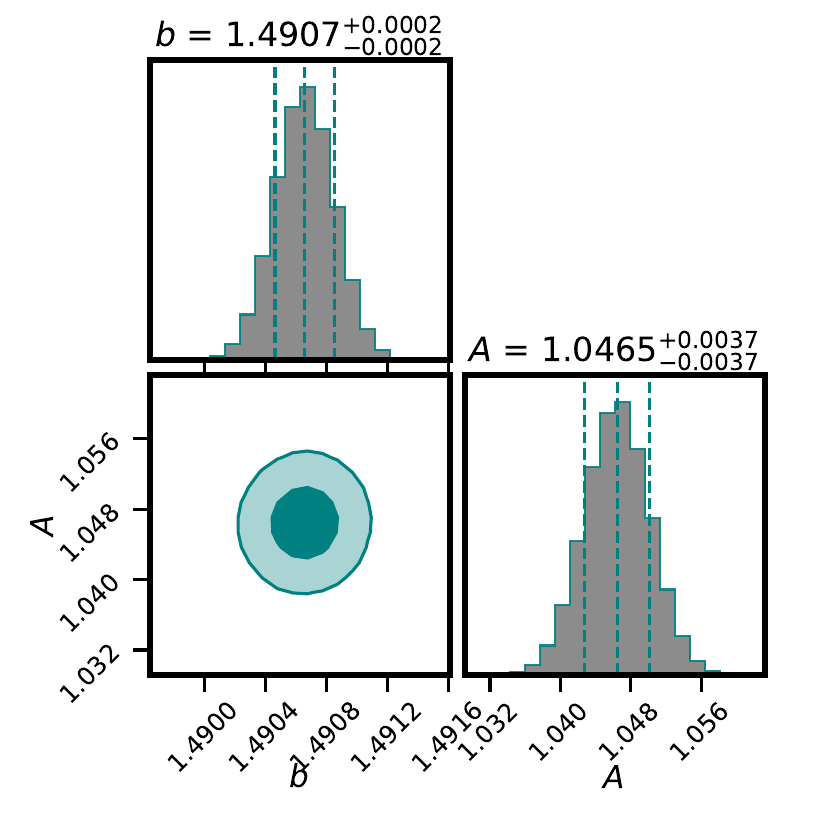}
        \captionsetup{labelformat=empty}
        \caption{\large{Bin 6 ($1.0\leq z<1.4$)}}
    \end{subfigure}%
    \begin{subfigure}[b]{0.33\linewidth}
        \centering
        \includegraphics[width=\linewidth]{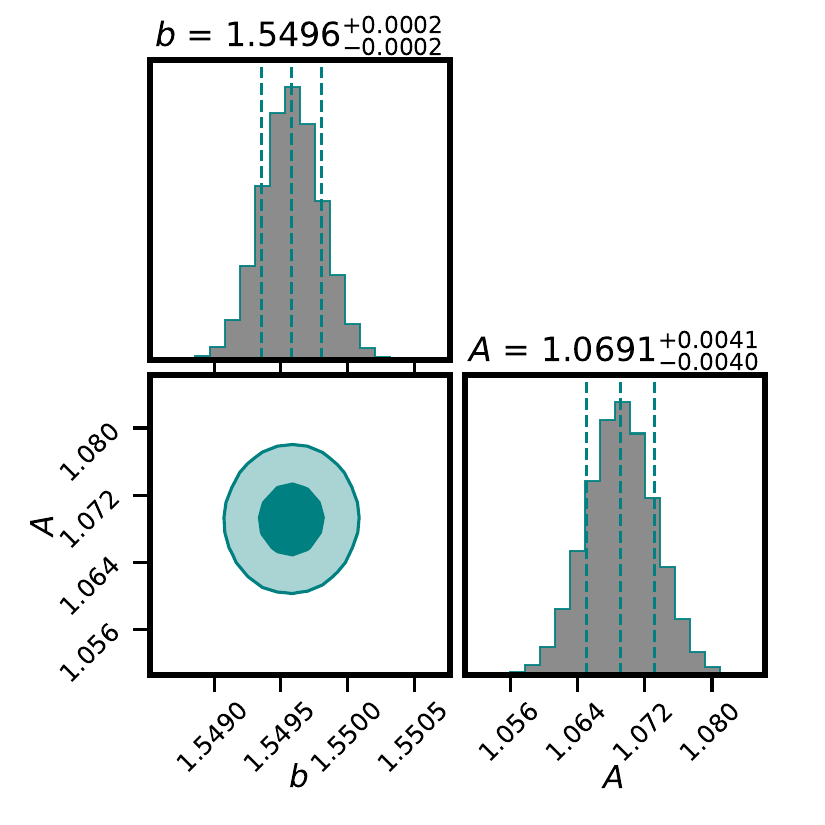}
        \captionsetup{labelformat=empty}
        \caption{\large{Bin 7 ($1.4\leq z<1.8$)}}
    \end{subfigure}\\
    \begin{subfigure}[b]{0.33\linewidth}
        \centering
        \includegraphics[width=\linewidth]{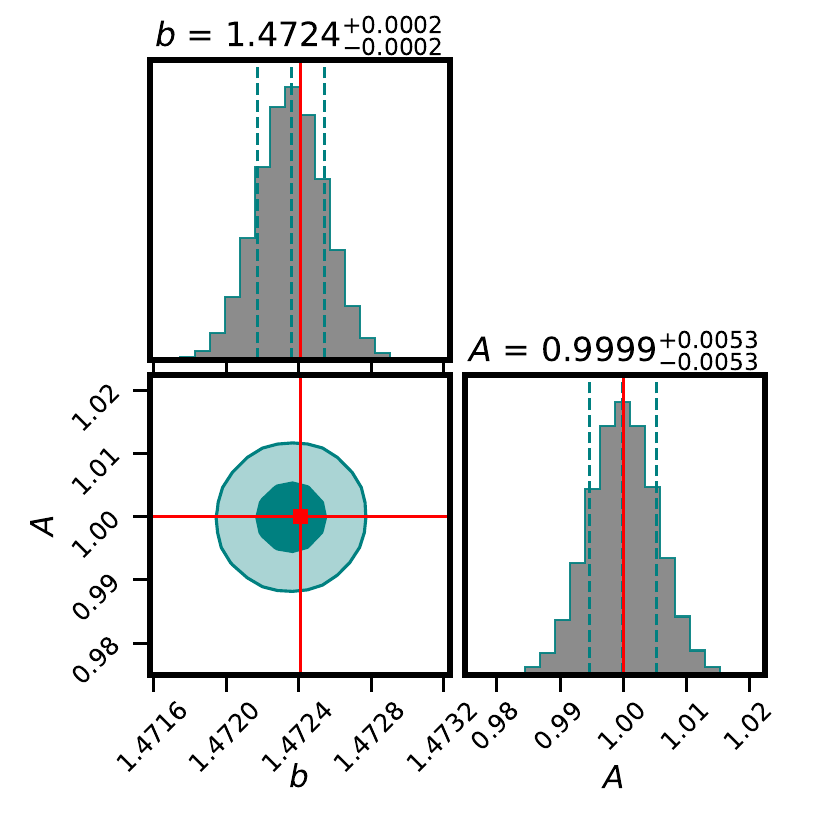}
        \captionsetup{labelformat=empty}
        \caption{\large{Bin 5 ($0.8\leq z<1.0$)}}
    \end{subfigure}%
    \begin{subfigure}[b]{0.33\linewidth}
        \centering
        \includegraphics[width=\linewidth]{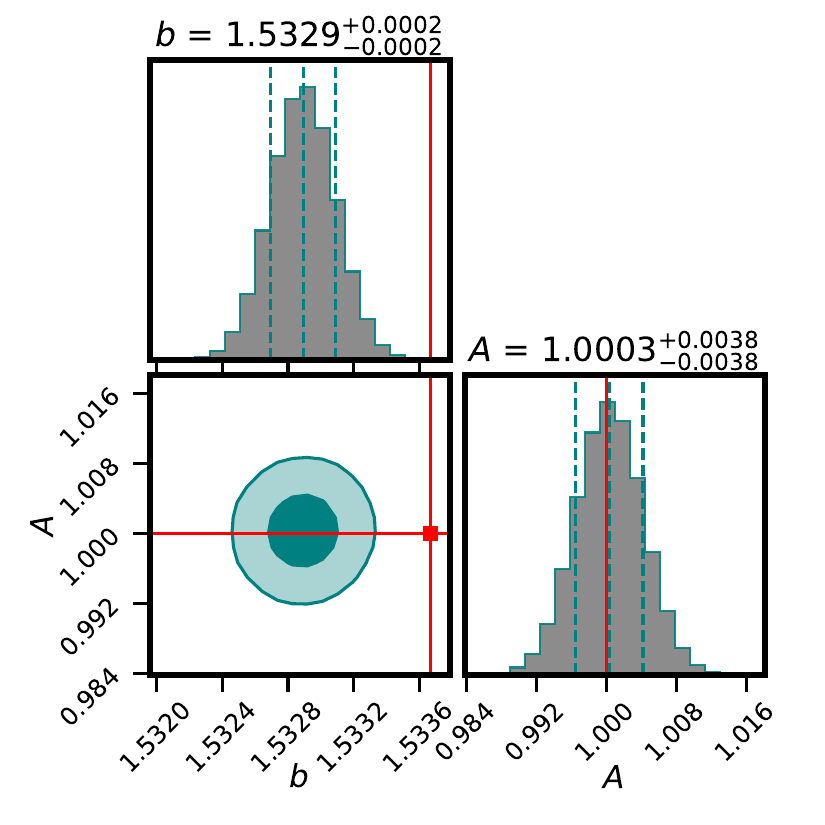}
        \captionsetup{labelformat=empty}
        \caption{\large{Bin 6 ($1.0\leq z<1.4$)}}
    \end{subfigure}%
    \begin{subfigure}[b]{0.33\linewidth}
        \centering
        \includegraphics[width=\linewidth]{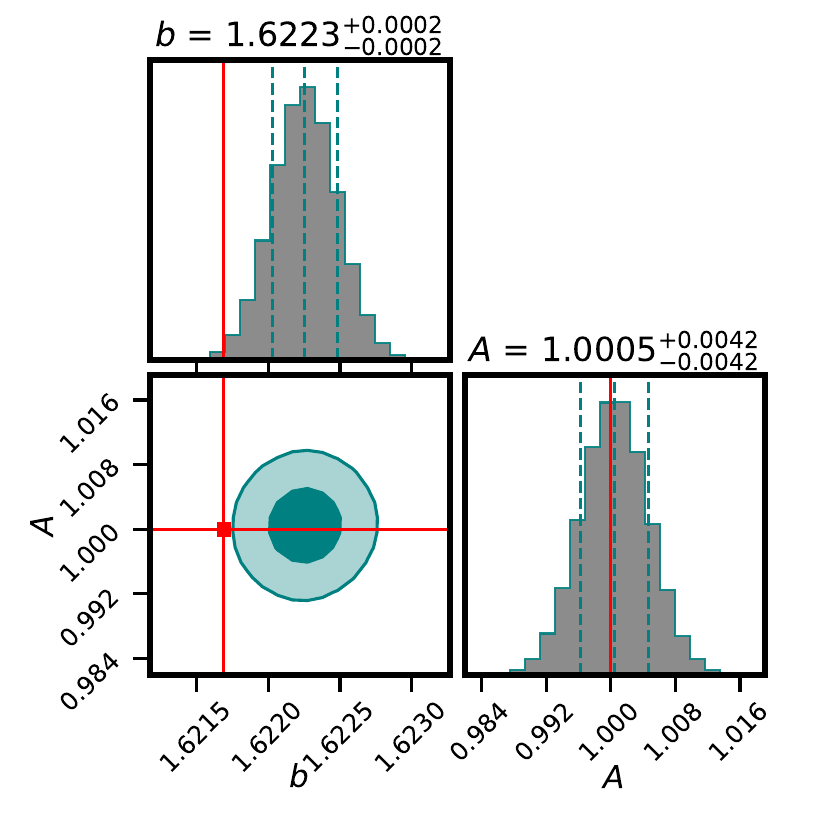}
        \captionsetup{labelformat=empty}
        \caption{\large{Bin 7 ($1.4\leq z<1.8$)}}
    \end{subfigure}
    \caption{Parameter posteriors obtained from the maximum likelihood estimation performed over the average power spectra from $300$ simulations with $\sigma_{0}=0.02$. The \textit{top} and \textit{bottom} panels show the posteriors before and after correction for leakage, respectively. The three vertical dashed lines are the median value of the posterior and $\pm 1\sigma$ errors. The $68\%$ and $95\%$ confidence contours are shown in darker and lighter shades, respectively. The red lines are the true values of parameters used for simulations.}
    \label{fig:plot_likeli_nsim_300_0.02}
\end{figure*}

\begin{figure*}
    \begin{subfigure}[b]{0.33\linewidth}
        \centering
        \includegraphics[width=\linewidth]{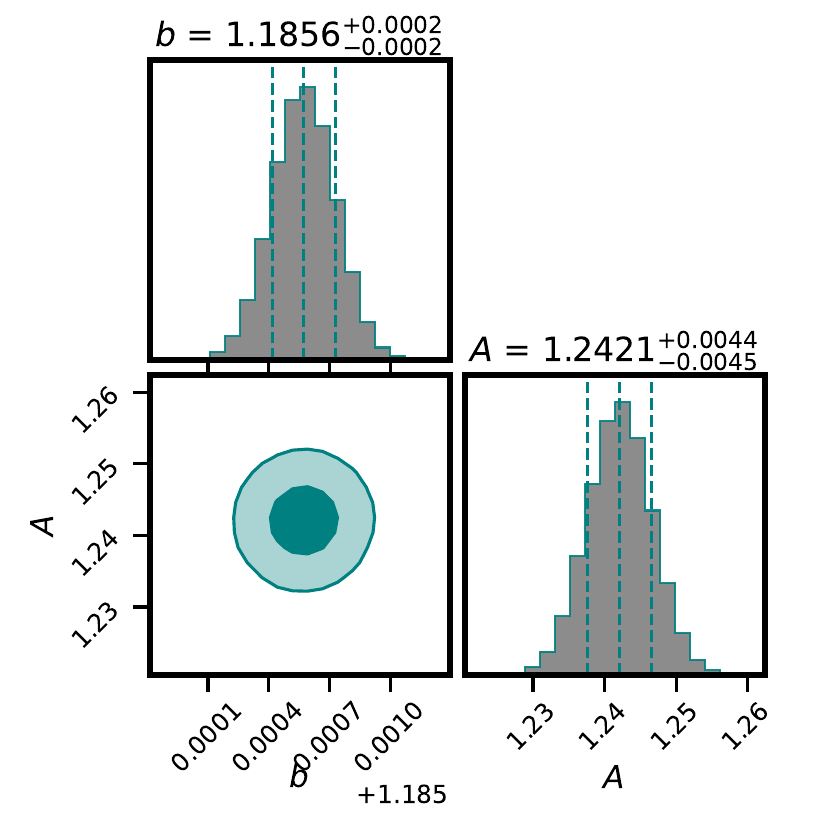}
        \captionsetup{labelformat=empty}
        \caption{\large{Bin 5 ($0.8\leq z<1.0$)}}
    \end{subfigure}%
    \begin{subfigure}[b]{0.33\linewidth}
        \centering
        \includegraphics[width=\linewidth]{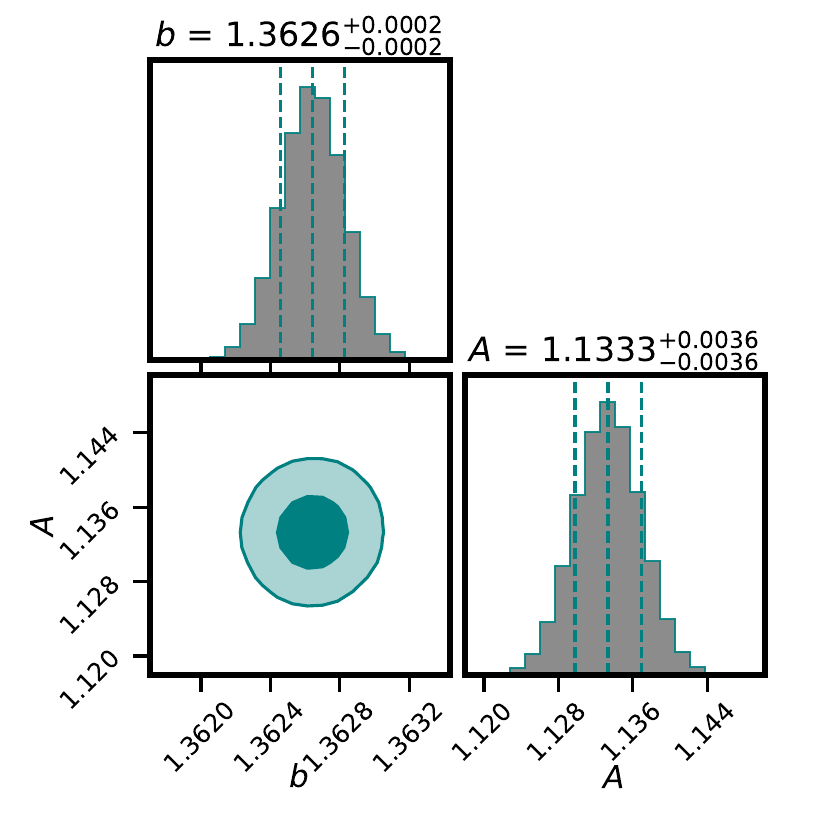}
        \captionsetup{labelformat=empty}
        \caption{\large{Bin 6 ($1.0\leq z<1.4$)}}
    \end{subfigure}%
    \begin{subfigure}[b]{0.33\linewidth}
        \centering
        \includegraphics[width=\linewidth]{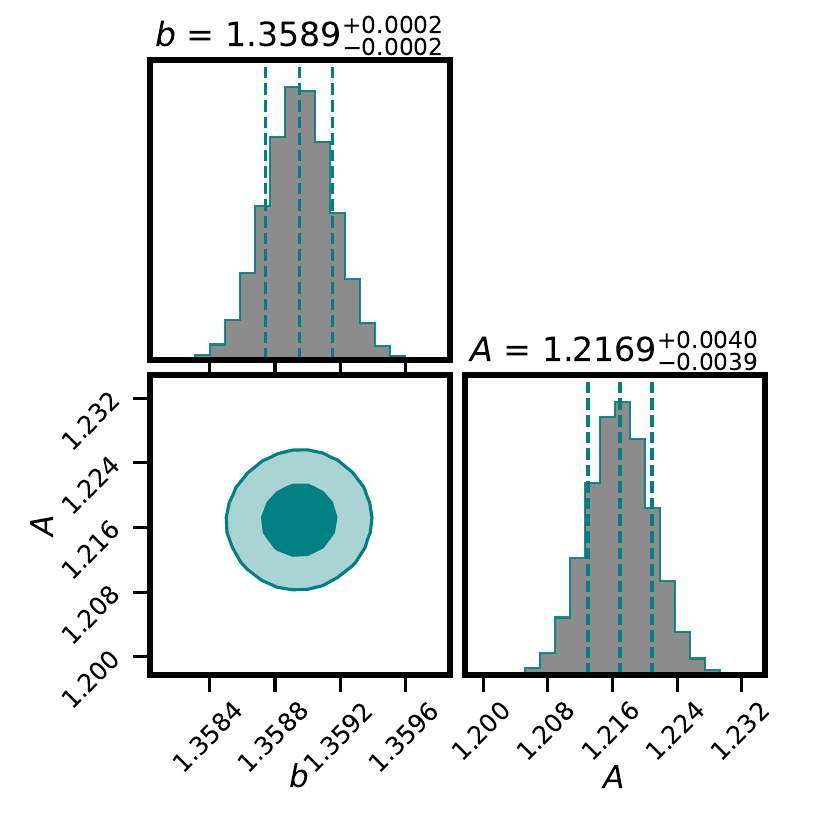}
        \captionsetup{labelformat=empty}
        \caption{\large{Bin 7 ($1.4\leq z<1.8$)}}
    \end{subfigure}\\
    \begin{subfigure}[b]{0.33\linewidth}
        \centering
        \includegraphics[width=\linewidth]{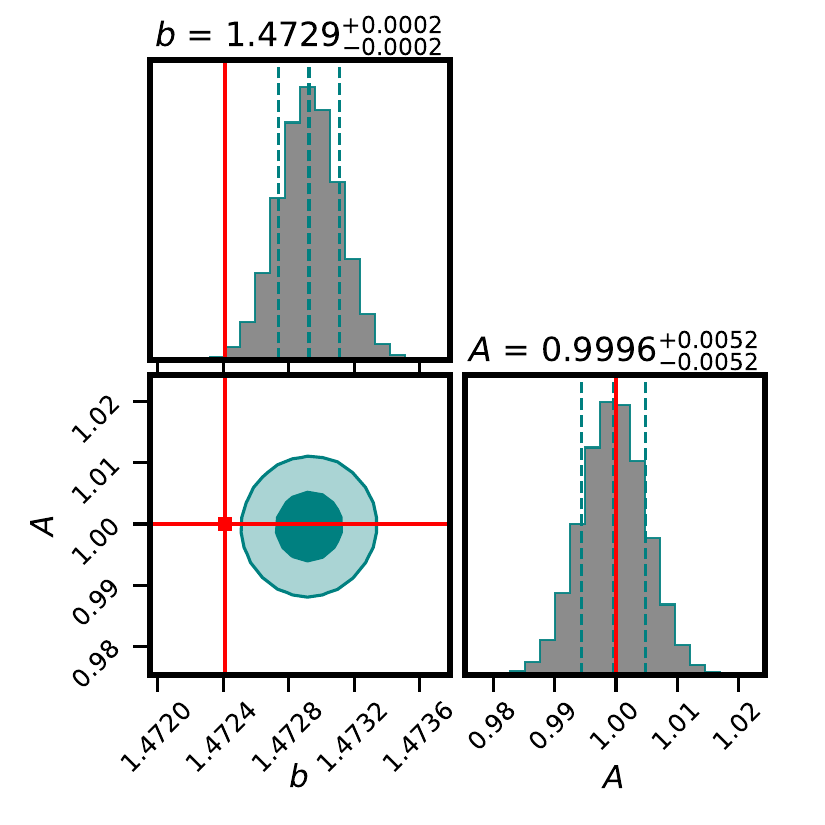}
        \captionsetup{labelformat=empty}
        \caption{\large{Bin 5 ($0.8\leq z<1.0$)}}
    \end{subfigure}%
    \begin{subfigure}[b]{0.33\linewidth}
        \centering
        \includegraphics[width=\linewidth]{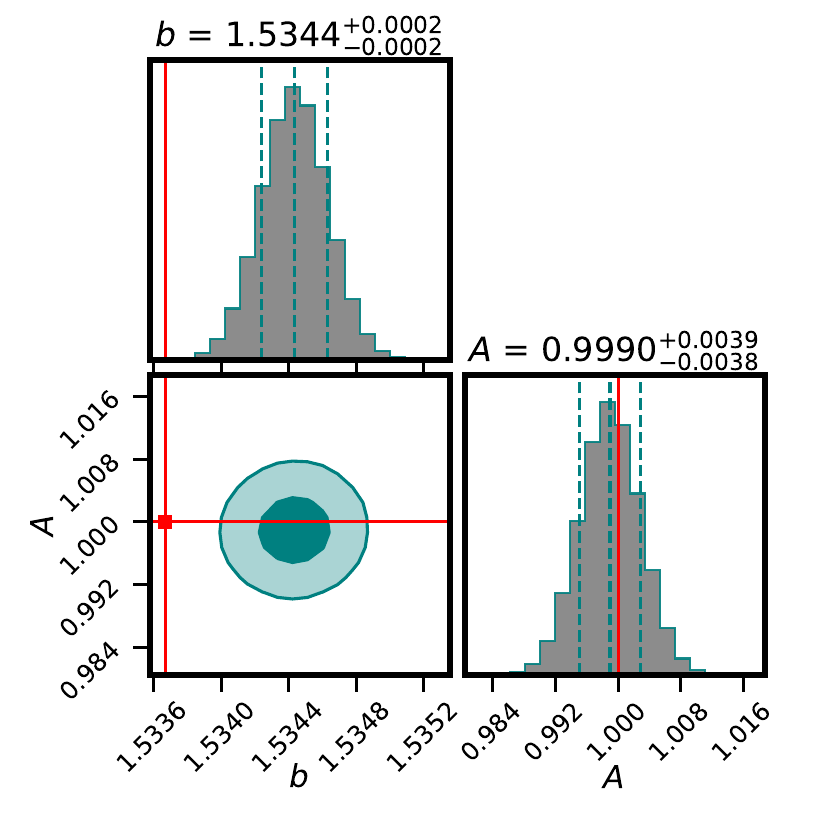}
        \captionsetup{labelformat=empty}
        \caption{\large{Bin 6 ($1.0\leq z<1.4$)}}
    \end{subfigure}%
    \begin{subfigure}[b]{0.33\linewidth}
        \centering
        \includegraphics[width=\linewidth]{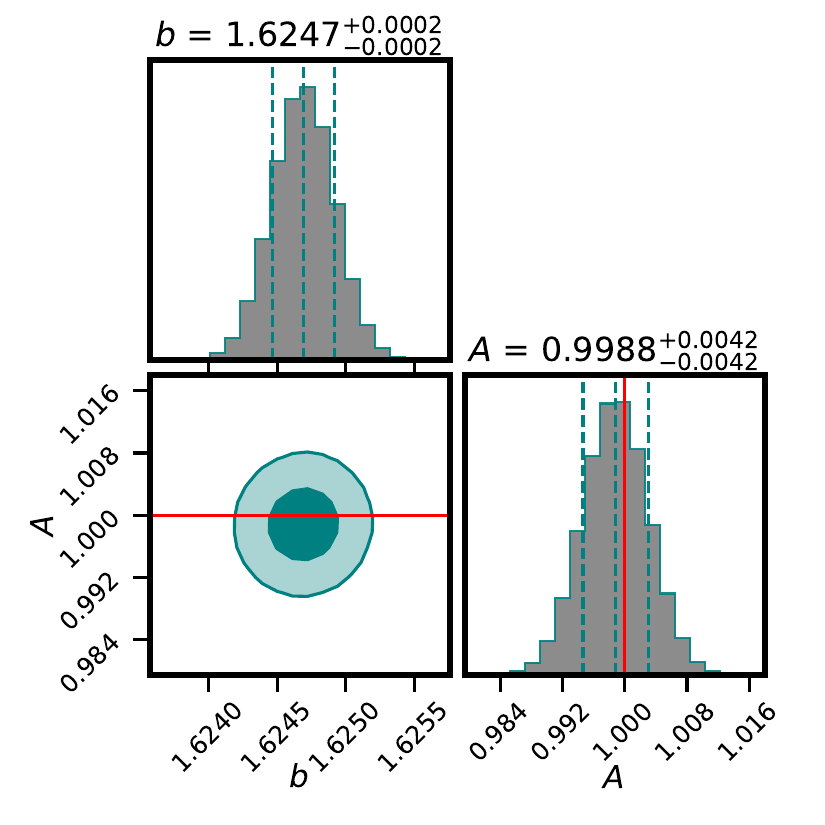}
        \captionsetup{labelformat=empty}
        \caption{\large{Bin 7 ($1.4\leq z<1.8$)}}
    \end{subfigure}
    \caption{Parameter posteriors obtained from the maximum likelihood estimation performed over the average power spectra from $300$ simulations with $\sigma_{0}=0.05$. The \textit{top} and \textit{bottom} panels show the posteriors before and after correction for leakage, respectively. The three vertical dashed lines are the median value of the posterior and $\pm 1\sigma$ errors. The $68\%$ and $95\%$ confidence contours are shown in darker and lighter shades, respectively. The red lines are the true values of the parameters used for simulations.}
    \label{fig:plot_likeli_nsim_300_0.05}
\end{figure*}

\begin{figure*}
    \begin{subfigure}[b]{0.33\linewidth}
        \centering
        \includegraphics[width=\linewidth]{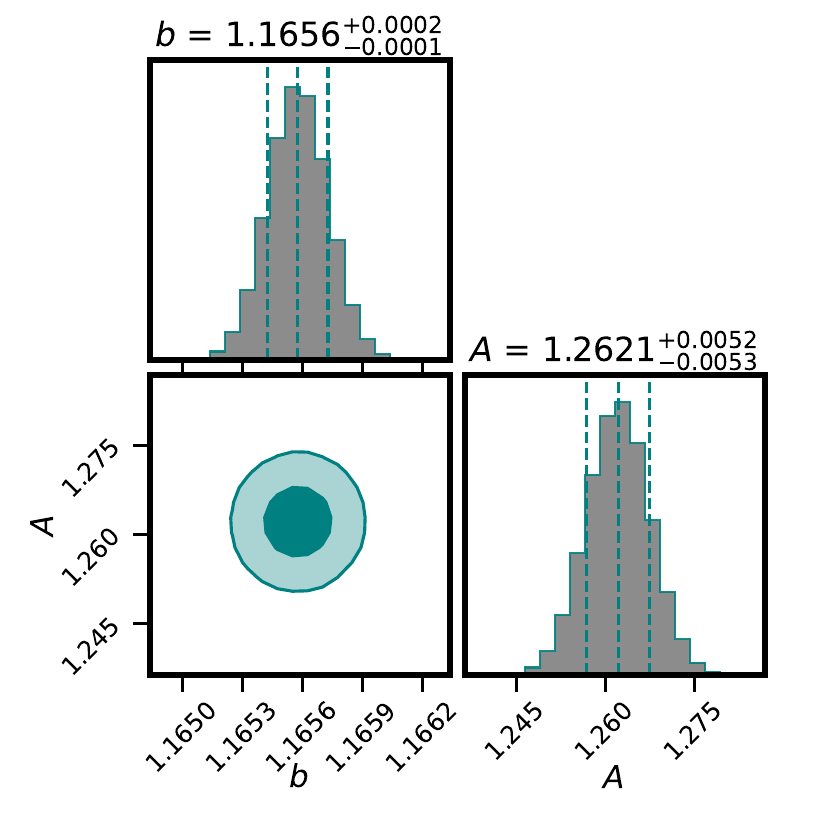}
        \captionsetup{labelformat=empty}
        \caption{\large{Bin 5 ($0.8\leq z<1.0$)}}
    \end{subfigure}%
    \begin{subfigure}[b]{0.33\linewidth}
        \centering
        \includegraphics[width=\linewidth]{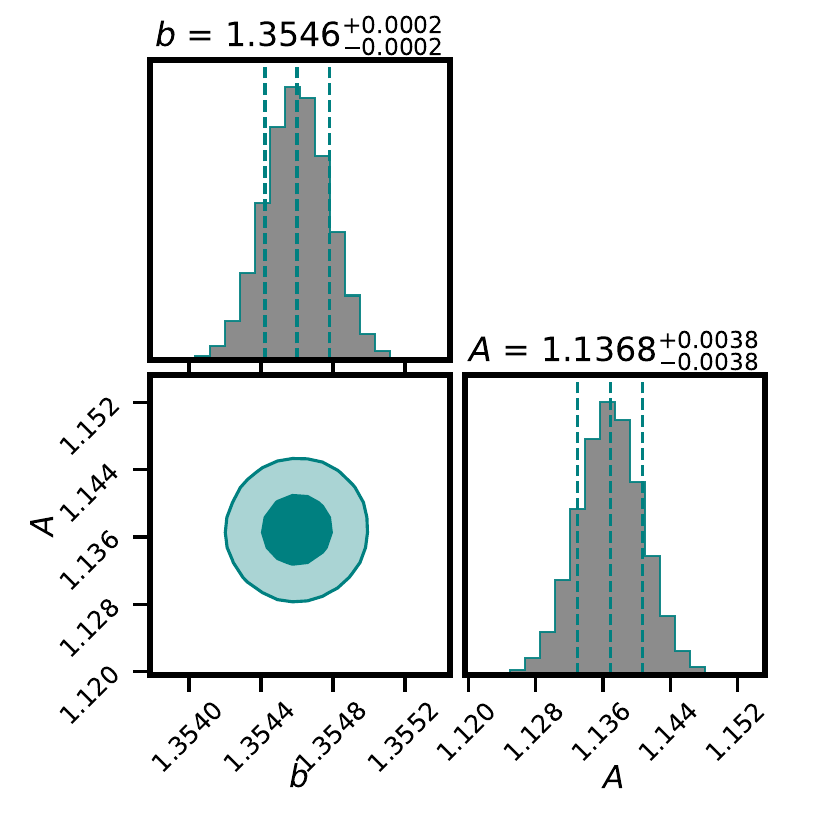}
        \captionsetup{labelformat=empty}
        \caption{\large{Bin 6 ($1.0\leq z<1.4$)}}
    \end{subfigure}%
    \begin{subfigure}[b]{0.33\linewidth}
        \centering
        \includegraphics[width=\linewidth]{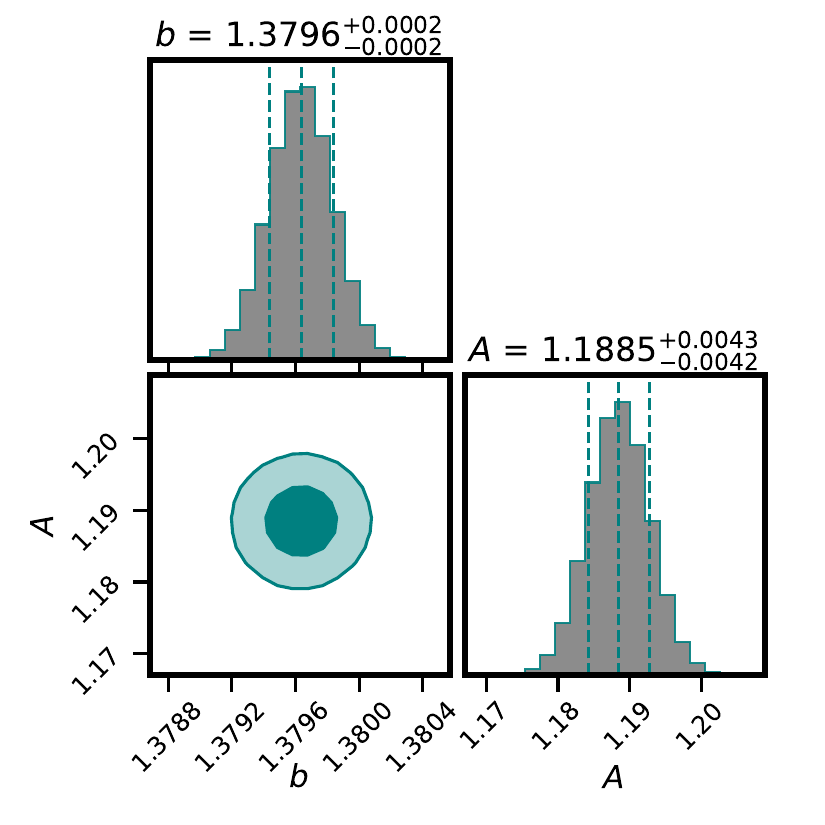}
        \captionsetup{labelformat=empty}
        \caption{\large{Bin 7 ($1.4\leq z<1.8$)}}
    \end{subfigure}\\
    \begin{subfigure}[b]{0.33\linewidth}
        \centering
        \includegraphics[width=\linewidth]{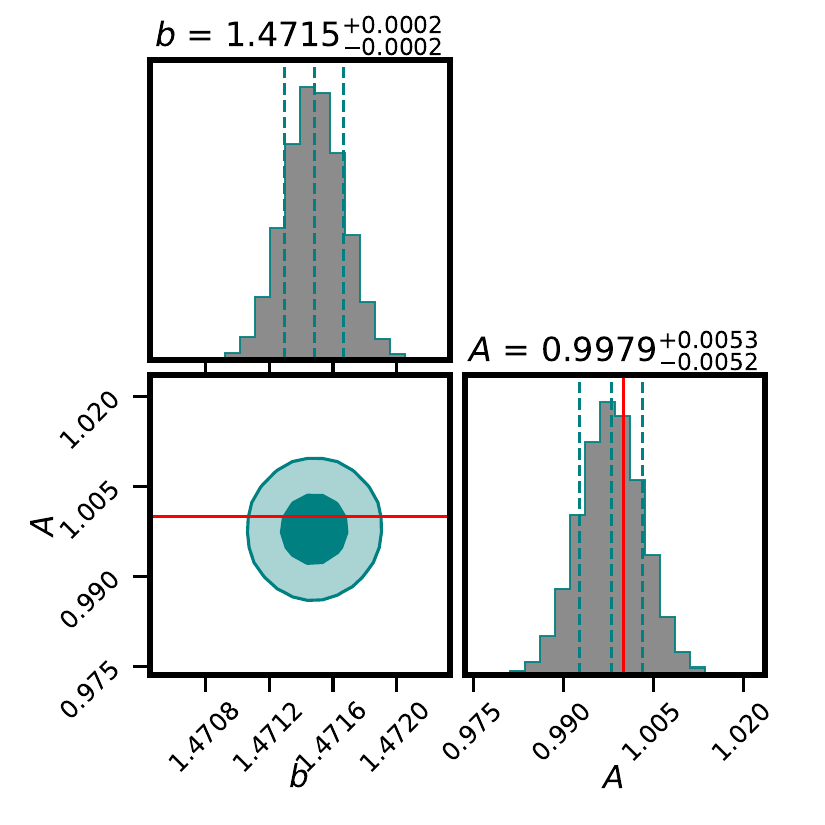}
        \captionsetup{labelformat=empty}
        \caption{\large{Bin 5 ($0.8\leq z<1.0$)}}
    \end{subfigure}%
    \begin{subfigure}[b]{0.33\linewidth}
        \centering
        \includegraphics[width=\linewidth]{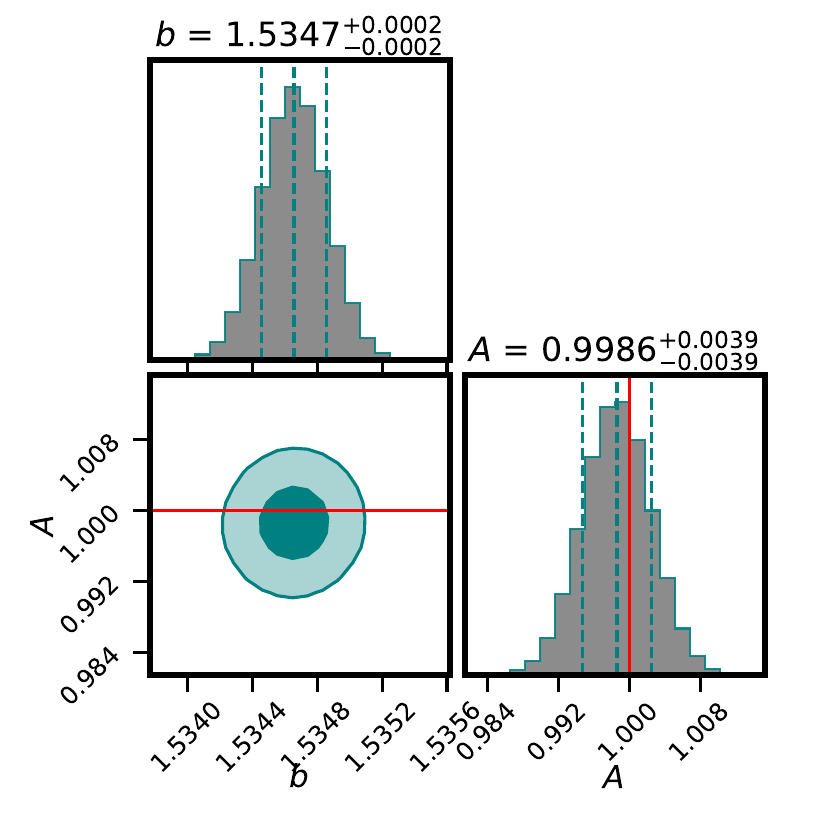}
        \captionsetup{labelformat=empty}
        \caption{\large{Bin 6 ($1.0\leq z<1.4$)}}
    \end{subfigure}%
    \begin{subfigure}[b]{0.33\linewidth}
        \centering
        \includegraphics[width=\linewidth]{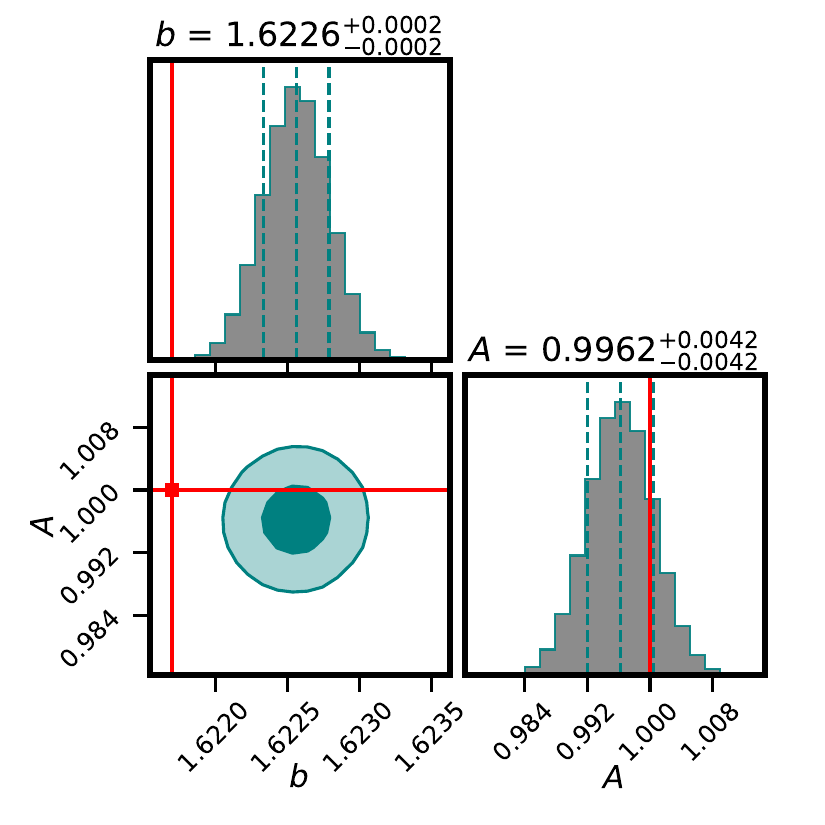}
        \captionsetup{labelformat=empty}
        \caption{\large{Bin 7 ($1.4\leq z<1.8$)}}
    \end{subfigure}
    \caption{{Parameter posteriors obtained from the maximum likelihood estimation performed over the average power spectra from $300$ simulations with a modified Lorentzian error distribution. The \textit{top} and \textit{bottom} panels show the posteriors before and after the correction for leakage, respectively. The three vertical dashed lines are the median value of the posterior and $\pm 1\sigma$ errors. The $68\%$ and $95\%$ confidence contours are shown in darker and lighter shades, respectively. The red lines are the true values of the parameters used for simulations.}}
    \label{fig:plot_likeli_nsim_300_gamma_0.02}
\end{figure*}

\begin{table*}[hbt!]
    \renewcommand{\arraystretch}{1.5}
    \centering
    \caption{Best-fit values of the linear galaxy bias $b$ and the amplitude of cross-correlation $A$ from simulations with $\sigma_{0}=0.02$.}
    \label{tab:likeli_params_comp_0.02}
    \begin{tabular}{cc||cc||cc} 
    \hline\hline
    $z$ & $b^{\text{true}}$ & \multicolumn{2}{c}{No correction} & \multicolumn{2}{c}{With correction}\\
    \cline{3-4}\cline{5-6}
    & & $b$ & $A$ & $b$ & $A$\\
    \hline\hline
        $[0.0,0.2)$ & $1.3241$ & $1.3552^{+0.0002}_{-0.0002}$ & $1.0243^{+0.0110}_{-0.0110}$ & $1.3357^{+0.0002}_{-0.0002}$ & $0.9952^{+0.0124}_{-0.0125}$\\
        $[0.2,0.4)$ & $1.3541$ & $1.3497^{+0.0002}_{-0.0002}$ & $1.0269^{+0.0064}_{-0.0064}$ & $1.3557^{+0.0002}_{-0.0002}$ & $1.0014^{+0.0069}_{-0.0069}$\\
        $[0.4,0.6)$ & $1.3909$ & $1.3623^{+0.0002}_{-0.0002}$ & $1.0440^{+0.0053}_{-0.0054}$ & $1.3911^{+0.0002}_{-0.0002}$ & $0.9990^{+0.0058}_{-0.0057}$\\
        $[0.6,0.8)$ & $1.4307$ & $1.3772^{+0.0002}_{-0.0002}$ & $1.0645^{+0.0050}_{-0.0050}$ & $1.4308^{+0.0002}_{-0.0002}$ & $0.9998^{+0.0054}_{-0.0053}$\\
        $[0.8,1.0)$ & $1.4724$ & $1.3909^{+0.0002}_{-0.0002}$ & $1.0816^{+0.0049}_{-0.0049}$ & $1.4724^{+0.0002}_{-0.0002}$ & $0.9999^{+0.0053}_{-0.0053}$\\
        $[1.0,1.4)$ & $1.5337$ & $1.4907^{+0.0002}_{-0.0002}$ & $1.0465^{+0.0037}_{-0.0037}$ & $1.5329^{+0.0002}_{-0.0002}$ & $1.0003^{+0.0038}_{-0.0038}$\\
        $[1.4,1.8)$ & $1.6217$ & $1.5496^{+0.0002}_{-0.0002}$ & $1.0691^{+0.0041}_{-0.0040}$ & $1.6223^{+0.0002}_{-0.0002}$ & $1.0005^{+0.0042}_{-0.0042}$\\
        $[1.8,2.2)$ & $1.7118$ & $1.5832^{+0.0003}_{-0.0003}$ & $1.0963^{+0.0047}_{-0.0047}$ & $1.7149^{+0.0003}_{-0.0003}$ & $0.9986^{+0.0048}_{-0.0048}$\\
        $[2.2,3.0)$ & $1.8277$ & $1.7318^{+0.0005}_{-0.0005}$ & $1.0668^{+0.0047}_{-0.0047}$ & $1.8137^{+0.0004}_{-0.0004}$ & $1.0050^{+0.0048}_{-0.0047}$\\
    \hline
    \end{tabular}
    \tablefoot{The galaxy bias $b$ and amplitude of cross-correlation $A$ are estimated from the average power spectra of $300$ simulations before the correction for leakage and after correction through the scattering matrix approach. $b^{\text{true}}$ is the true value of the galaxy bias for the tomographic bins. The error bars on the parameters correspond to the average of $300$ realisations.}
\end{table*}

\begin{table*}[hbt!]
    \renewcommand{\arraystretch}{1.5}
    \centering
    \caption{Best-fit values of the linear galaxy bias $b$ and the amplitude of cross-correlation $A$ from simulations with $\sigma_{0}=0.05$.}
    \label{tab:likeli_params_comp_0.05}
    \begin{tabular}{cc||cc||cc} 
    \hline\hline
    $z$ & $b^{\text{true}}$ & \multicolumn{2}{c}{No correction} & \multicolumn{2}{c}{With correction}\\
    \cline{3-4}\cline{5-6}
    & & $b$ & $A$ & $b$ & $A$\\
    \hline\hline
        $[0.0,0.2)$ & $1.3241$ & $1.2172^{+0.0002}_{-0.0002}$ & $1.2081^{+0.0100}_{-0.0100}$ & $1.3686^{+0.0002}_{-0.0002}$ & $0.9872^{+0.0123}_{-0.0123}$\\
        $[0.2,0.4)$ & $1.3541$ & $1.2288^{+0.0002}_{-0.0002}$ & $1.0941^{+0.0059}_{-0.0058}$ & $1.3555^{+0.0002}_{-0.0002}$ & $1.0019^{+0.0069}_{-0.0069}$\\
        $[0.4,0.6)$ & $1.3909$ & $1.2194^{+0.0002}_{-0.0002}$ & $1.1354^{+0.0049}_{-0.0049}$ & $1.3911^{+0.0002}_{-0.0002}$ & $0.9986^{+0.0057}_{-0.0057}$\\
        $[0.6,0.8)$ & $1.4307$ & $1.2088^{+0.0002}_{-0.0002}$ & $1.1849^{+0.0045}_{-0.0046}$ & $1.4311^{+0.0002}_{-0.0002}$ & $0.9998^{+0.0054}_{-0.0054}$\\
        $[0.8,1.0)$ & $1.4724$ & $1.1856^{+0.0002}_{-0.0002}$ & $1.2421^{+0.0044}_{-0.0045}$ & $1.4729^{+0.0002}_{-0.0002}$ & $0.9996^{+0.0052}_{-0.0052}$\\
        $[1.0,1.4)$ & $1.5337$ & $1.3626^{+0.0002}_{-0.0002}$ & $1.1333^{+0.0036}_{-0.0036}$ & $1.5344^{+0.0002}_{-0.0002}$ & $0.9990^{+0.0039}_{-0.0038}$\\
        $[1.4,1.8)$ & $1.6217$ & $1.3589^{+0.0002}_{-0.0002}$ & $1.2169^{+0.0040}_{-0.0039}$ & $1.6247^{+0.0002}_{-0.0002}$ & $0.9988^{+0.0042}_{-0.0042}$\\
        $[1.8,2.2)$ & $1.7118$ & $1.3692^{+0.0003}_{-0.0003}$ & $1.2791^{+0.0046}_{-0.0046}$ & $1.7138^{+0.0003}_{-0.0003}$ & $0.9994^{+0.0048}_{-0.0048}$\\
        $[2.2,3.0)$ & $1.8277$ & $1.6220^{+0.0005}_{-0.0005}$ & $1.1614^{+0.0047}_{-0.0047}$ & $1.8078^{+0.0004}_{-0.0004}$ & $1.0082^{+0.0047}_{-0.0048}$\\
    \hline
    \end{tabular}
    \tablefoot{The galaxy bias $b$ and amplitude of cross-correlation $A$ are estimated from the average power spectra of $300$ simulations before the correction for leakage and after correction through the scattering matrix approach. $b^{\text{true}}$ is the true value of the galaxy bias for the tomographic bins. The error bars on the parameters correspond to the average of $300$ realisations.}
\end{table*}

\begin{table*}[hbt!]
    \renewcommand{\arraystretch}{1.5}
    \centering
    \caption{Best-fit values of the linear galaxy bias $b$ and the amplitude of cross-correlation $A$ from simulations with a modified Lorentzian error distribution.}
    \label{tab:likeli_params_comp_gamma_0.02}
    \begin{tabular}{cc||cc||cc} 
    \hline\hline
    $z$ & $b^{\text{true}}$ & \multicolumn{2}{c}{No correction} & \multicolumn{2}{c}{With correction}\\
    \cline{3-4}\cline{5-6}
    & & $b$ & $A$ & $b$ & $A$\\
    \hline\hline
        $[0.0,0.2)$ & $1.3241$ & $1.0381^{+0.0002}_{-0.0002}$ & $1.2866^{+0.0124}_{-0.0126}$ & $1.3191^{+0.0002}_{-0.0002}$ & $0.9953^{+0.0124}_{-0.0125}$\\
        $[0.2,0.4)$ & $1.3541$ & $1.1456^{+0.0002}_{-0.0002}$ & $1.1743^{+0.0069}_{-0.0069}$ & $1.3534^{+0.0002}_{-0.0002}$ & $1.0015^{+0.0069}_{-0.0069}$\\
        $[0.4,0.6)$ & $1.3909$ & $1.1593^{+0.0002}_{-0.0002}$ & $1.1939^{+0.0057}_{-0.0057}$ & $1.3903^{+0.0002}_{-0.0002}$ & $0.9971^{+0.0058}_{-0.0057}$\\
        $[0.6,0.8)$ & $1.4307$ & $1.1654^{+0.0002}_{-0.0002}$ & $1.2277^{+0.0054}_{-0.0054}$ & $1.4307^{+0.0002}_{-0.0002}$ & $0.9993^{+0.0053}_{-0.0054}$\\
        $[0.8,1.0)$ & $1.4724$ & $1.1656^{+0.0002}_{-0.0002}$ & $1.2621^{+0.0052}_{-0.0053}$ & $1.4715^{+0.0002}_{-0.0002}$ & $0.9979^{+0.0053}_{-0.0052}$\\
        $[1.0,1.4)$ & $1.5337$ & $1.3546^{+0.0002}_{-0.0002}$ & $1.1368^{+0.0038}_{-0.0038}$ & $1.5347^{+0.0002}_{-0.0002}$ & $0.9986^{+0.0039}_{-0.0038}$\\
        $[1.4,1.8)$ & $1.6217$ & $1.3796^{+0.0002}_{-0.0002}$ & $1.1885^{+0.0043}_{-0.0042}$ & $1.6226^{+0.0002}_{-0.0002}$ & $0.9962^{+0.0042}_{-0.0042}$\\
        $[1.8,2.2)$ & $1.7118$ & $1.4059^{+0.0003}_{-0.0003}$ & $1.2331^{+0.0049}_{-0.0049}$ & $1.7125^{+0.0003}_{-0.0003}$ & $0.9994^{+0.0048}_{-0.0048}$\\
        $[2.2,3.0)$ & $1.8277$ & $1.6290^{+0.0004}_{-0.0004}$ & $1.1511^{+0.0049}_{-0.0048}$ & $1.8279^{+0.0004}_{-0.0005}$ & $0.9974^{+0.0047}_{-0.0047}$\\
    \hline
    \end{tabular}
    \tablefoot{The galaxy bias $b$ and amplitude of cross-correlation $A$ are estimated from the average power spectra of $300$ simulations before the correction for leakage and after correction through the scattering matrix approach. $b^{\text{true}}$ is the true value of the galaxy bias for the tomographic bins. The error bars on the parameters correspond to the average of $300$ realisations.}
\end{table*}

\begin{figure*}
    \begin{subfigure}[b]{0.33\linewidth}
        \centering
        \includegraphics[width=0.9\textwidth]{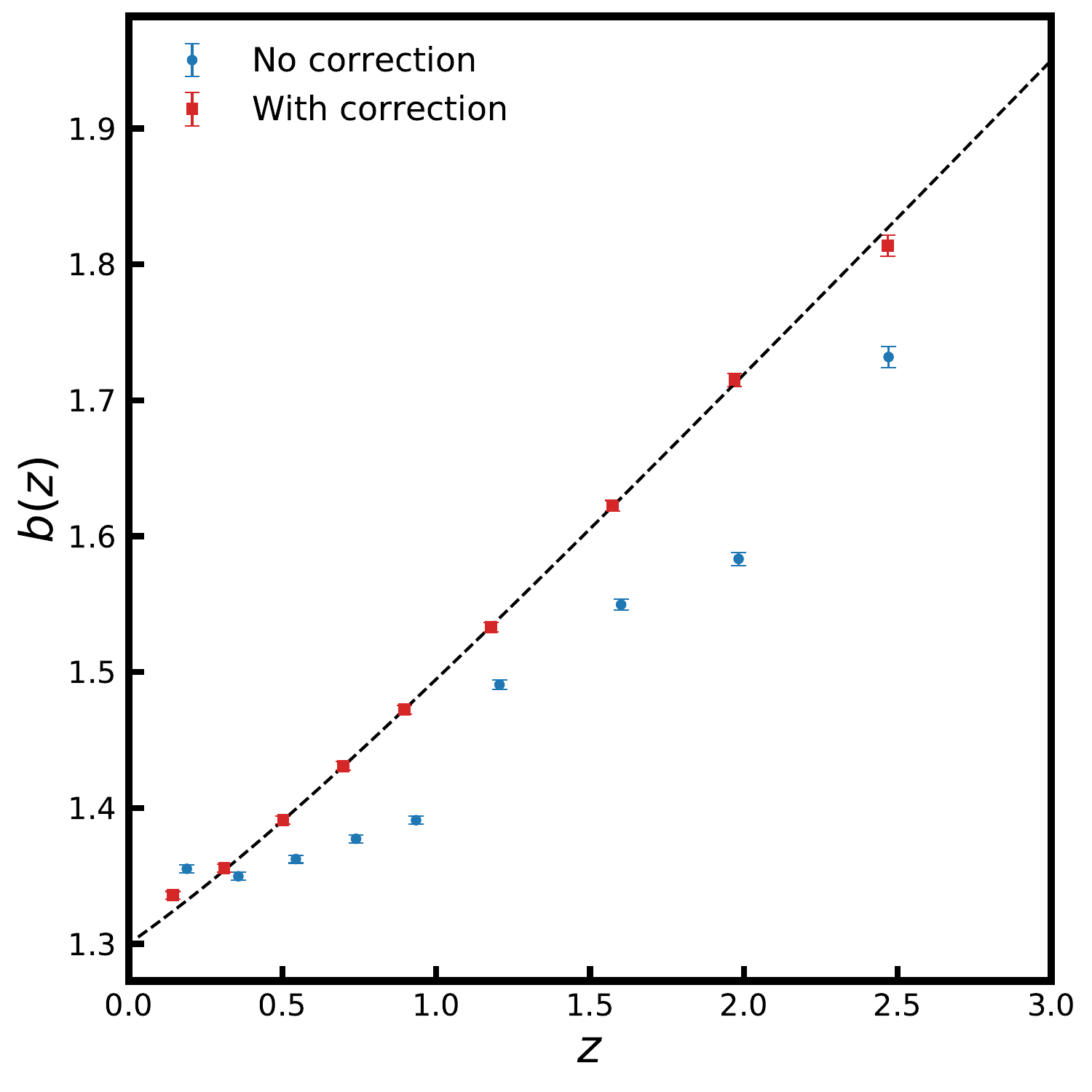}
        \caption{}
    \end{subfigure}%
    \begin{subfigure}[b]{0.33\linewidth}
        \centering
        \includegraphics[width=0.9\textwidth]{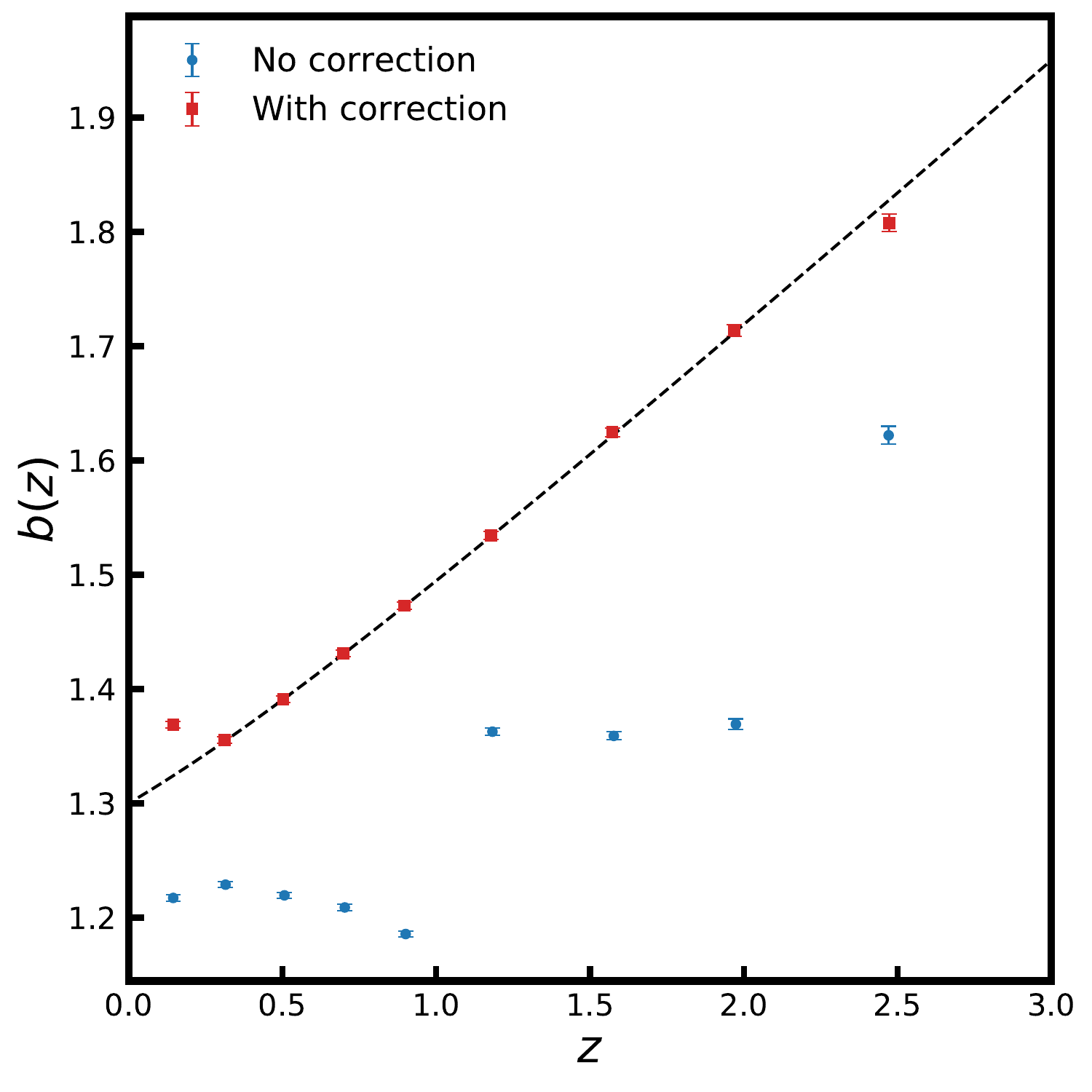}
        \caption{}
    \end{subfigure}%
    \begin{subfigure}[b]{0.33\linewidth}
        \centering
        \includegraphics[width=0.9\textwidth]{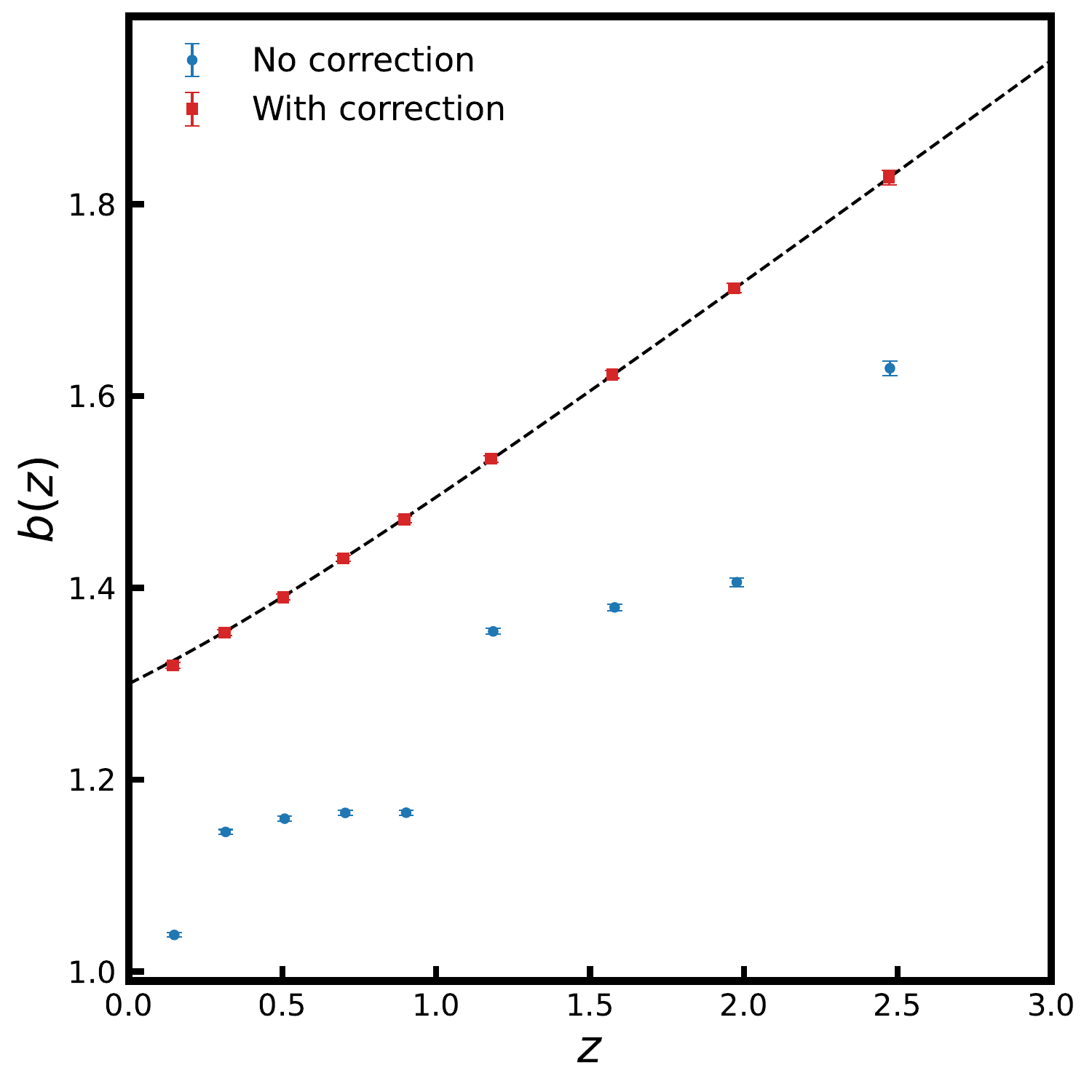}
        \caption{}
    \end{subfigure}    
    \caption{{Comparison of the linear galaxy bias parameter estimated from the average of $300$ simulations before (blue circles) and after (red squares) the correction for the redshift bin mismatch of objects, with (a) the Gaussian error distribution with $\sigma_{0}=0.02$, (b) the Gaussian error distribution with $\sigma_{0}=0.05$, and (c) the modified Lorentzian error distribution with $\gamma_{0}=0.02$. The error bars on the parameters correspond to a single realisation.}}
    \label{fig:lsst_tomography_gal_bias_parameter}
\end{figure*}

{In Fig. \ref{fig:lsst_tomography_parameters_zscore}, we show the relative difference between the mean and the fiducial value (in terms of the standard deviation for a single realisation) of the linear galaxy bias and the amplitude of the cross correlation for $\sigma_{0}=0.02$ (left column), $\sigma_{0}=0.05$ (middle column), and $\gamma_{0}=0.02$ (right column).} The blue circles and red squares represent the parameter estimates before and after the correction for leakage, respectively. The error bars on the data points correspond to the average estimated power spectra. The top panel of Fig. \ref{fig:lsst_tomography_parameters_zscore} shows the relative difference for the linear galaxy bias parameter. {When the scatter of objects across redshift bins is not properly accounted for, the galaxy bias can deviate between $5- 30\,\sigma$ when $\sigma_{0}=0.02$, and by $25-120\,\sigma$ when $\sigma_{0}=0.05$ and $\gamma_{0}=0.02$.} These large deviations on the linear galaxy bias are visible because the errors from the likelihood estimation {(quoted in Tables \ref{tab:likeli_params_comp_0.02}-\ref{tab:likeli_params_comp_gamma_0.02})} are between $0.15-0.5\%$ for a single realisation. This shows that the estimates for the galaxy bias are very tightly constrained. {The bottom panel of Fig. \ref{fig:lsst_tomography_parameters_zscore} shows the difference values for the amplitude of the cross correlation, which can deviate by up to $\sim 1.2\,\sigma$ with $\sigma_{0}=0.02$, by up to $\sim 3.5\,\sigma$ when $\sigma_{0}=0.05$, and by up to $\sim 3.0\,\sigma$ with $\gamma_{0}=0.02$.} As clearly conveyed by Fig. \ref{fig:lsst_tomography_parameters_zscore}, the parameters of the linear galaxy bias and of the cross-correlation amplitude become consistent with their expected values after we corrected for the effect of the redshift bin mismatch of the objects through our scattering matrix formalism.

\begin{figure*}
    \begin{subfigure}[b]{0.33\linewidth}
        \centering
        \includegraphics[width=0.9\linewidth]{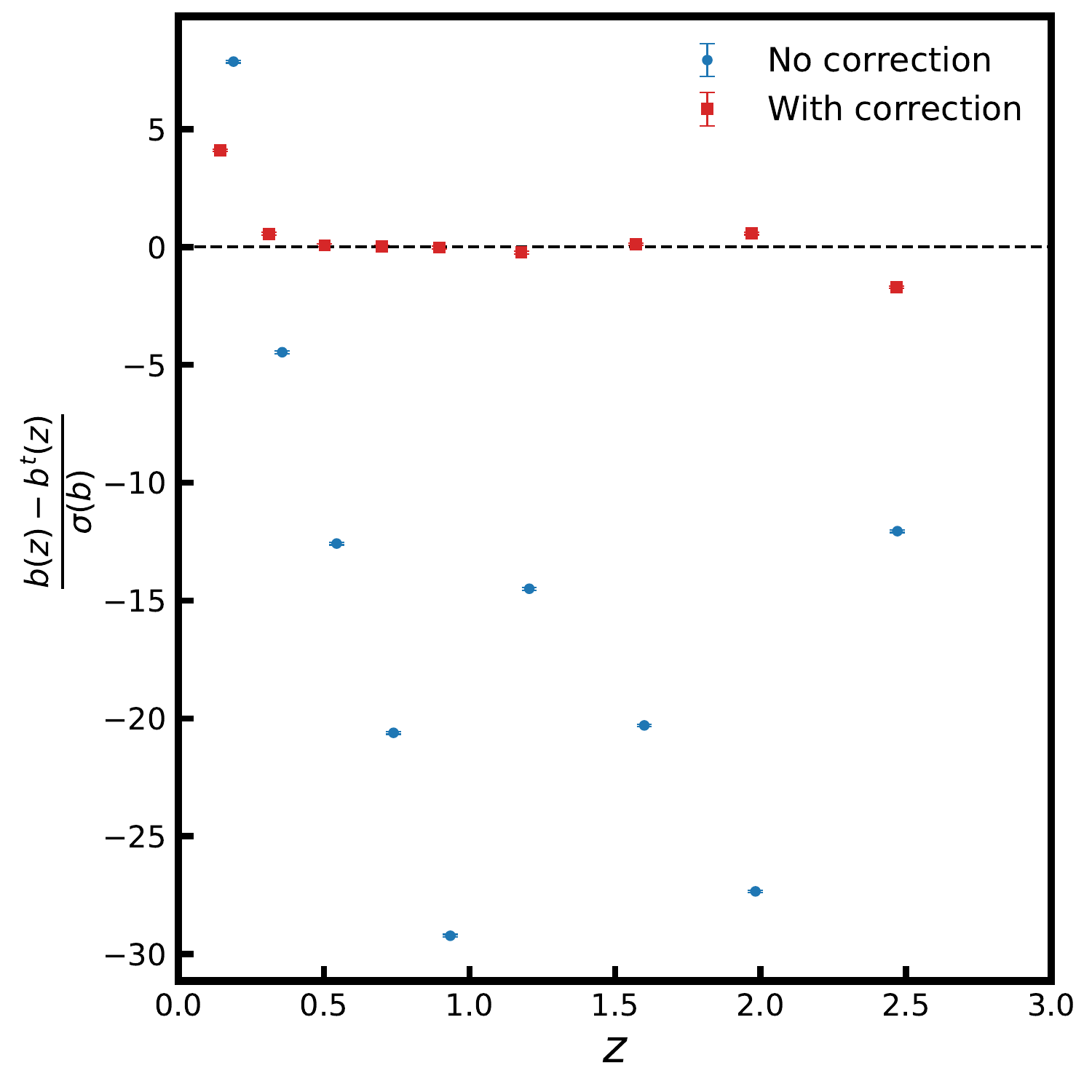}
        \captionsetup{labelformat=empty}
    \end{subfigure}%
    \begin{subfigure}[b]{0.33\linewidth}
        \centering
        \includegraphics[width=0.9\linewidth]{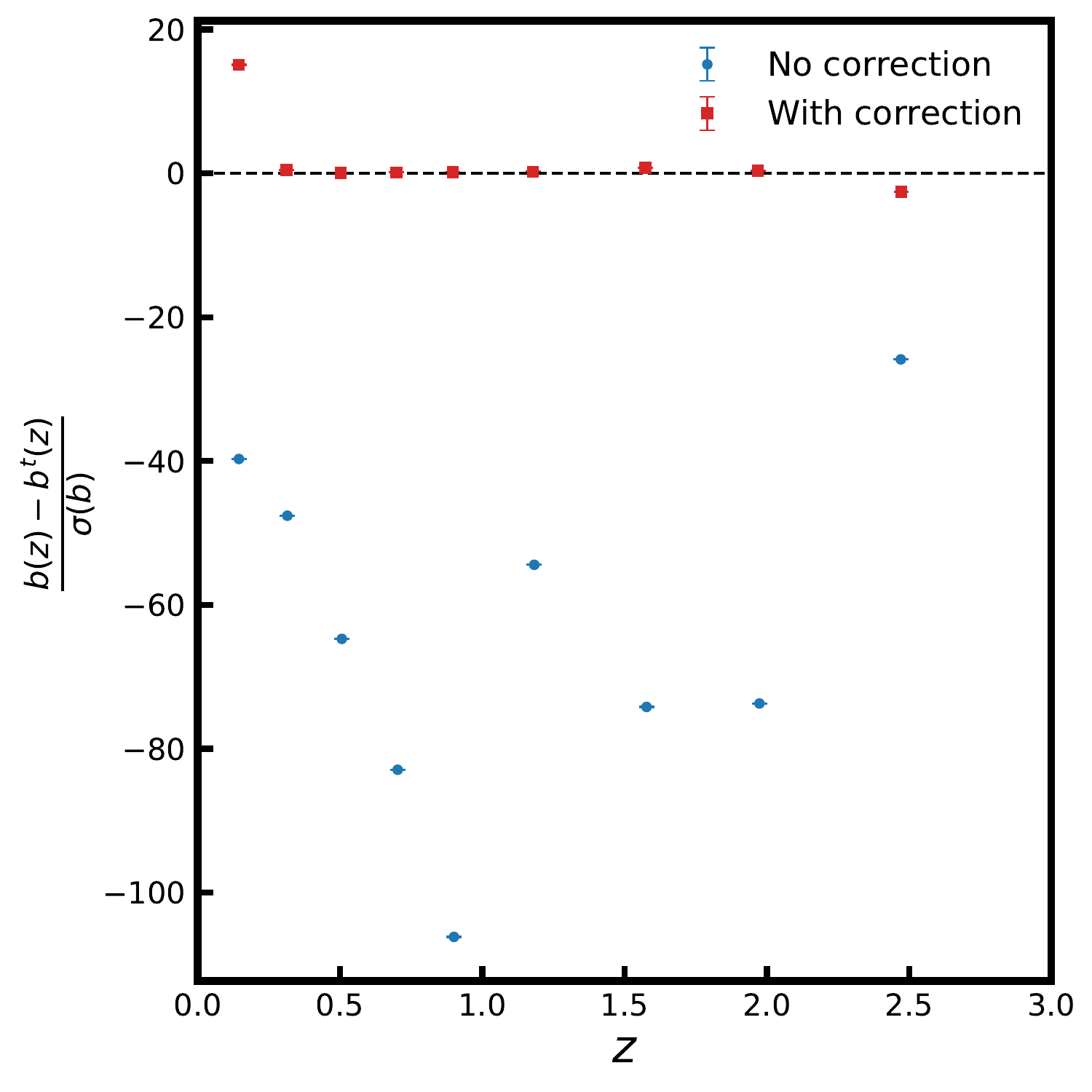}
        \captionsetup{labelformat=empty}
    \end{subfigure}%
    \begin{subfigure}[b]{0.33\linewidth}
        \centering
        \includegraphics[width=0.9\linewidth]{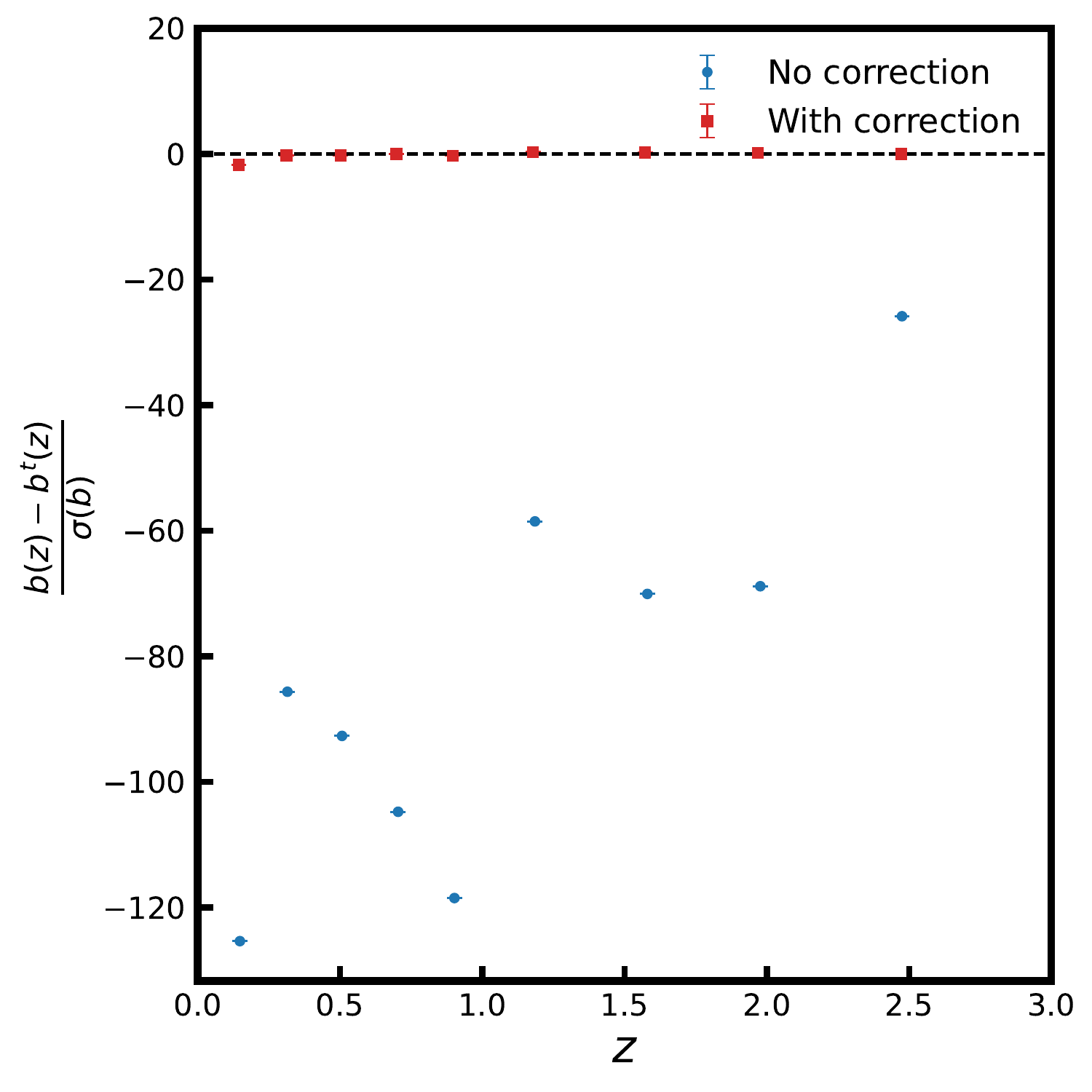}
        \captionsetup{labelformat=empty}
    \end{subfigure}\\
    \begin{subfigure}[b]{0.33\linewidth}
        \centering
        \includegraphics[width=0.9\linewidth]{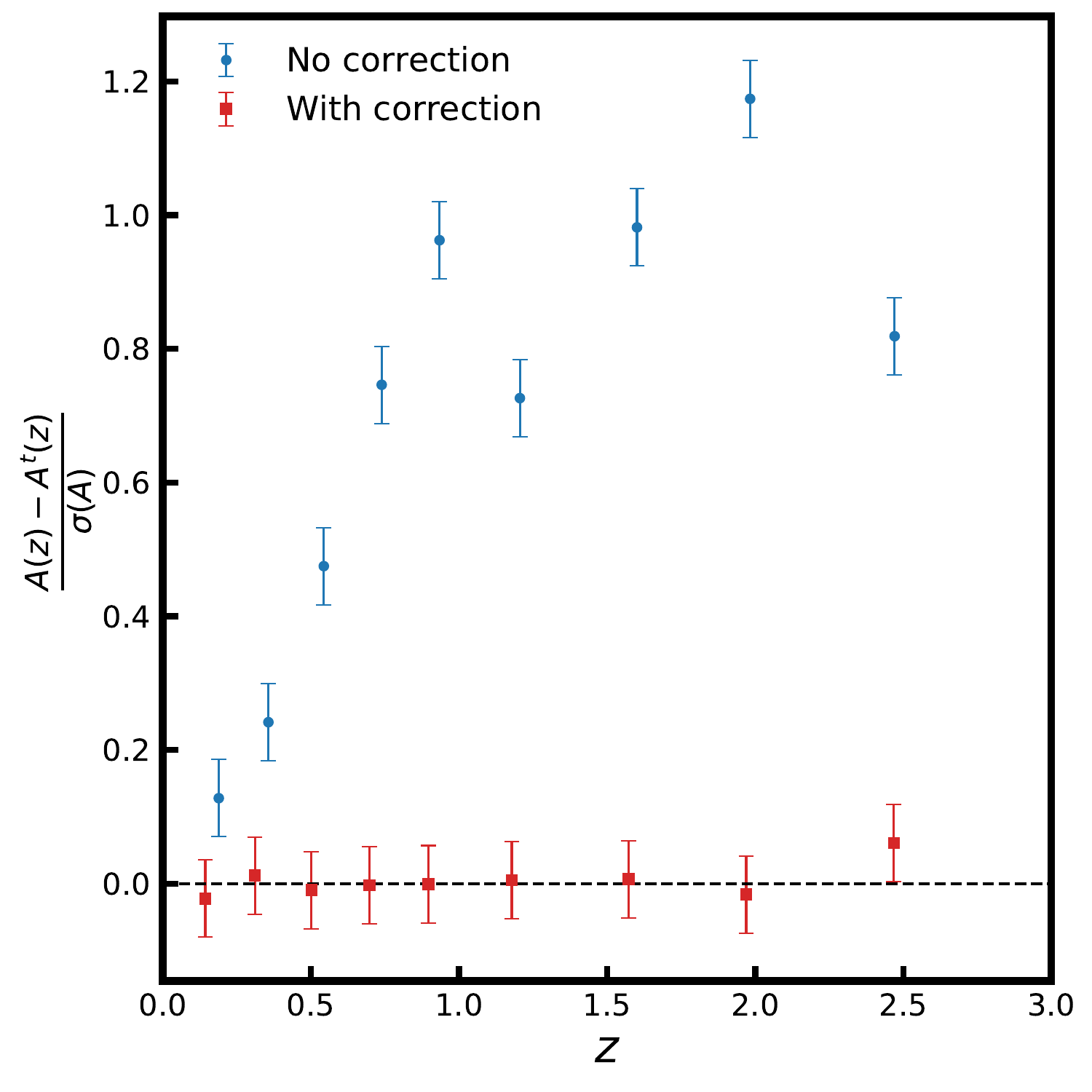}
        \caption{}
    \end{subfigure}%
    \begin{subfigure}[b]{0.33\linewidth}
        \centering
        \includegraphics[width=0.9\linewidth]{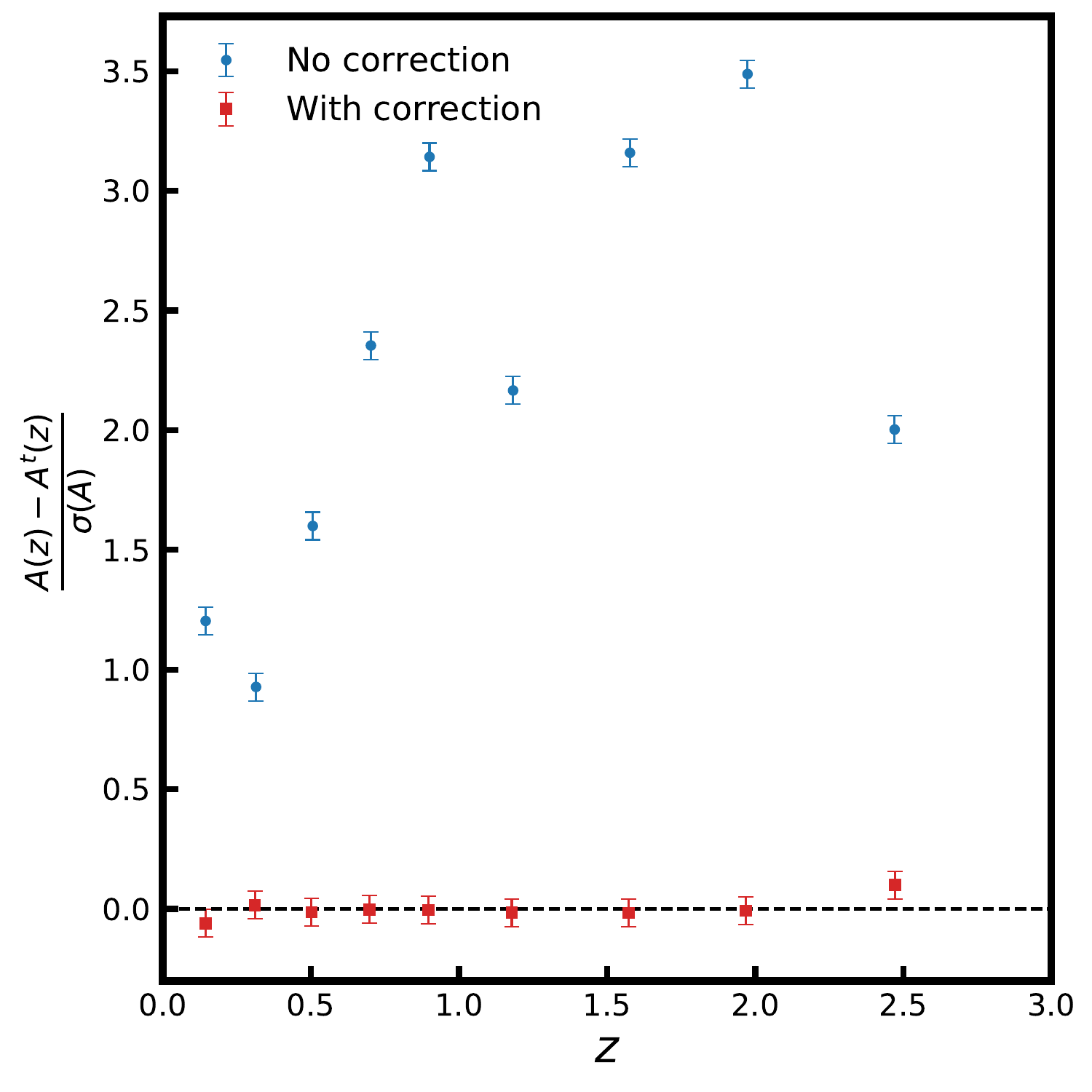}
        \caption{}
    \end{subfigure}%
    \begin{subfigure}[b]{0.33\linewidth}
        \centering
        \includegraphics[width=0.9\linewidth]{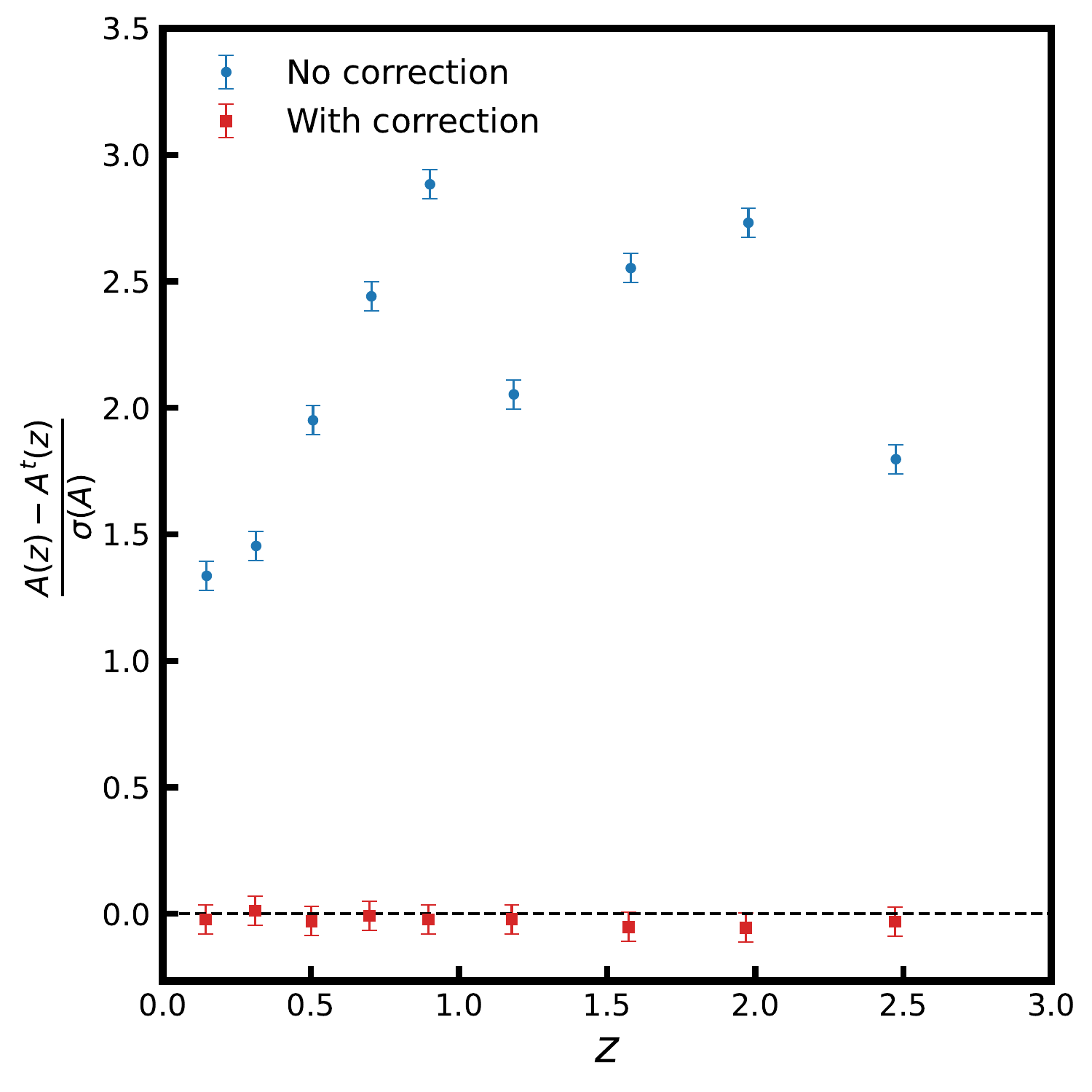}
        \caption{}
    \end{subfigure}
    \caption{{Comparison of the relative difference for linear galaxy bias (top) and for the cross-correlation amplitude estimated from the average power spectra (bottom) before (blue circles) and after (red squares) the correction for the redshift bin mismatch with (a) a Gaussian error distribution with $\sigma_{0}=0.02$, (b) a Gaussian error distribution with $\sigma_{0}=0.05$, and (c) a modified Lorentzian error distribution with $\gamma_{0}=0.02$. The error bars on the data points correspond to the average spectra, and the standard deviations $\sigma(b)$ and $\sigma(A)$ correspond to a single realisation.}}
    \label{fig:lsst_tomography_parameters_zscore}
\end{figure*}


\section{Note on estimating $\sigma_{8}$}\label{sec:sigma8}

We showed in section \ref{sec:parameter_estimation} that the scatter of objects across redshift bins can lead to a biased parameter estimation and can thus alter our inferences about the cosmological model. In this section, we estimate the effects of leakage on the $\sigma_{8}$ parameter. \cite{Peacock&Bilicki2018} proposed a method for computing the $\sigma_{8}$ parameter from the cross-correlation amplitude using the relation
\begin{equation}
    \sigma_{8}(z) = A(z)\,\sigma_{8,0}\,D(z),
    \label{eq:sigma8_evolution}
\end{equation}
where $\sigma_{8,0}$ is the value of the $\sigma_{8}$ parameter at redshift $z=0$, and $D(z)$ is the linear growth function given by Eq.\,(\ref{eq:growth function}). We computed the value of $\sigma_{8,0}$ for our assumed background cosmology using the software \texttt{CAMB}. {In Fig \ref{fig:comp_sigma8}, we show the impact of the scattering of objects on the $\sigma_{8}$ parameter computed using Eq.\,(\ref{eq:sigma8_evolution}) for Case-I (in the left and middle panels) and Case-II (right panel).} The dashed black lines are the fiducial evolution of the $\sigma_{8}$ parameter with redshift. We measure a higher than expected value of $\sigma_{8}$ up to $\sim 1\,\sigma$ for $\sigma_{0}=0.02$ before correcting for the redshift bin mismatch of the objects. {With $\sigma_{0}=0.05$ and $\gamma_{0}=0.02$, the $\sigma_{8}$ parameter is biased between $1-3\,\sigma$.} As expected, the biases on the amplitude of the cross correlation are directly reflected in the $\sigma_{8}$ parameter. {Because of the broad wings of the modified Lorentzian error distribution, the deviation in the $\sigma_8$ parameter even for a smaller redshift scatter $\gamma_0=0.02$ is as strong as for a Gaussian error distribution with a larger redshift scatter, that is, $\sigma_0 = 0.05$. Hence, a precise modelling of the error distribution is necessary to quantify the effects of the redshift scatter.} We obtain an unbiased estimate of the $\sigma_{8}$ parameter after leakage correction, and thus ,it becomes crucial to correct for the mismatch between different redshift bins in a tomographic analysis to obtain unbiased parameter estimates.

\begin{figure*}
    \begin{subfigure}[b]{0.33\linewidth}
        \centering
        \includegraphics[width=\textwidth]{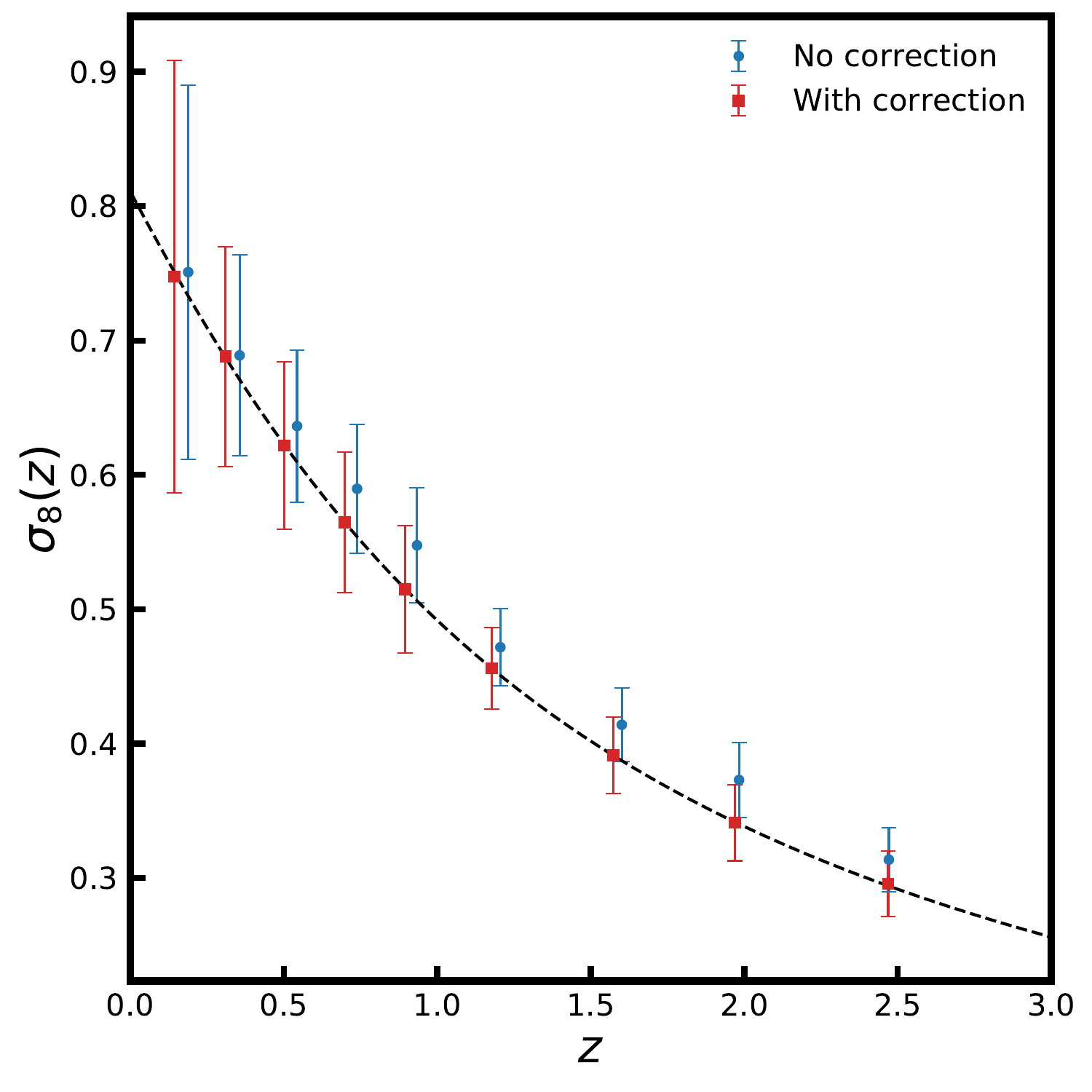}
        \caption{}
    \end{subfigure}%
    \begin{subfigure}[b]{0.33\linewidth}
        \centering
        \includegraphics[width=\textwidth]{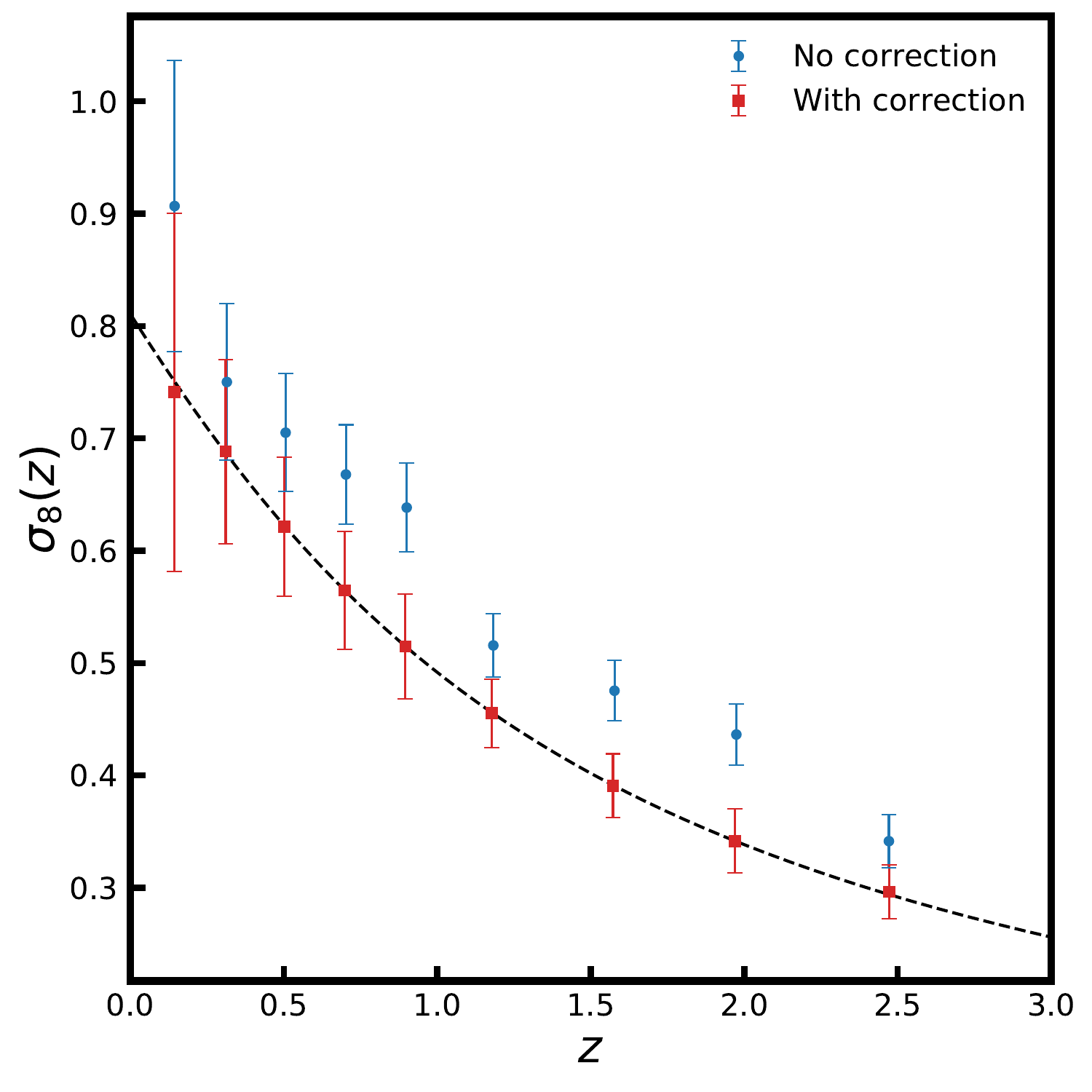}
        \caption{}
    \end{subfigure}%
    \begin{subfigure}[b]{0.33\linewidth}
        \centering
        \includegraphics[width=\textwidth]{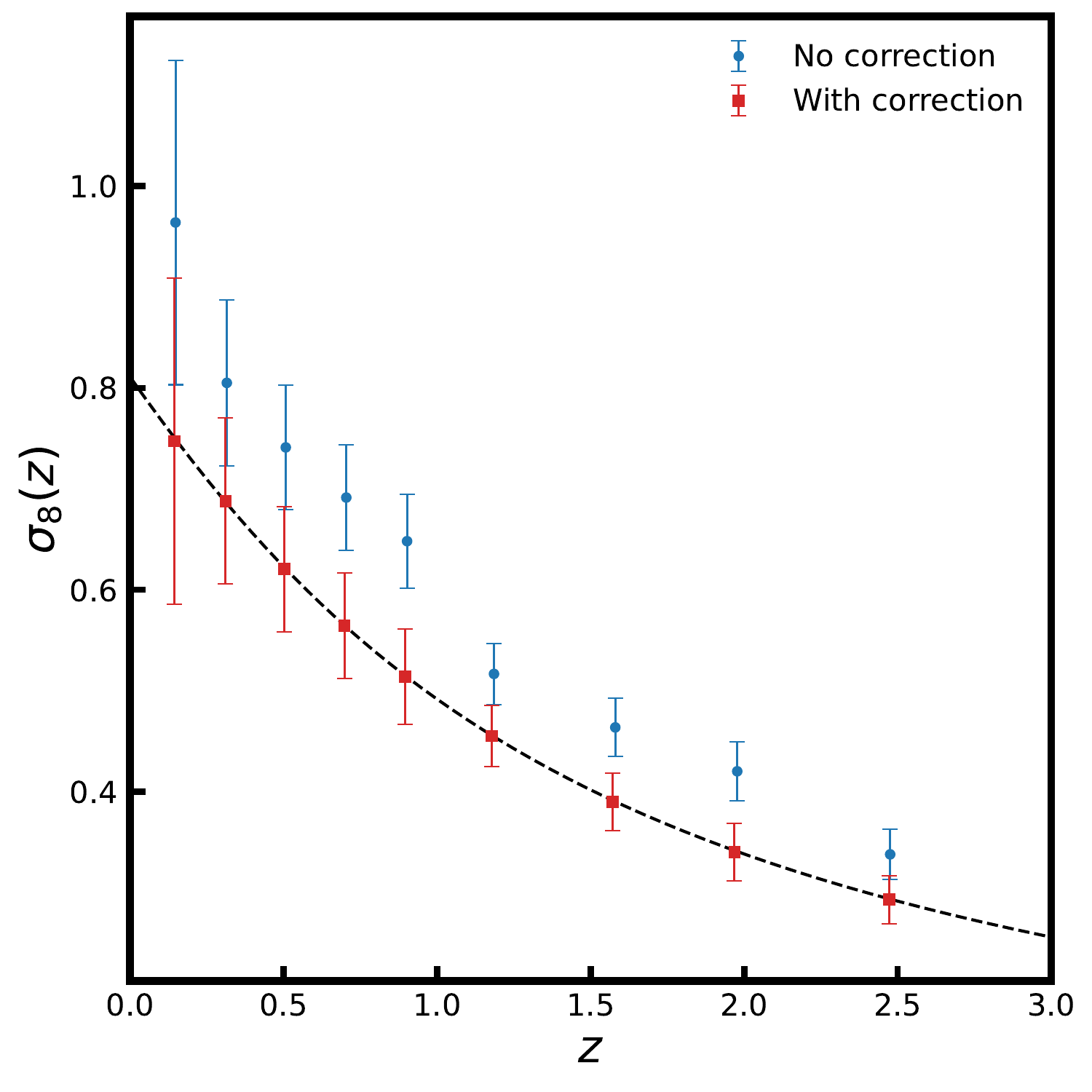}
        \caption{}
    \end{subfigure}
    \caption{{Comparison of the $\sigma_{8}$ parameter estimated from the average power spectra of $300$ simulations before and after the correction for leakage for (a) a Gaussian error distribution with $\sigma_{0}=0.02$, (b) a Gaussian error distribution with $\sigma_{0}=0.05$, and (c) a modified Lorentzian error distribution with $\gamma_{0}=0.02$. The dashed black line is the fiducial evolution of the $\sigma_{8}$ parameter used during simulations. The error bars on the parameters correspond to a single realisation.}}
    \label{fig:comp_sigma8}
\end{figure*}


\section{Summary and conclusions}\label{sec:summary}

We presented the tomographic study of cross correlations by performing simulations of the \textit{Planck} CMB lensing convergence and galaxy density field mimicking properties of the LSST photometric survey. We used the code \texttt{FLASK} to simulate log-normal fields and divided the galaxies into nine redshift bins. {We considered Gaussian photometric redshift errors with a standard deviation of $\sigma(z)=0.02(1+z)$ and $\sigma(z)=0.05(1+z)$ and a modified Lorentzian photometric redshift error with a deviation of $0.02(1+z)$, but we did not include catastrophic redshift errors or photometric calibration errors, which we keep for future studies.} In this way, we generated an ideal observational scenario in our simulations that was free from other systematics, which is crucial for demonstrating the importance of the redshift bin mismatch of the objects. We computed the galaxy auto-power spectrum and the cross-power spectrum between the galaxy over-density and the CMB convergence fields, and we used these power spectra to estimate two parameters, the redshift-dependent linear galaxy bias $b$, and amplitude of the cross correlation $A$, employing the maximum likelihood method.

We estimated the true redshift distribution from a simulated photometric redshift distribution by the convolution method (section \ref{sec:convolution}). In addition, we also estimated the true redshift distribution with the deconvolution method (section \ref{sec:deconvolution}). The most important quantity for an accurate recovery of the true redshift distribution is the precise estimation of the error distributions $p(z_{t}-z_{p}|z_{p})$ for the convolution and $p(z_{p}-z_{t}|z_{t})$ for the deconvolution methods. {We estimated the error functions with sub-percent accuracy ($<0.25\%$) by fitting a single Gaussian to $p(z_{p}-z_{t}|z_{t})$ and summing of three Gaussians to $p(z_{t}-z_{p}|z_{p})$ (shown in Fig. \ref{fig:err_func_fit_true_and_photo}) for the Gaussian error distributions.} We find that the sum of the Gaussians accurately capturse the peculiarities of the error function $p(z_{t}-z_{p}|z_{p})$ such as non-Gaussian tails and higher peaks in the centre. {For the modified Lorentzian error distribution, we fit both $p(z_{p}-z_{t}|z_{t})$ and $p(z_{t}-z_{p}|z_{p})$ with a single modified Lorentzian function.}

The galaxy auto-power spectra measured from photometric datasets are found to be consistently smaller in every bin than their fiducial predictions. {The offsets vary between $2-15\%$ for simulations with $\sigma(z)=0.02(1+z)$ and between $15-40\%$ for $\sigma(z)=0.05(1+z)$ as well as for the modified Lorentzian error distribution.} The measured cross-power spectra are also biased with smaller deviations ($<5\%$) for both cases. The measured power spectra are inconsistent with their expectations because the objects scatter from one redshift bin to the next due to the photometric redshift errors. This conclusion is consistent with the fact that the deviations are larger in the case of photometric redshift scatter $\sigma(z)=0.05(1+z)$. {The offsets in the power spectra also vary when we changed from a Gaussian error distribution to a modified Lorentzian error distribution due to the broad wings of the modified Lorentzian function.}

To alleviate the differences in the power spectra, we implemented the scattering matrix approach introduced by \cite{Zhang2010} to counter the effect of the redshift bin mismatch of the objects. The scattering matrix describes the fraction of objects in a photometric redshift bin that comes from different true redshift bins. The power spectra in photometric redshift bins then transform as a linear combination of the power spectra from different true redshift bins (Eqs.\,\ref{eq:scattering_relation_gg_matrix} and \ref{eq:scattering_relation_kg_matrix}). \cite{Zhang2017} proposed an algorithm based on the non-negative matrix factorisation method to solve similar numerical problems, but this method is computationally challenging for the many data points in the power spectra and the number of tomographic bins. To circumvent these challenges, we proposed an alternative method for a fast and accurate computation of the scattering matrix based on the reconstruction of the true redshift distribution by the deconvolution method (see section \ref{sec:scattering_matrix}). We showed in Fig. \ref{fig:performance_scattering_matrix} that our new method for computing the scattering matrix is robust and only proves inefficient in the first tomographic bin because of a cut in the redshift distribution at boundary $z=0$. With a precise estimation of the scattering matrix, we corrected the theoretical power spectra for the tomographic bins to compare with the estimated galaxy power spectra from simulated photometric datasets. Fig. \ref{fig:plot_photo_from_scat_mat_dist_nsim_300_0.02} and \ref{fig:plot_photo_from_scat_mat_dist_nsim_300_0.05} showed that the scattering matrix method makes the estimated power spectra consistent with the leakage-corrected theoretical power spectra.

We quantified the impact of the redshift bin mismatch of the objects on the estimation of the linear galaxy bias and on the amplitude of the cross correlation. To estimate the parameters after the correction for leakage, we transformed the estimated photometric power spectra into estimated true power spectra by inverting Eqs.\,(\ref{eq:scattering_relation_gg_matrix}) and (\ref{eq:scattering_relation_kg_matrix}) (as described in section \ref{sec:parameter_estimation}). {The best-fit values of these parameters before and after the correction for leakage are quoted in Tables \ref{tab:likeli_params_comp_0.02} and \ref{tab:likeli_params_comp_0.05} for Case-I ($\sigma(z)=0.02(1+z)$ and $\sigma(z)=0.05(1+z)$) and in Table \ref{tab:likeli_params_comp_gamma_0.02} for Case-II.} {Without accounting for the leakage, we estimate lower values for the linear galaxy bias by $5- 30\,\sigma$ when $\sigma_{0}=0.02$, and by $25-110\,\sigma$ when $\sigma_{0}=0.05$ and $\gamma_{0}=0.02$. The amplitude of the cross correlation, on the other hand, is estimated to be higher than its fiducial value of unity up to $\sim 1.2\,\sigma$ with $\sigma_{0}=0.02$, $\sim 3.5\,\sigma$ when $\sigma_{0}=0.05$, and up to $\sim 3.0\,\sigma$ when $\gamma_{0}=0.02$.} It is important to note here that the estimates of a lower galaxy bias and higher amplitude are not to be generalised for every tomographic analysis. The offsets of the power spectra and parameters estimated from photometric datasets in a tomographic study strongly depend on the photometric redshift error distributions and on the redshift distribution of the objects. After correcting for leakage by using the scattering matrix, both parameters are very well constrained with their expected values.

The amplitude of the cross correlation is an indicator of the validity of the background cosmological model. Thus, when the bias resulting from photometric redshift errors is not corrected for, incorrect inferences are inevitable when testing cosmological models with tomographic analyses. Other estimators frequently used to test the cosmological models, such as the $D_{G}$ \citep{Giannantonio2016} or $E_{G}$ (\citealt{Pullen2016}; \citealt{Zhang2007}) statistics, also employ the ratio of the cross-power spectra to the galaxy auto-power spectra, and they are therefore prone to similar systematics. We studied the relation between the amplitude of the cross correlation and the more familiar $\sigma_{8}$ parameter in section \ref{sec:sigma8}. {The $\sigma_{8}$ parameter deviates by up to $\sim 1\,\sigma$ when $\sigma(z)=0.02(1+z)$, and up to $\sim 3\,\sigma$ with $\sigma(z)=0.05(1+z)$ and with a modified Lorentzian error distribution.} We showed that the offsets in the amplitude that aer due to the scatter of the objects are synonymous with the deviations in the $\sigma_{8}$ parameter. With next-generation galaxy surveys such as the Vera C. Rubin Observatory Legacy Survey of Space and Time (LSST; \citealt{Ivezi2019}; \citealt{LSST2009}), Euclid \citep{Euclid2011}, and the Dark Energy Spectroscopic Instrument (DESI; \citealt{DESI2019}), the tomographic approach will emerge as a powerful tool for placing stringent constraints on the validity of cosmological models. We therefore propose that the scattering matrix approach developed and presented in this paper be strictly used for future tomographic studies.

\begin{acknowledgements}
The authors would like to thank Agnieszka Pollo and Maciej Bilicki for their valuable comments and discussion. CSS thanks Deepika Bollimpalli and Swayamtrupta Panda for discussions on the deconvolution method. The work has been supported by the Polish Ministry of Science and Higher Education grant DIR/WK/2018/12. The authors acknowledge the use of \texttt{CAMB}, HEALPix, \texttt{EMCEE} and \texttt{FLASK} software packages.
\end{acknowledgements}

%
   \bibliographystyle{aa} 
   \bibliography{references} 
%

\FloatBarrier
\onecolumn
\begin{appendix} 
\section{Validation of the deconvolution method}\label{appndx:validate_deconvolution}
In this section, we present the performance of the deconvolution method as described in section \ref{sec:deconvolution} using a toy example. We generated a fiducial redshift distribution by random sampling from the LSST photometric redshift distribution profile (\citealt{Ivezi2019}; \citealt{LSST2009}) with a mean redshift $0.9$, which we call the true distribution. We convolved the true distribution with a Gaussian distribution with $\mu = 0.1,\sigma = 0.02$ and call the result the observed distribution. In the left column of Fig. \ref{fig:test_deconv_plot1}, we show the true and observed distributions by solid blue and red lines, respectively. We show the robustness of our deconvolution method by applying it on unsmoothed distributions. In the right column of Fig. \ref{fig:test_deconv_plot1}, we compare the true distribution (solid blue line) with the distribution recovered using our deconvolution method (dashed red line). The recovered distribution agrees well with the true redshift distribution, which validates our deconvolution method for reconstructing the true redshift distribution.
\begin{figure}[h]
\begin{subfigure}[b]{0.5\linewidth}
    \centering
    \includegraphics[width=\linewidth]{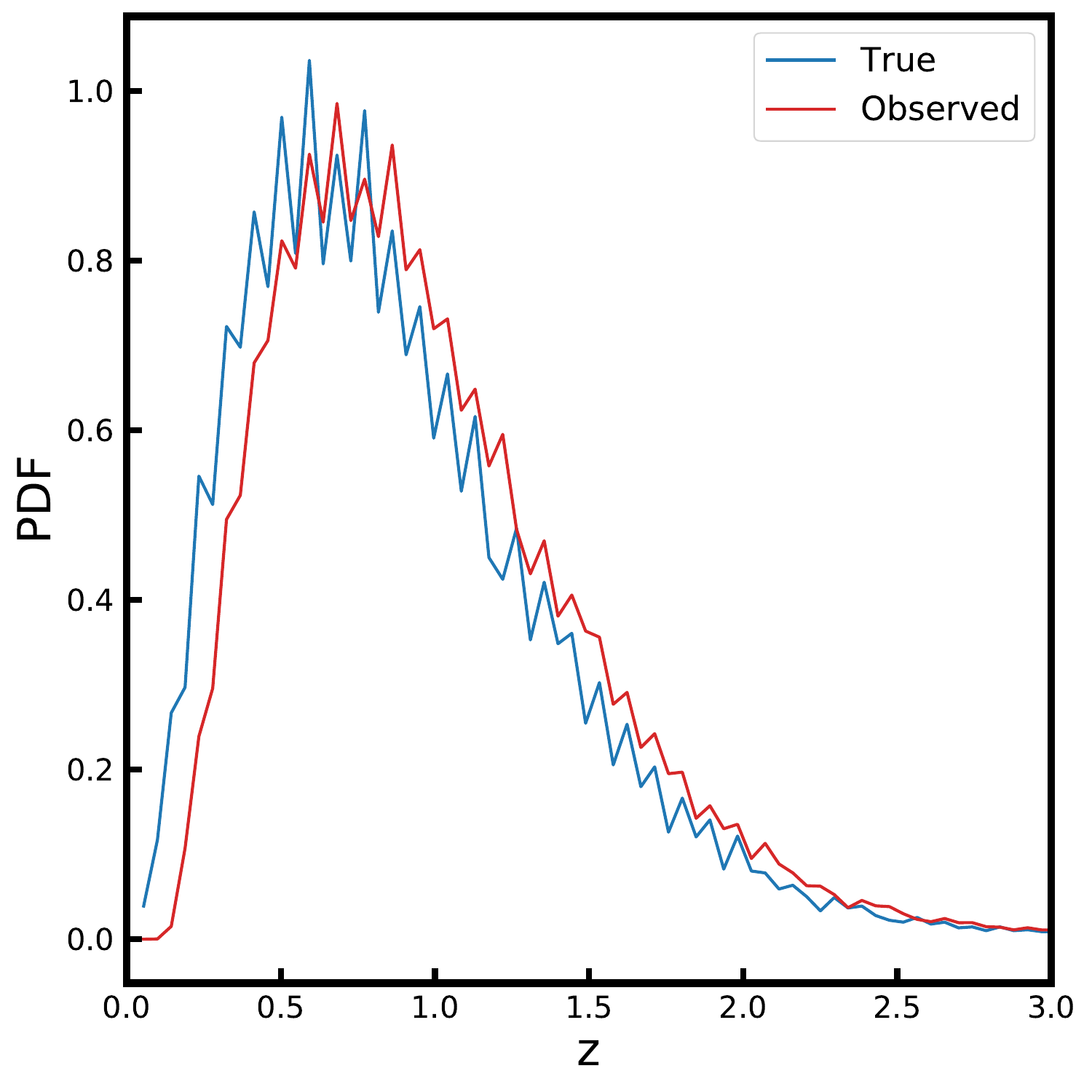}
\end{subfigure}%
\begin{subfigure}[b]{0.5\linewidth}
    \centering
    \includegraphics[width=\linewidth]{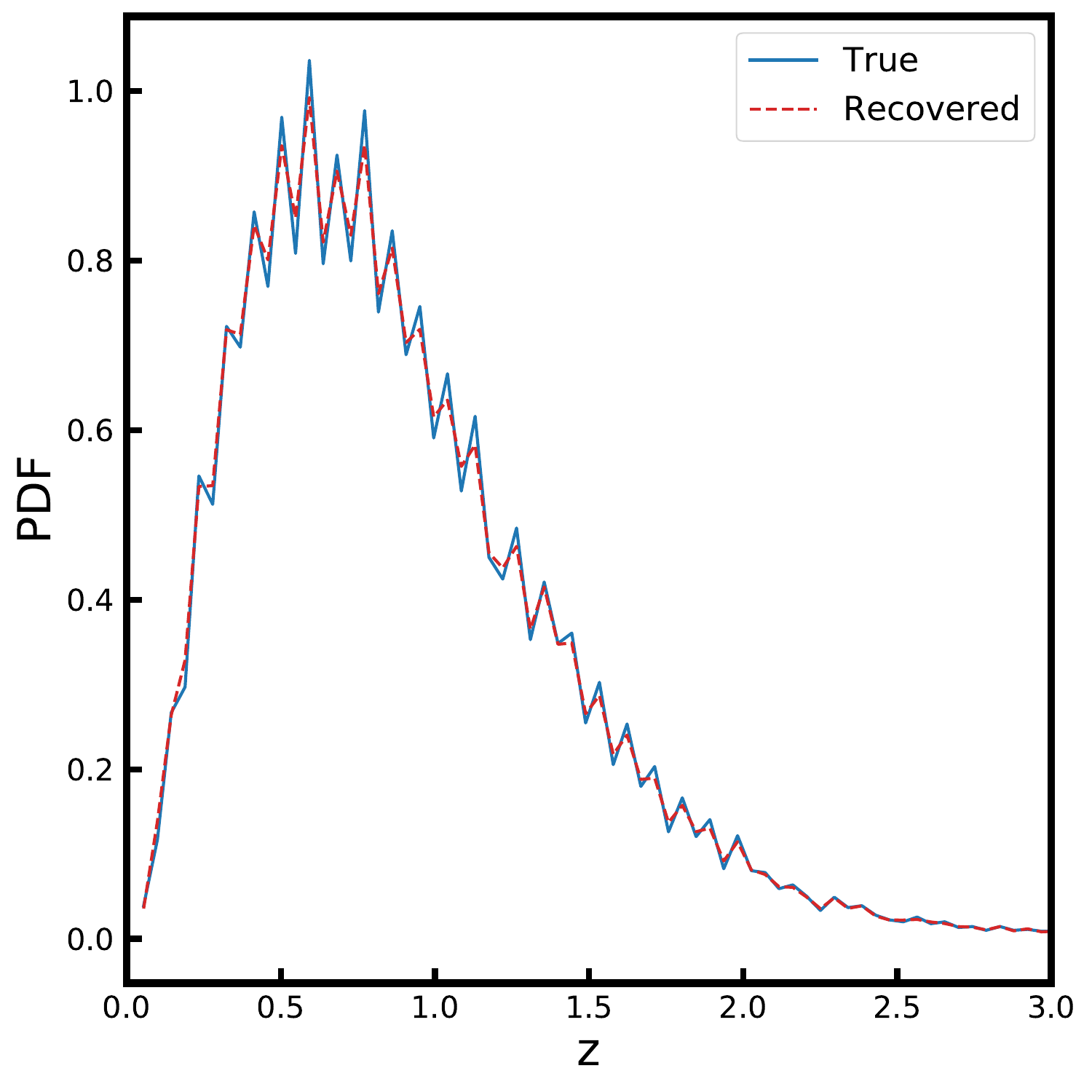}
\end{subfigure}
\caption{Performance of the deconvolution method for estimating the true redshift distribution. \textit{Left}: True (solid blue line) and observed (solid red line) distributions for our toy example. \textit{Right}: Comparison of the true redshift distribution (solid blue line) with the distribution recovered from the deconvolution method (dashed red line). The recovered distribution agrees excellently with the true redshift distribution.}
\label{fig:test_deconv_plot1}
\end{figure}

\section{Power spectra from true datasets}\label{appndx:validate_true_flask}

It is important to check for any systematics that may arise when preparing true datasets with the code \texttt{FLASK}. These systematics, if significant, also affect any inferences made from photometric datasets. In Fig. \ref{fig:plot_gg_true_from_flask_nsim_300} we show for every tomographic bin the relative difference between noise-subtracted average galaxy auto-power spectra estimated before adding photometric redshift errors and their theoretical expectation with error bars computed from Eq.\,(\ref{eq:err_simul}). Fig. \ref{fig:plot_kg_true_from_flask_nsim_300} and \ref{fig:plot_kk_true_from_flask_nsim_300} present the relative difference for the cross-power spectrum and CMB convergence auto-power spectrum, respectively. The power spectra are consistent with their theoretical expectations in all tomographic bins, and thus, the \texttt{FLASK} simulations are free from any internal systematics.

\begin{figure*}
    \centering
    \includegraphics[width=\linewidth]{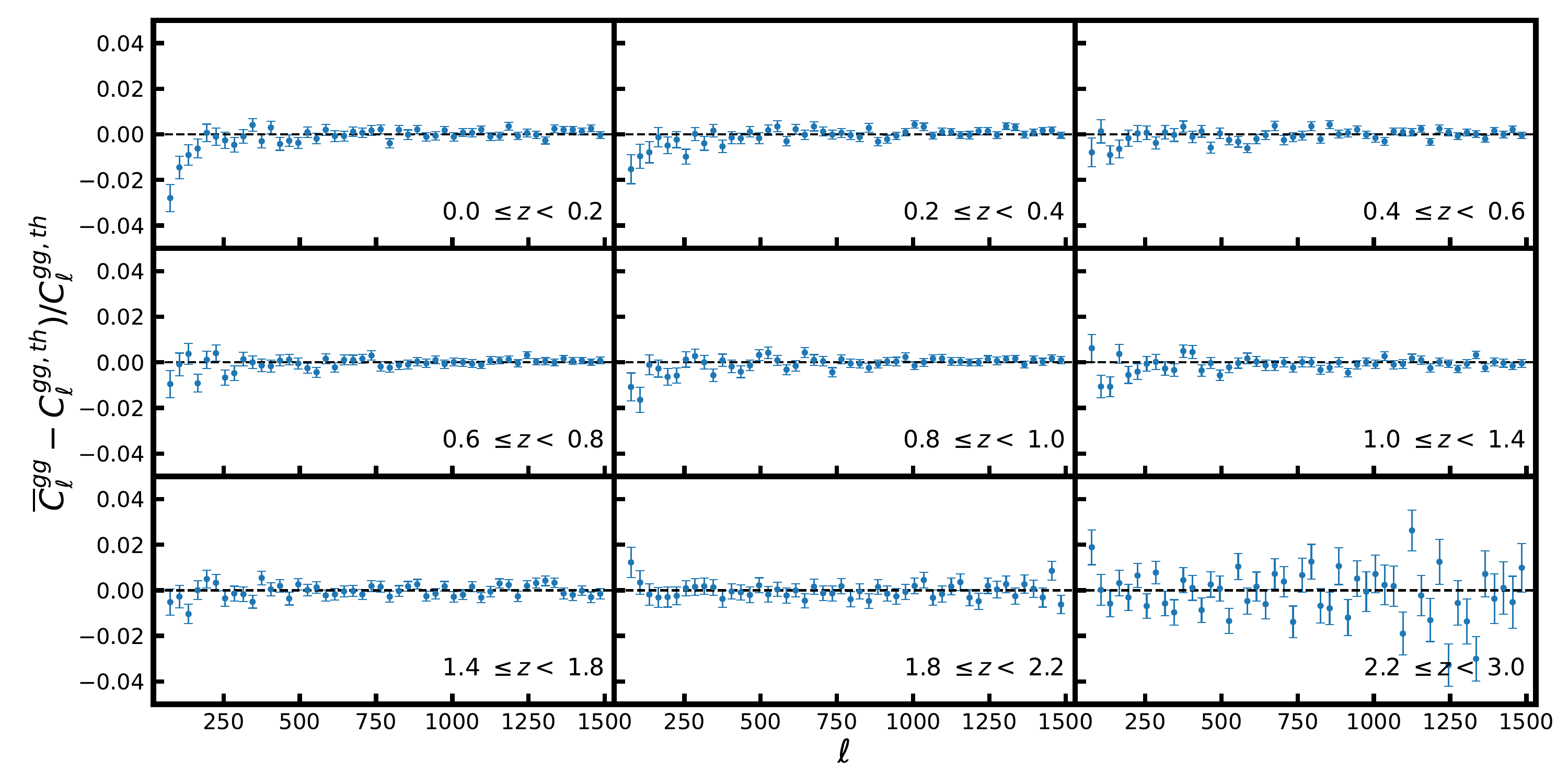}
    \caption{Relative errors on the average galaxy auto-power spectrum reconstructed from $300$ simulations generated by the code \texttt{FLASK} without adding photometric redshift errors. The error bars are computed from the covariance matrix of simulations using Eq.\,(\ref{eq:err_simul}).}
    \label{fig:plot_gg_true_from_flask_nsim_300}
\end{figure*}

\begin{figure*}
    \centering
    \includegraphics[width=\linewidth]{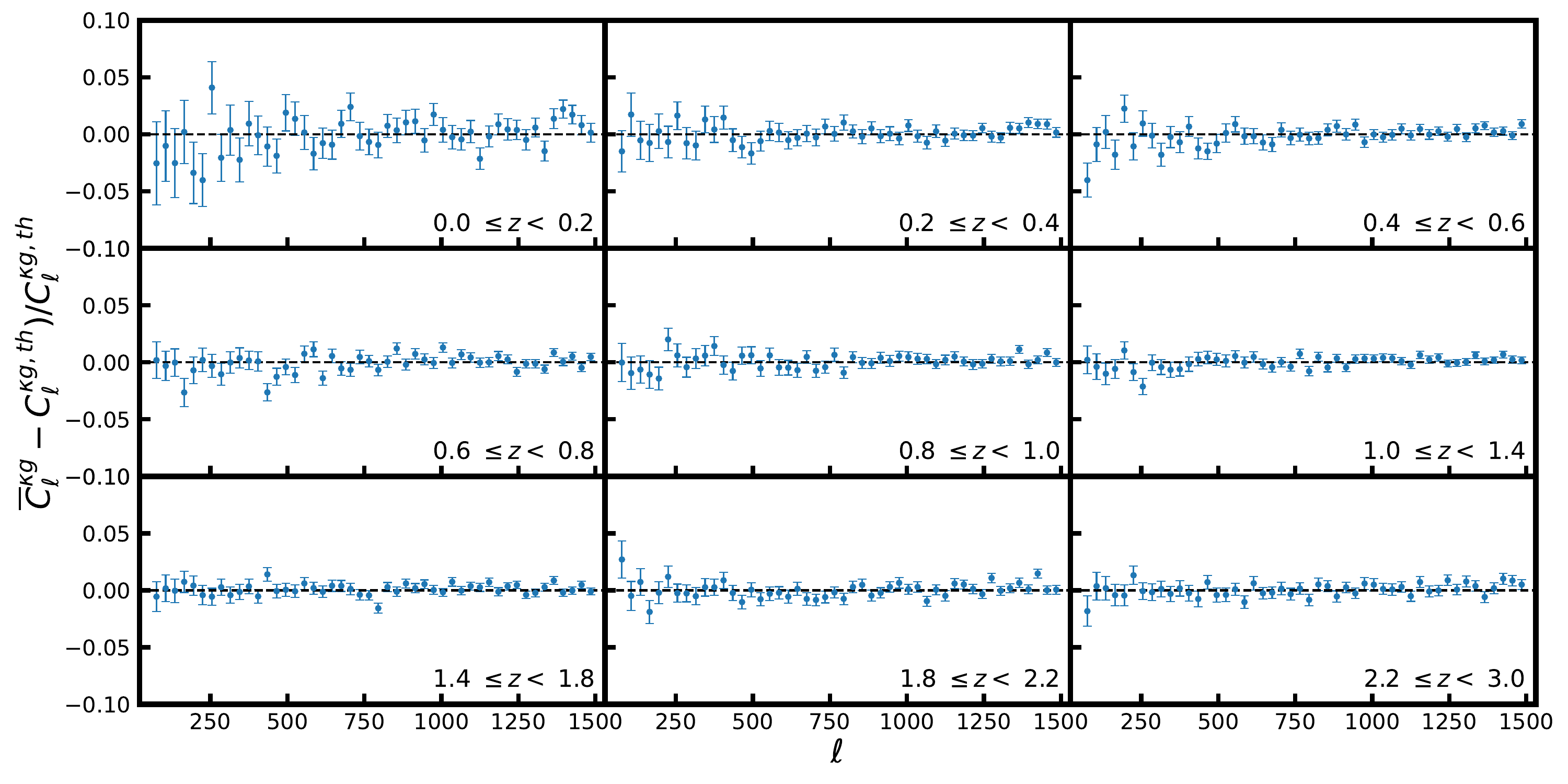}
    \caption{Relative errors on the average cross-power spectrum between galaxy over-density and CMB lensing reconstructed from $300$ simulations without adding photometric redshift errors. The error bars are computed from the covariance matrix of simulations using Eq.\,(\ref{eq:err_simul}).}
    \label{fig:plot_kg_true_from_flask_nsim_300}
\end{figure*}

\begin{figure*}
    \centering
    \includegraphics[width=0.7\linewidth]{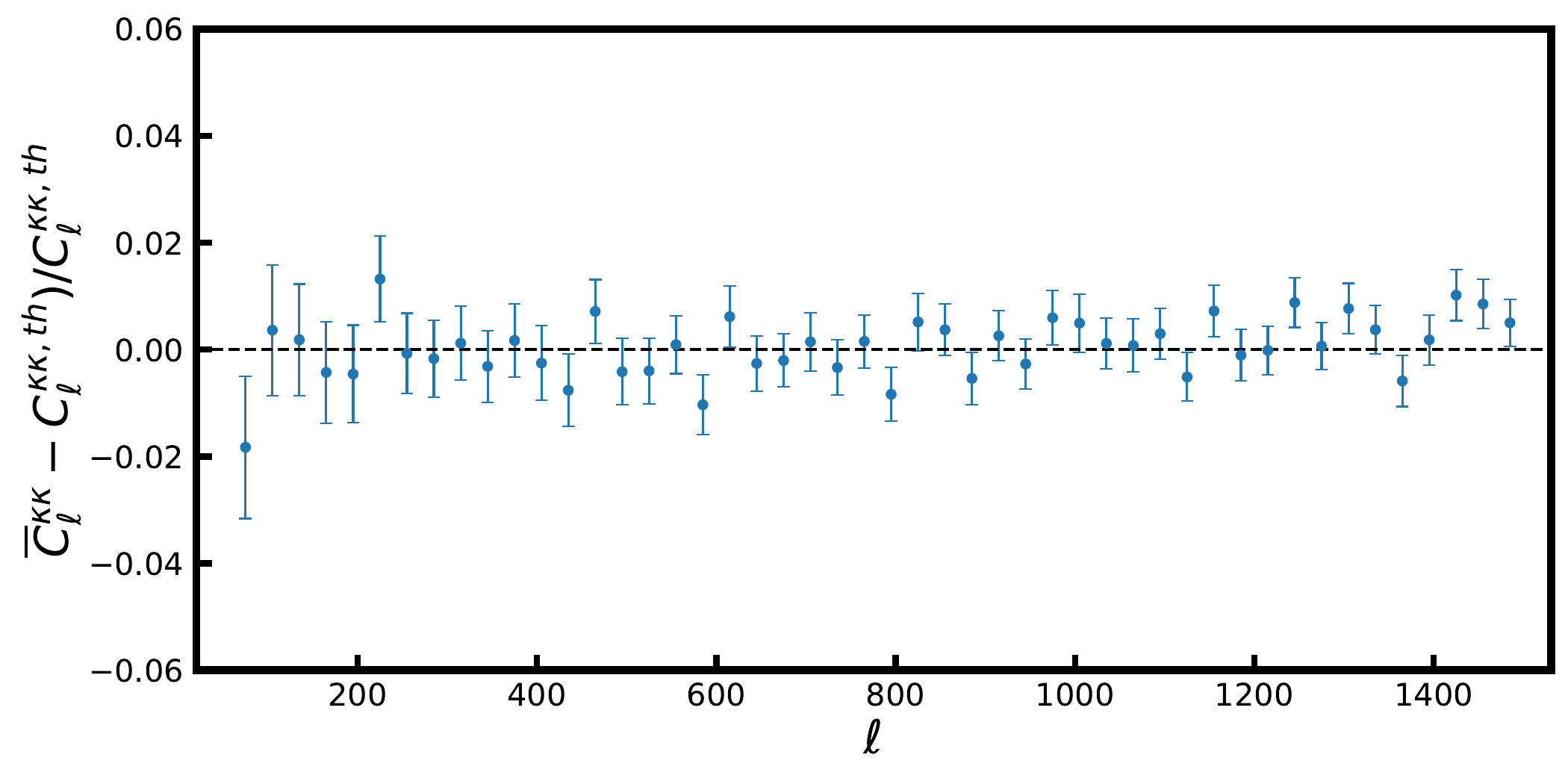}
    \caption{Relative errors on the average CMB convergence auto-power spectrum reconstructed from $300$ simulations. The error bars are computed from the covariance matrix of simulations using Eq.\,(\ref{eq:err_simul}).}
    \label{fig:plot_kk_true_from_flask_nsim_300}
\end{figure*}

~\newpage
~\newpage
\section{Power spectra from photometric datasets}\label{sec_apndx:power_spectra_true_err_conv}

\begin{figure*}[h]
    \centering
    \includegraphics[width=\linewidth]{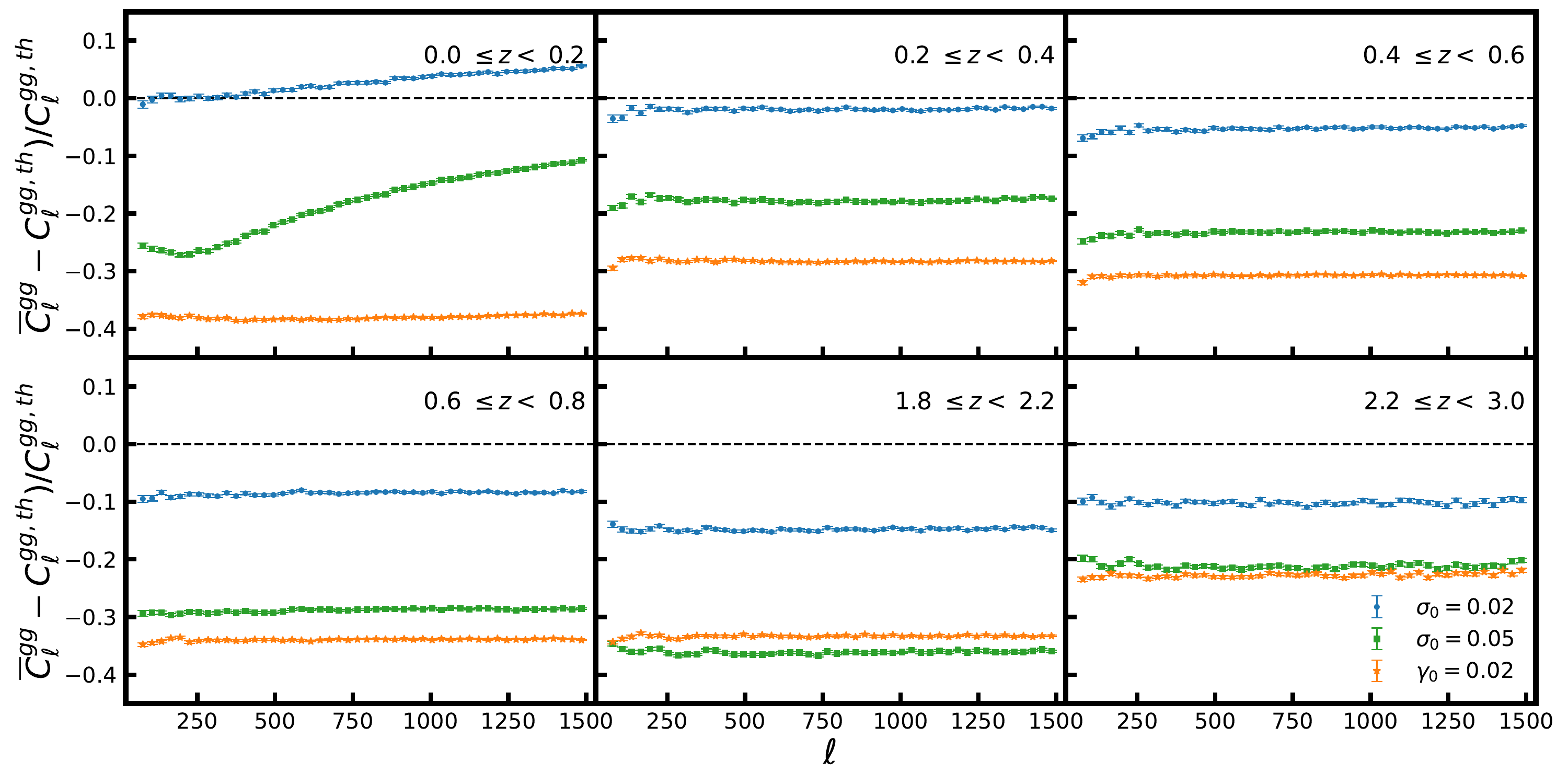}
    \caption{{Average galaxy auto-power spectrum reconstructed from $300$ simulations after adding photometric redshift errors (complementary to Fig. \ref{fig:plot_photo_from_true_conv_err_nsim_300}). The error bars are computed from the covariance matrix of simulations using Eq.\,(\ref{eq:err_simul}).}}
    \label{fig_apndx:plot_gg_photo_auto_bin_from_true_conv_err_nsim_300}
\end{figure*}

\begin{figure*}
    \centering
    \includegraphics[width=\linewidth]{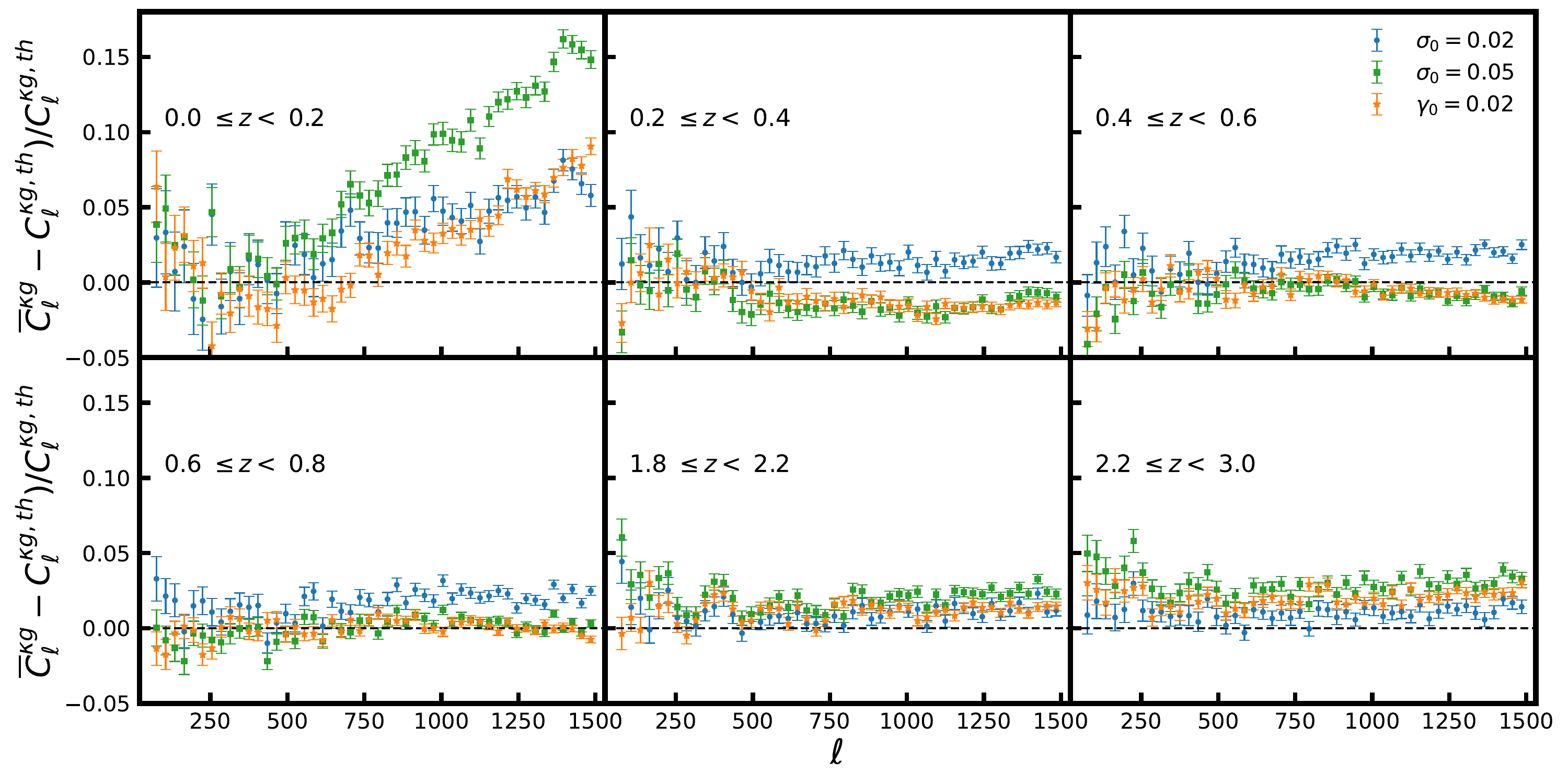}
    \caption{{Average cross-power spectrum between galaxy over-density and CMB convergence reconstructed from $300$ simulations after adding photometric redshift errors (complementary to Fig. \ref{fig:plot_photo_from_true_conv_err_nsim_300}). The error bars are computed from the covariance matrix of simulations using Eq.\,(\ref{eq:err_simul}).}}
    \label{fig_apndx:plot_kg_photo_from_true_conv_err_nsim_300}
\end{figure*}

~\newpage
\section{Power spectra after correction for leakage}\label{sec_apndx:power_spectra_scat_mat}

\begin{figure*}[h]
    \centering
    \includegraphics[width=\linewidth]{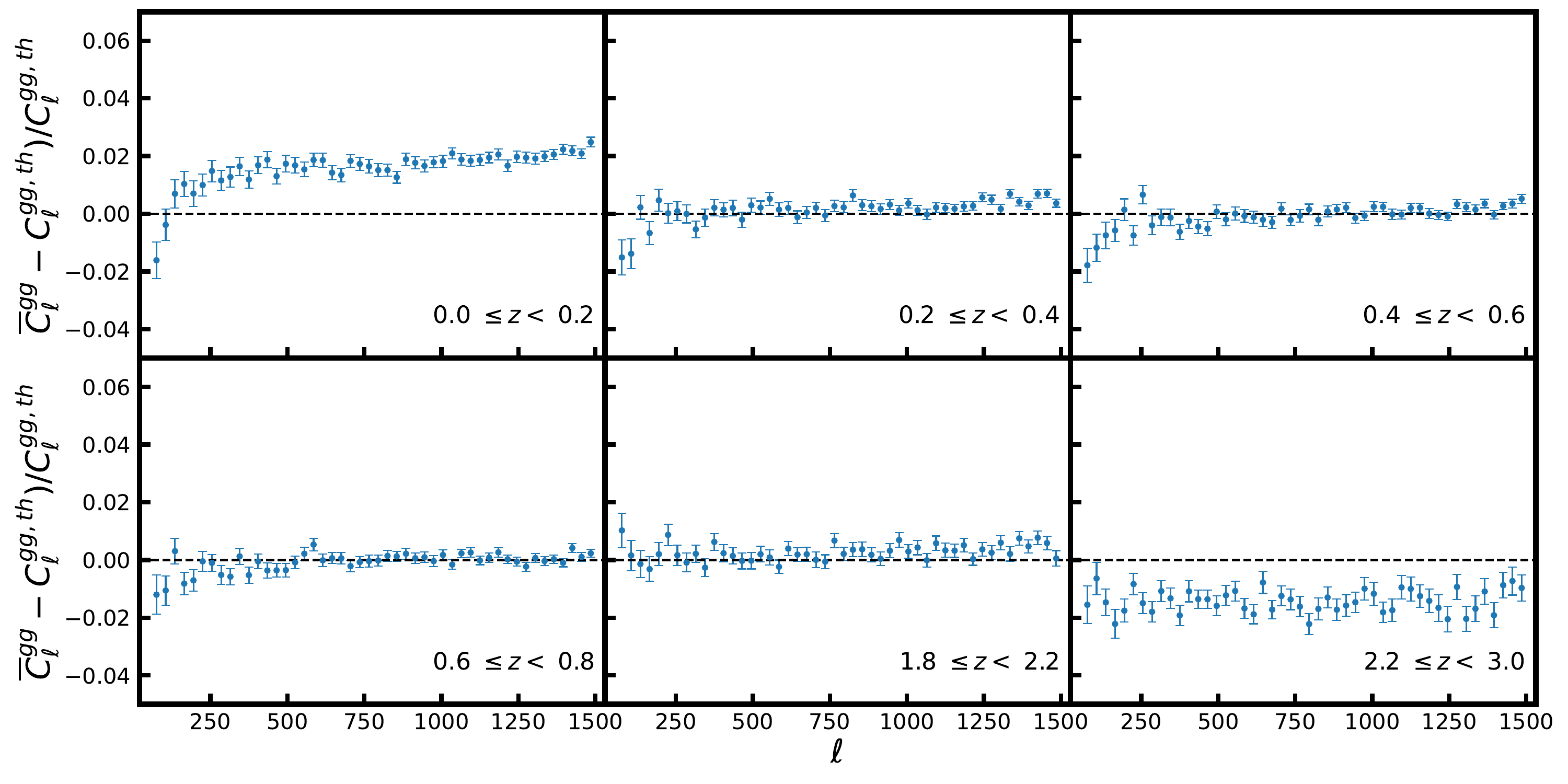}
    \caption{Average galaxy auto-power spectrum reconstructed from $300$ simulations of photometric datasets with $\sigma_{0}=0.02$, computed through the scattering matrix (complementary to Fig. \ref{fig:plot_photo_from_scat_mat_dist_nsim_300_0.02}). The error bars are computed from the covariance matrix of simulations using Eq.\,(\ref{eq:err_simul}).}
    \label{fig_apndx:plot_gg_photo_auto_bin_from_scat_mat_dist_nsim_300_0.02}
\end{figure*}

\begin{figure*}
    \centering
    \includegraphics[width=\linewidth]{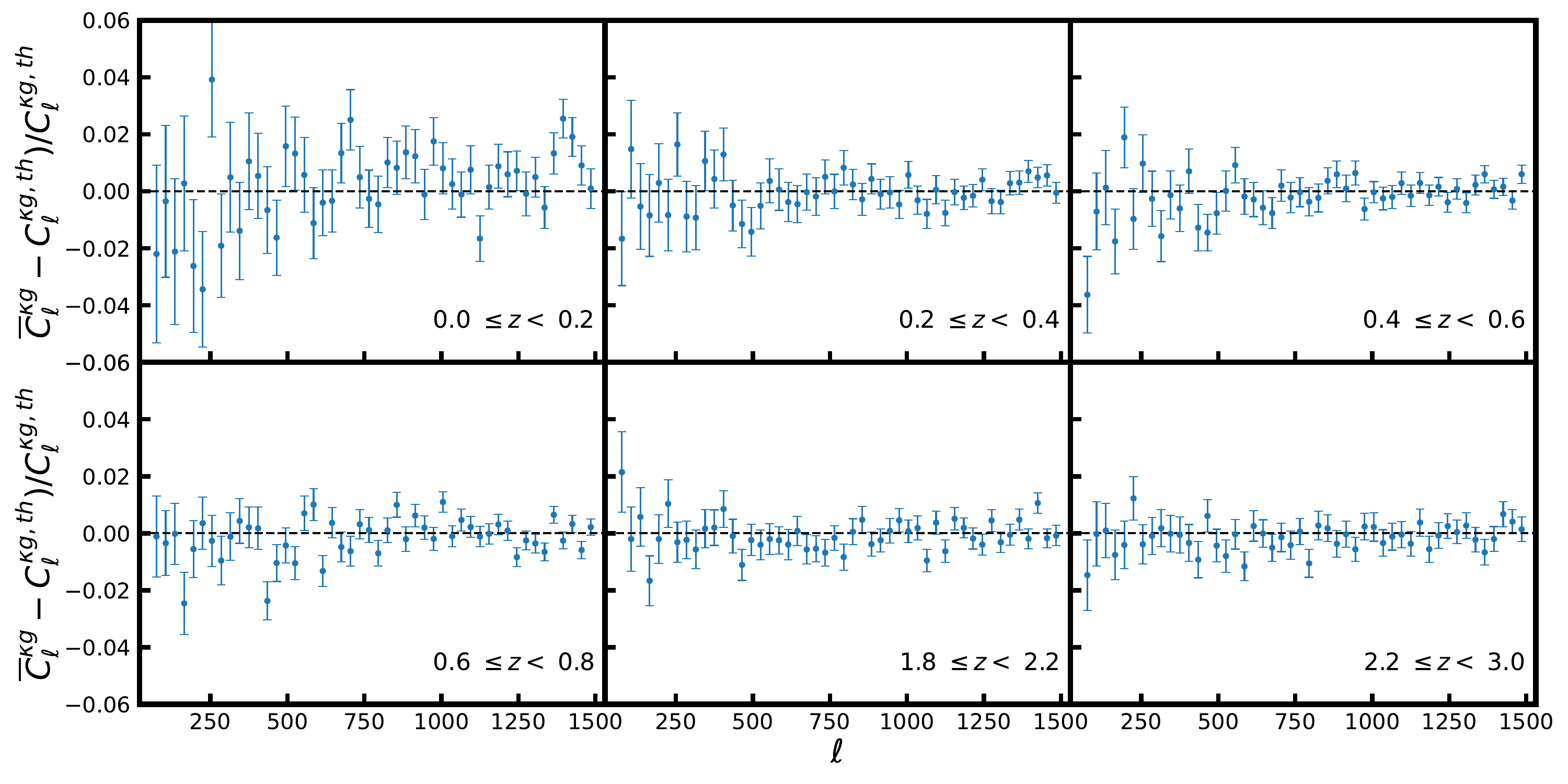}
    \caption{Average cross-power spectrum between galaxy over-density and CMB convergence reconstructed from $300$ simulations of photometric datasets with $\sigma_{0}=0.02$, shown for three tomographic bins computed through the scattering matrix (complementary to Fig. \ref{fig:plot_photo_from_scat_mat_dist_nsim_300_0.02}). The error bars are computed from the covariance matrix of simulations using Eq.\,(\ref{eq:err_simul}).}
    \label{fig_apndx:plot_kg_photo_from_scat_mat_dist_nsim_300_0.02}
\end{figure*}

\begin{figure*}
    \centering
    \includegraphics[width=\linewidth]{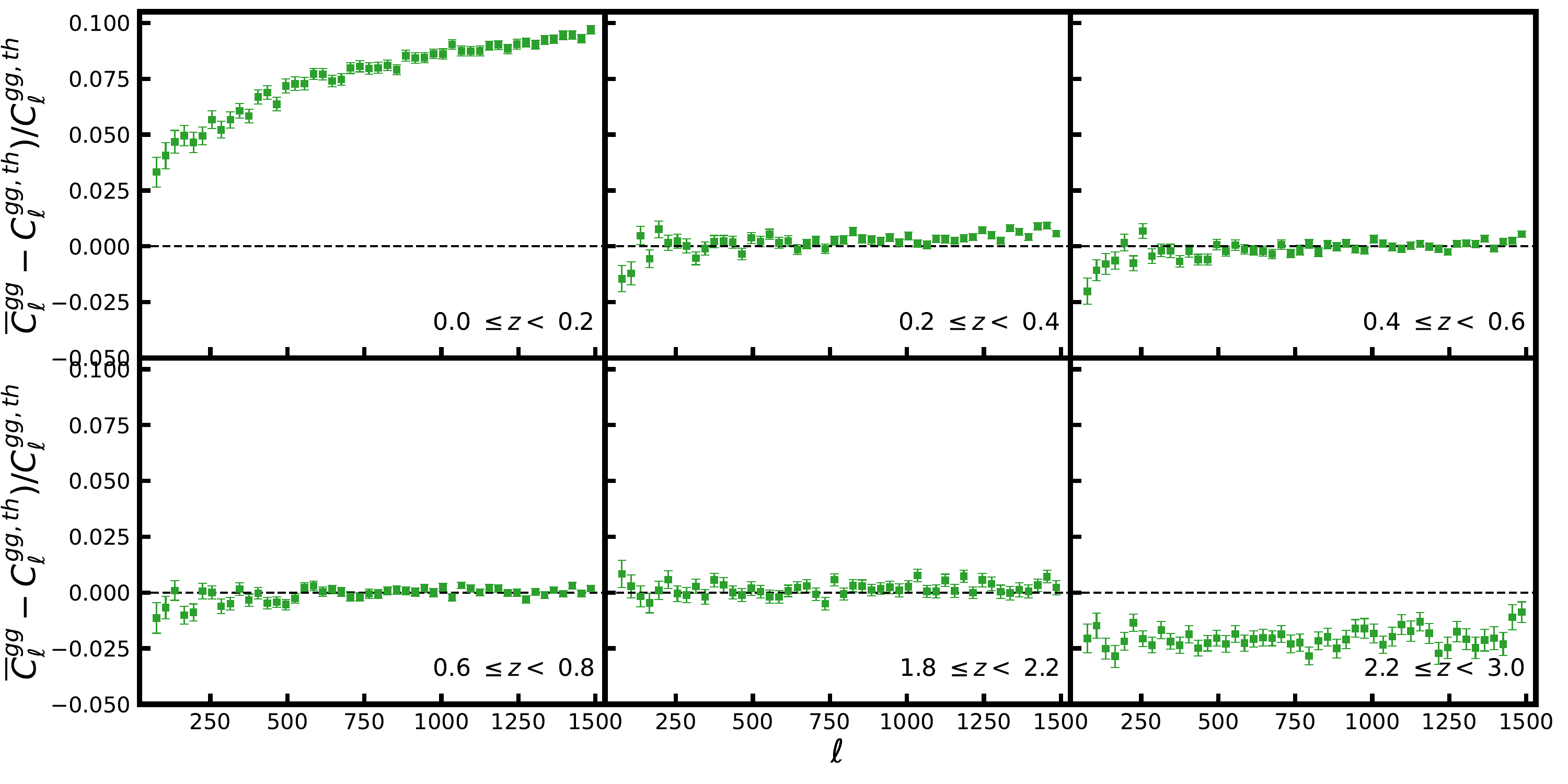}
    \caption{Average galaxy auto-power spectrum reconstructed from $300$ simulations of photometric datasets with $\sigma_{0}=0.05$, computed through the scattering matrix (complementary to Fig. \ref{fig:plot_photo_from_scat_mat_dist_nsim_300_0.05}). The error bars are computed from the covariance matrix of simulations using Eq.\,(\ref{eq:err_simul}).}
    \label{fig_apndx:plot_gg_photo_auto_bin_from_scat_mat_dist_nsim_300_0.05}
\end{figure*}

\begin{figure*}
    \centering
    \includegraphics[width=\linewidth]{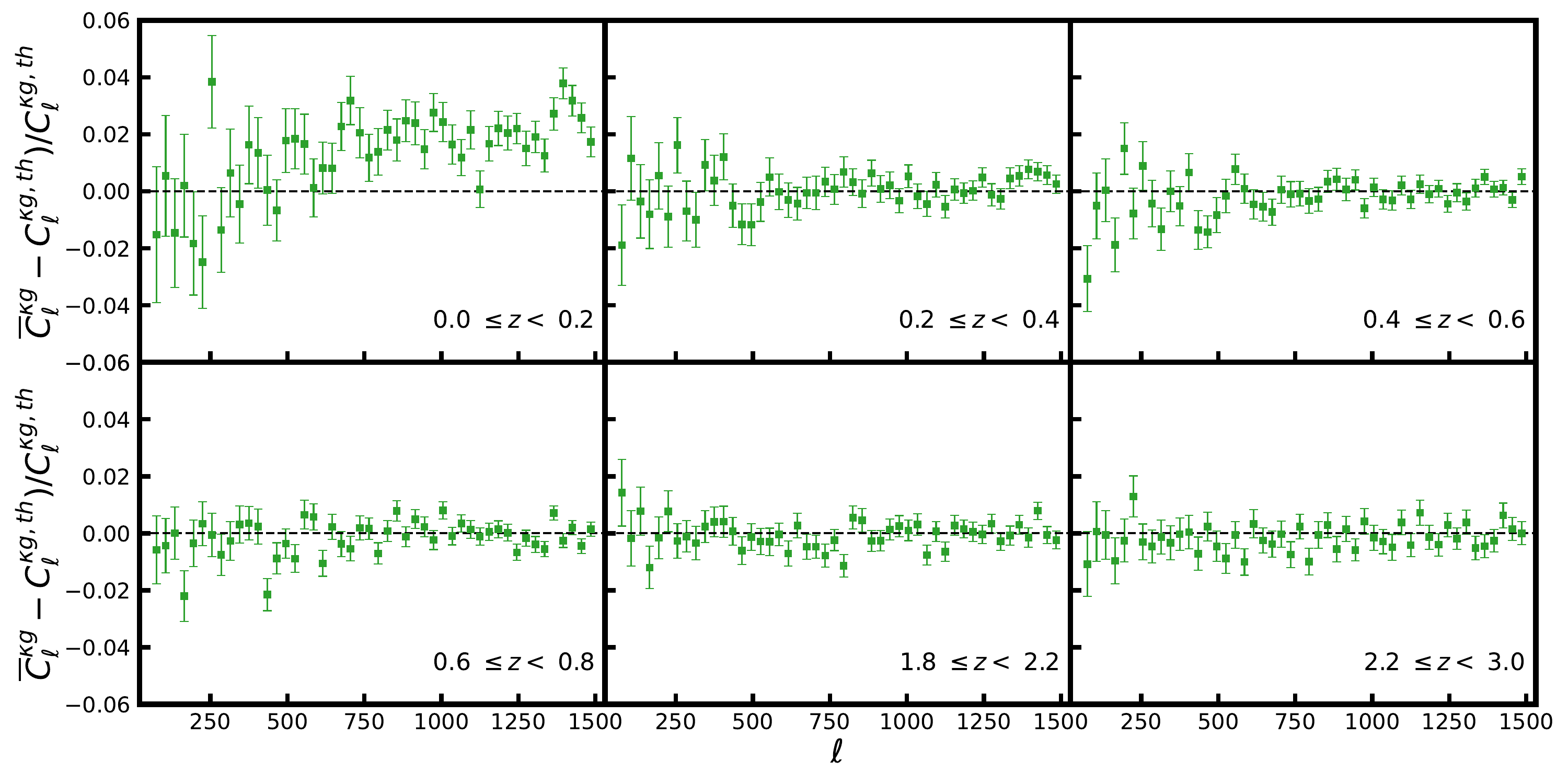}
    \caption{Average cross-power spectrum between galaxy over-density and CMB convergence reconstructed from $300$ simulations of photometric datasets with $\sigma_{0}=0.05$, shown for three tomographic bins computed through the scattering matrix (complementary to Fig. \ref{fig:plot_photo_from_scat_mat_dist_nsim_300_0.05}). The error bars are computed from the covariance matrix of simulations using Eq.\,(\ref{eq:err_simul}).}
    \label{fig_apndx:plot_kg_photo_from_scat_mat_dist_nsim_300_0.05}
\end{figure*}

\begin{figure*}
    \centering
    \includegraphics[width=\linewidth]{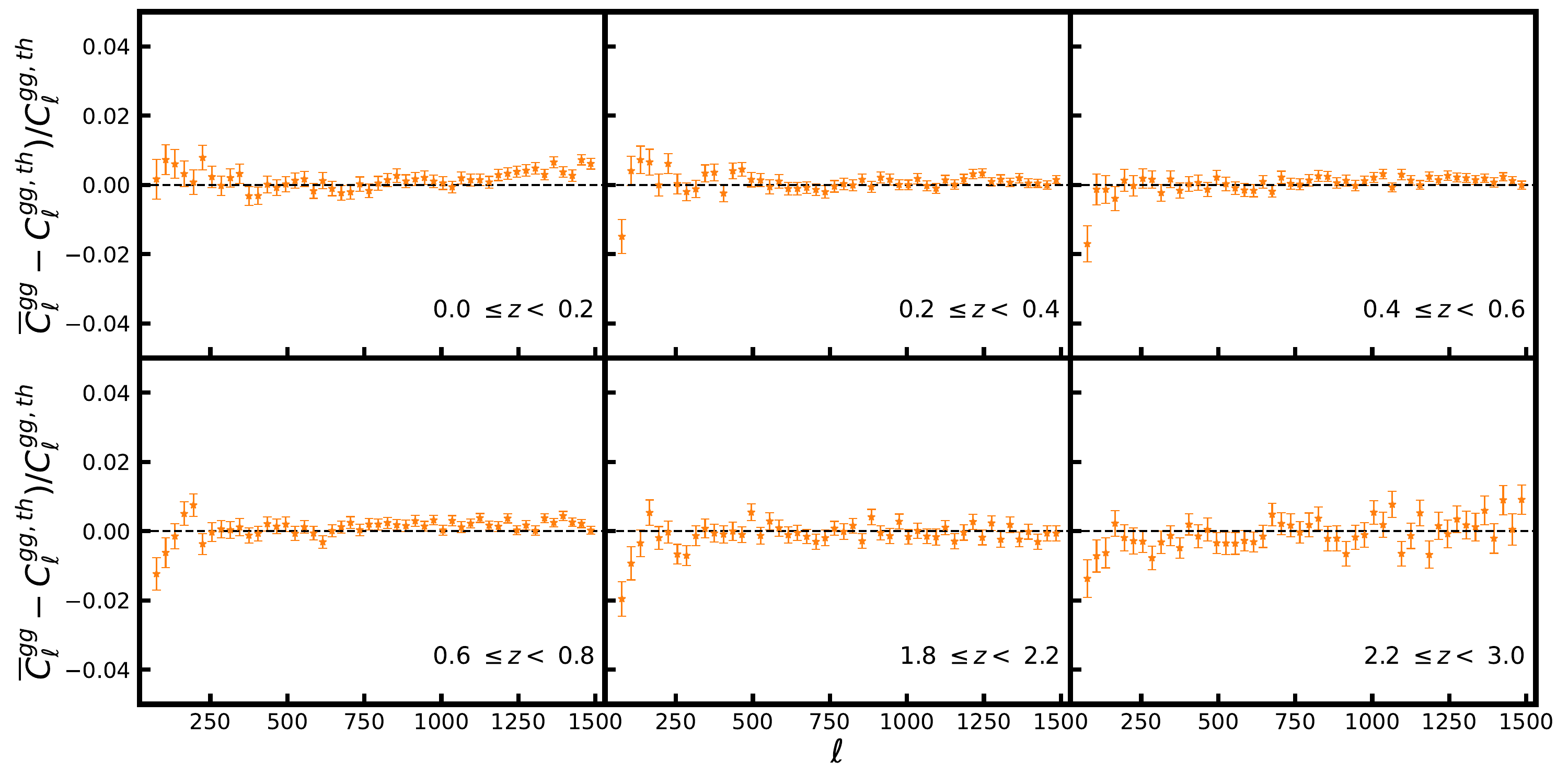}
    \caption{{Average galaxy auto-power spectrum reconstructed from $300$ simulations of photometric datasets with a modified Lorentzian error distribution, computed through the scattering matrix (complementary to Fig. \ref{fig:plot_photo_from_scat_mat_dist_nsim_300_gamma_0.02}). The error bars are computed from the covariance matrix of simulations using Eq.\,(\ref{eq:err_simul}).}}
    \label{fig_apndx:plot_gg_photo_auto_bin_from_scat_mat_dist_nsim_300_gamma_0.02}
\end{figure*}

\begin{figure*}
    \centering
    \includegraphics[width=\linewidth]{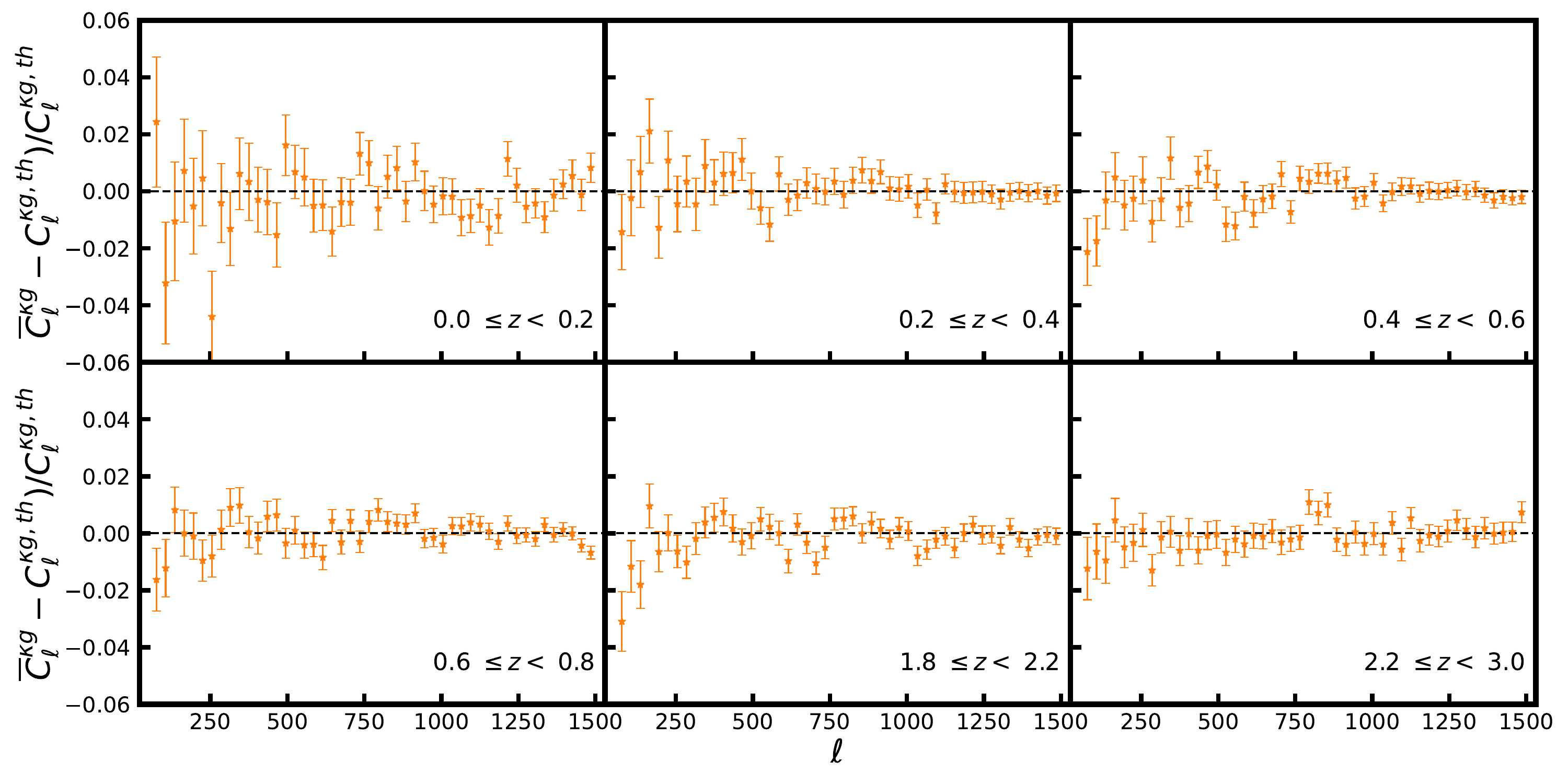}
    \caption{{Average cross-power spectrum between galaxy over-density and CMB convergence reconstructed from $300$ simulations of photometric datasets with a modified Lorentzian error distribution, shown for three tomographic bins computed through the scattering matrix (complementary to Fig. \ref{fig:plot_photo_from_scat_mat_dist_nsim_300_gamma_0.02}). The error bars are computed from the covariance matrix of simulations using Eq.\,(\ref{eq:err_simul}).}}
    \label{fig_apndx:plot_kg_photo_from_scat_mat_dist_nsim_300_gamma_0.02}
\end{figure*}
\end{appendix}

\end{document}